\documentclass[11pt]{article}
\usepackage{jheppub}
\usepackage{amsmath,amssymb,amsfonts,graphicx,slashed,amsthm,mathtools,upgreek, enumerate, tensor}
\usepackage[dvipsnames]{xcolor}
\usepackage{arydshln}

\usepackage{booktabs}

\usepackage{float}

\usepackage{tikz}
\usetikzlibrary{calc,arrows,decorations.pathreplacing}
\usepackage{caption}

\usepackage{comment}
\usepackage{hyperref}
\usepackage[utf8]{inputenc}
\usepackage[titletoc]{appendix}

\usepackage{cleveref}

\usepackage{placeins}

\usepackage[normalem]{ulem}

\DeclareMathOperator{\arcosh}{arcosh}

\usepackage{xparse}

\NewDocumentCommand{\GM}{o}{%
  \mathrm{GM}\IfValueT{#1}{^{(#1)}}%
}

\usepackage{braket}
\usepackage{bm}
\usepackage{cancel}
\usepackage{multirow}

\usepackage{subcaption}
\numberwithin{equation}{section}

\allowdisplaybreaks

\title{The Junction Law for Multipartite Entanglement in Confining Holographic Backgrounds}

\author[a, b]{Norihiro Iizuka}
 
\author[b]{Akihiro Miyata}

\affiliation[a]{\it Department of Physics, National Tsing Hua University, Hsinchu 300044, Taiwan}

\affiliation[b]{\it Yukawa Institute for Theoretical Physics, Kyoto University, Kyoto 606-8502, Japan}

\emailAdd{iizuka@phys.nthu.edu.tw}
\emailAdd{akihiro.miyata@yukawa.kyoto-u.ac.jp}

\abstract{We investigate how the junction law for multipartite entanglement is realized in confining holographic backgrounds, using genuine multi-entropy (GM) as our main diagnostic. We first study an AdS$_3$ hard-wall toy model as an analytic benchmark, where multi-way cuts and junction geometries can be analyzed explicitly. In this setup, we classify the relevant saddles, determine the dominant phases, and show that the genuinely multipartite contribution diagnosed by GM is localized near the junction. We also examine how this structure depends on subsystem sizes, asymmetry, and the confinement scale, including phase transitions between competing saddles. We then move beyond the hard-wall benchmark to smooth confining geometries, focusing on the D4-soliton and D3-soliton backgrounds and formulating the corresponding framework also for the Klebanov--Strassler background. In the smooth-cap examples, we find that the junction picture persists, while the detailed phase structure differs from the hard-wall case: in particular, the hard-wall plateau does not survive, and GM instead decreases monotonically and vanishes at a finite critical scale. We also find that the short-distance behavior is background-dependent, with $\GM^{(3)}\sim L^{-4}$ in the D4-soliton background, $\GM^{(3)}\sim L^{-2}$ in the D3-soliton background, and $\GM^{(3)}\sim L^{-2}\cdot (\log L)^{2}$ in the Klebanov--Strassler background. 
These results clarify which features of the junction-law picture are robust in confining holography and which features of the phase structure and short-distance scaling are background-dependent.}
\keywords{}

\preprint{}

\begin{document}

\maketitle

\parskip=10pt%To change the spacing between paragraphs. 

\section{Introduction}

Entanglement entropy has become one of the most useful probes of holographic geometry \cite{Ryu:2006bv,Ryu:2006ef,Hubeny:2007xt}. 
In confining backgrounds, in particular, the competition between connected and disconnected extremal surfaces provides a sharp diagnostic of how infrared structure is encoded in entanglement \cite{Klebanov:2007ws}. 
This viewpoint has been especially fruitful in the bipartite setting, where strip entanglement entropy distinguishes geometries with a smooth cap from those with an effectively sharp infrared cutoff.

However, bipartite entanglement is not sufficient to characterize the full structure of multipartite correlations. In holography, several recent developments have made this limitation increasingly clear \cite{Akers:2019gcv,Hayden:2021gno,Iizuka:2025ioc}. This motivates the search for diagnostics that are intrinsically sensitive to multipartite entanglement, rather than to combinations of pairwise correlations alone.

A natural quantity for this purpose is genuine multi-entropy (GM), introduced in \cite{Iizuka:2025ioc,Iizuka:2025caq}. 
The basic idea is to start from the multi-entropy
$S^{(\mathtt{q})}(A_1\!:\!A_2\!:\!\cdots\!:\!A_{\mathtt{q}})$,
introduced in \cite{Gadde:2022cqi,Penington:2022dhr,Gadde:2023zzj,Gadde:2023zni}, and subtract the contributions attributable to lower-partite entanglement. 
In the tripartite ($\mathtt{q}=3$) case, this gives
\begin{equation}
\GM^{(3)}(A:B:C)
=
S^{(3)}(A:B:C)
-\frac{1}{2}\Bigl(S(A)+S(B)+S(C)\Bigr),
\label{eq:GM3}
\end{equation}
which isolates the genuinely tripartite part of the entanglement structure. 
Related combinations of this form have also appeared in field-theoretic discussions of the tripartite case \cite{Harper:2024ker,Liu:2024ulq}. 
In our earlier analysis, this quantity was shown to be naturally associated with a junction law: the essential multipartite contribution is localized near a bulk junction in a gapped system \cite{Iizuka:2026qqg}. 
Recently, the mathematical structure underlying the construction of genuine multi-entropy was clarified in terms of Möbius inversion on the partition lattice~\cite{Gadde:2026msg}.

The purpose of the present paper is to investigate how this picture is realized in confining holographic backgrounds. 
More specifically, we ask to what extent the junction-localized genuinely multipartite contribution persists once one moves away from idealized toy models and into geometries with more realistic infrared structure. Here, by a junction we mean the region where the different subsystems meet, or more generally coexist in close proximity \cite{Iizuka:2026qqg}.
This leads to two closely related questions: what aspects of the junction law are universal across confining backgrounds, and which features depend sensitively on the detailed way in which the infrared region ends.

From the point of view of holographic entanglement, this is a particularly natural question, because confining geometries come in qualitatively different infrared forms. 
Some are effectively modeled by a sharp hard wall, while others end smoothly through a cigar cap or a deformed-conifold tip. 
Even in the bipartite problem, these differences are already visible in the competition of extremal surfaces \cite{Klebanov:2007ws}. 
It is therefore far from obvious that genuinely multipartite quantities should behave in the same way in all such backgrounds.

To address this problem, we begin with an AdS$_3$ hard-wall model as an analytic benchmark. 
Hard-wall constructions have long been used as simple holographic models of confinement and low-energy QCD \cite{Erlich:2005qh}. 
Our use of the hard wall here is not primarily as a phenomenological model of hadron physics, but rather as a particularly simple holographic toy model for a gapped or confining system. 
This model is simple enough that the relevant multi-way cuts and candidate saddles can be analyzed explicitly, while still retaining the essential infrared competition between connected and wall-assisted configurations. 
In this setting, one can classify the dominant saddles, study the effect of asymmetry, and determine the phase structure of both the tripartite multi-entropy and the genuine multi-entropy. 
In particular, the hard-wall model makes it possible to see explicitly how wall-assisted saddles and partially disconnected saddles modify the behavior of $\GM^{(3)}$. At the same time, it provides an explicit benchmark for identifying which structural features of the junction-law picture persist more generally, and which may instead be specific to a sharp infrared wall.

We then move beyond the hard-wall benchmark and study three smooth confining geometries: the D4-soliton background \cite{Witten:1998zw}, the D3-soliton (AdS-soliton) background \cite{Horowitz:1998ha}, and the Klebanov--Strassler background \cite{Klebanov:2000hb}. 
These provide increasingly nontrivial examples of holographic confinement in which the infrared region ends smoothly rather than abruptly. 
For these backgrounds, the multipartite problem can be reformulated as an effective one-dimensional variational problem for strip-like branches, together with a universal Steiner-type force-balance condition at a symmetric junction. 
This allows us to compare the connected $Y$-type multipartite saddle and the cap-assisted disconnected saddle in a unified way across different confining geometries. 
This in turn allows us to distinguish which structural aspects of the junction-law picture persist, how the detailed phase structure is modified, and which short-distance features remain background-dependent.
These backgrounds have also played an important role in the study of entanglement as a probe of confinement \cite{Klebanov:2007ws}.

Before proceeding, let us briefly comment on several closely related recent developments. 
Recent work has started to uncover new constraints on holographic multipartite entanglement beyond the original definitions of multi-entropy and genuine multi-entropy. 
In particular, it was argued that holographic tripartite entanglement obeys a relation incompatible with purely GHZ-like structure \cite{Balasubramanian:2025hxg}, and a related argument was used to show that tripartite entanglement in the HaPPY code is not holographic \cite{Akella:2025owv}. 
On both the holographic and field-theory sides, multi-entropy, genuine multi-entropy, and related multipartite quantities have also been studied further in \cite{Harper:2025uui,Berthiere:2025toi,Yuan:2025dgx,Ju:2025eyn,Anegawa:2025prn,Chen:2026xtx,DelZotto:2026fpw}. 
These developments are complementary to the present work, whose main focus is the realization of the junction-law picture in confining holographic backgrounds.

The organization of this paper is as follows. 
In Sec.~\ref{GMholographicsetup}, we review the definition of genuine multi-entropy and summarize the holographic setup used throughout the paper. 
In Sec.~\ref{sec:hardwall_benchmark}, we analyze the AdS$_3$ hard-wall model as an analytic benchmark and determine the relevant candidate saddles and phase structure explicitly. 
In Sec.~\ref{sec:smooth_general_framework}, we extract the common variational framework that underlies the smooth confining backgrounds studied later. 
We then study the D4-soliton, D3-soliton, and Klebanov--Strassler backgrounds in Secs.~\ref{sec:D4_soliton}, \ref{sec:D3_soliton}, and \ref{sec:KS_background}. 
Finally, in Sec.~\ref{sec:universal_vs_background_dependent}, we compare the results and discuss which aspects of the junction law are universal and which are background-dependent.

%%%%%%%%%%%%%%%%%%%%%%%%%%%%%%%%%%%

\section{Genuine multi-entropy and holographic setup}
\label{GMholographicsetup}

In this section, we summarize the multipartite entanglement quantities used in this paper and fix the holographic setup.
We begin with a brief review of multi-entropy and then specialize to the genuine tripartite combination that will play the central role below. 
Our purpose here is twofold. 
First, we would like to make clear what quantity is being computed and why its magnitude has a natural quantitative meaning. 
Second, we would like to separate this conceptual definition from the explicit geometric competition among candidate saddles, which will first be introduced in the hard-wall benchmark of Sec.~\ref{sec:hardwall_benchmark}.

\subsection{Multi-entropy}

We begin with the multipartite multi-entropy introduced in \cite{Gadde:2022cqi,Penington:2022dhr,Gadde:2023zzj,Gadde:2023zni}. 
For a pure state $\ket{\psi}$ on $\mathtt{q}$ subsystems $A_1,\dots,A_{\mathtt q}$, the $\mathtt q$-partite R\'enyi multi-entropy is defined by
\begin{align}
S^{(\mathtt q)}_n(A_1:\cdots:A_{\mathtt q})
:=
\frac{1}{1-n}\frac{1}{n^{\mathtt q-2}}
\log
\frac{Z_n^{(\mathtt q)}}{\bigl(Z_1^{(\mathtt q)}\bigr)^{n^{\mathtt q-1}}},
\label{eq:multi_Renyi_def}
\end{align}
where
\begin{align}
Z_n^{(\mathtt q)}
:=
\bra{\psi}^{\otimes n^{\mathtt q-1}}
\Sigma_1(g_1)\Sigma_2(g_2)\cdots\Sigma_{\mathtt q}(g_{\mathtt q})
\ket{\psi}^{\otimes n^{\mathtt q-1}}
\label{eq:multi_partition_function}
\end{align}
is defined by appropriate twist-operator insertions $\Sigma_\mathtt{k}(g_\mathtt{k})$ implementing the replica permutations. 
The corresponding von Neumann-type multi-entropy is obtained by the $n\to1$ limit,
\begin{align}
S^{(\mathtt q)}(A_1:\cdots:A_{\mathtt q})
:=
\lim_{n\to1} S_n^{(\mathtt q)}(A_1:\cdots:A_{\mathtt q}).
\label{eq:multi_vN_def}
\end{align}
Although \eqref{eq:multi_vN_def} is formally defined by the $n\to1$ limit, the required analytic continuation in $n$ is in general nontrivial. 
In particular, while multi-entropy is naturally defined for integer replica number $n$, extending it away from integer $n$ and justifying the limit $n\to1$ is a nontrivial assumption\footnote{This issue has also been emphasized in the holographic context. The proposed holographic dual of R\'enyi multi-entropy is known to be subtle, and explicit counterexamples for the naive R\'enyi-level proposal at integer $n\ge 3$ have been discussed in AdS/CFT \cite{Penington:2022dhr}. It may nevertheless still be the case that the $n\to1$ continuation relevant for multi-entropy is well defined, at least in restricted classes of states such as fixed-area states or tensor-network-like models \cite{Hayden:2016cfa,Pastawski:2015qua,Akers:2018fow,Dong:2018seb}. In the present paper, we will simply assume that the analytic continuation to $n\to1$ exists and that the resulting multi-entropy is well defined.}.

For the present paper, the detailed replica construction will not be needed beyond the following general facts, which make multi-entropy a natural multipartite quantity:
\begin{itemize}
\item it is invariant under local unitary transformations on each subsystem;
\item it is symmetric under permutations of the subsystems;
\item it is additive under tensor products of pure states.
\end{itemize}
In particular, if
\begin{equation}
\ket{\psi}=\ket{\psi_A}\otimes\ket{\psi_B},
\end{equation}
with the total Hilbert space partitioned as
\begin{equation}
H_{A_1B_1\,A_2B_2\,\cdots\,A_{\mathtt q}B_{\mathtt q}}
=
H^{\psi_A}_{A_1A_2\cdots A_{\mathtt q}}
\otimes
H^{\psi_B}_{B_1B_2\cdots B_{\mathtt q}},
\end{equation}
then
\begin{align}
S_n^{(\mathtt q)}(A_1B_1:\cdots:A_{\mathtt q}B_{\mathtt q})_{\ket\psi}
=
S_n^{(\mathtt q)}(A_1:\cdots:A_{\mathtt q})_{\ket{\psi_A}}
+
S_n^{(\mathtt q)}(B_1:\cdots:B_{\mathtt q})_{\ket{\psi_B}}.
\label{eq:multi_additivity}
\end{align}
The same additivity therefore also holds after taking the $n\to1$ limit.

This additivity is important conceptually. 
It means that when independent multipartite-entangled resources are combined by tensor product, the multi-entropy increases additively. 
Thus its magnitude is not merely a yes/no diagnostic, but has a genuine quantitative meaning.

At the same time, $S^{(\mathtt q)}$ by itself is not yet a clean measure of \emph{genuine} $\mathtt q$-partite entanglement, because it is also sensitive to lower-partite structures. 
This naturally leads to the genuine multi-entropy.

\subsection{Genuine multi-entropy}

The role of genuine multi-entropy is to isolate the irreducibly multipartite part of the multi-entropy by subtracting contributions that can already be attributed to lower-partite entanglement. 
In the general formalism of \cite{Iizuka:2025ioc}, this is achieved by taking appropriate symmetric linear combinations of multi-entropies associated with different partitions. 
The resulting quantity vanishes for states that factorize across a nontrivial partition and, because it is built linearly from additive quantities, it is itself additive under tensor products.

In the present paper we will only need the tripartite case. 
For $\mathtt q=3$, the genuine R\'enyi multi-entropy is
\begin{align}
\GM^{(3)}_n(A:B:C)
=
S^{(3)}_n(A:B:C)
-\frac12\Bigl(
S^{(2)}_n(AB:C)+S^{(2)}_n(AC:B)+S^{(2)}_n(BC:A)
\Bigr),
\label{eq:GM3_Renyi_definition}
\end{align}
and the corresponding von Neumann-limit quantity is
\begin{align}
\GM^{(3)}(A:B:C)
:=
\lim_{n\to1}\GM^{(3)}_n(A:B:C).
\label{eq:GM3_vN_limit}
\end{align}
Since for a pure state one has
\begin{equation}
S^{(2)}(AB:C)=S(C),\qquad
S^{(2)}(AC:B)=S(B),\qquad
S^{(2)}(BC:A)=S(A),
\end{equation}
this reduces to
\begin{equation}
\GM^{(3)}(A:B:C)
=
S^{(3)}(A:B:C)
-\frac12\Bigl(S(A)+S(B)+S(C)\Bigr).
\label{eq:GM3_definition}
\end{equation}

This is the basic quantity studied throughout the present paper. 
Its significance is twofold. 
First, it removes purely bipartite contributions from the tripartite multi-entropy. 
Second, because it is a linear combination of additive quantities, it also remains additive under tensor products.

This interpretation is especially natural in holography. 
The multipartite multi-entropy $S^{(\mathtt q)}$ is conjecturally dual to the area of a minimal $\mathtt q$-way cut \cite{Gadde:2022cqi,Gadde:2023zzj}, while ordinary entanglement entropies are given by the usual RT/HRT surfaces. 
For $\mathtt q=3$, the combination \eqref{eq:GM3_definition} therefore compares a genuine tripartite bulk cut against the sum of ordinary bipartite contributions. 
In our previous work \cite{Iizuka:2025ioc}, this quantity was shown, in holographic settings, to isolate a junction-localized contribution and accordingly to be non-negative. 
The main question of the present paper is how that junction-localized genuinely tripartite contribution behaves in confining holographic backgrounds.

\subsection{Boundary setup and holographic ingredients}

Our main configuration throughout this paper is the $BC$-symmetric tripartition
\begin{equation}
A=\left[-\frac{L}{2},\frac{L}{2}\right],\qquad
B=\left(-\infty,-\frac{L}{2}\right),\qquad
C=\left(\frac{L}{2},\infty\right).
\label{eq:tripartition_BCsymmetric}
\end{equation}
Here $A$ is a finite interval (or strip), while $B$ and $C$ are semi-infinite. 
This is the configuration that will be used for the comparison across the hard-wall benchmark, the D4-soliton background, the D3-soliton background, and the Klebanov--Strassler background.

This $BC$-symmetric configuration may also be viewed as a limiting case of the more general $AB$-asymmetric tripartition in the hard-wall model,
\begin{equation}
A=\{-L_A<x<0\},\qquad
B=\{0<x<L_B\},\qquad
C=\{x<-L_A,\;x>L_B\},
\label{eq:tripartition_asymmetric}
\end{equation}
by taking one interval much larger than the other. 
More generally, the asymmetric configuration is useful for exposing the full multipartite phase structure in the hard-wall benchmark. 
In the AdS$_3$ hard-wall model, both \eqref{eq:tripartition_BCsymmetric} and \eqref{eq:tripartition_asymmetric} can be analyzed explicitly. 
By contrast, in the D4-soliton, D3-soliton, and Klebanov--Strassler backgrounds, we will mainly focus on the symmetric configuration \eqref{eq:tripartition_BCsymmetric}.

To construct \eqref{eq:GM3_definition}, one must evaluate two logically distinct quantities.

First, the tripartite multi-entropy $S^{(3)}(A:B:C)$ is obtained holographically by minimizing an appropriate tripartite bulk cut that separates the three boundary subsystems. 
In the simplest cases, this minimization is realized by a Steiner-type network with a bulk junction, although in confining backgrounds it can also compete with partially disconnected or cap-assisted configurations. 
This is the multipartite ingredient in the construction of $\GM^{(3)}$.

Second, one must compute the ordinary entanglement entropies $S(A)$, $S(B)$, and $S(C)$ from the usual extremal surfaces associated with the individual regions. 
For the $BC$-symmetric setup, the semi-infinite regions $B$ and $C$ are represented by the standard disconnected cap-reaching surfaces.

Thus, $\GM^{(3)}$ is governed by two distinct minimization problems:
\begin{enumerate}
\item the minimization defining the tripartite multi-entropy $S^{(3)}(A:B:C)$,
\item the minimization defining the ordinary entanglement entropy of the finite region $A$.
\end{enumerate}

This distinction is important because the transition scale in the multipartite sector does not in general coincide with that in the bipartite sector. 
As a result, the behavior of $\GM^{(3)}$ is controlled not by a single transition, but by the interplay of these two minimization problems.

\subsection{Hard wall versus smooth confining caps}

A central theme of this paper is the contrast between a sharp infrared cutoff and a smooth confining cap. 
Our strategy is to use the hard-wall model as an analytic benchmark and then compare it with several standard smooth confining geometries that realize confinement in qualitatively different ways.

In the AdS$_3$ hard-wall model, the geometry terminates abruptly at $z=z_0$. 
Precisely because the infrared cutoff is sharp, the multipartite competition becomes especially transparent: the relevant saddles can be visualized directly as geodesic networks, their lengths can be written explicitly, and the associated transition loci can be analyzed essentially in closed form. 
In this sense, the hard-wall benchmark is useful not so much for realism as for clarity.

By contrast, in smooth confining backgrounds the infrared region ends without a sharp wall. 
Instead, the bulk develops a smooth cap, and the multipartite problem is more naturally reformulated as an effective one-dimensional variational problem for strip-like branches in an appropriate optical metric. 
This is the setting relevant for the D4-soliton \cite{Witten:1998zw}, the D3-soliton (AdS-soliton) background \cite{Horowitz:1998ha}, and the Klebanov--Strassler background \cite{Klebanov:2000hb}.

The main role of the hard-wall benchmark is therefore both conceptual and technical: it provides the clearest setting in which the multipartite candidates can first be identified explicitly and in which the full phase structure can be seen analytically. 
The smooth confining backgrounds can then be viewed as more realistic settings in which the same basic multipartite minimization logic survives, while the detailed behavior of $\GM^{(3)}$ and the associated phase structure can change.

%%%%%%%%%%%%%%%%%%%%%%%%%%%%%%%%%%%%%%%%%%%%%%%%%%%%

%%%%%%%%%%%%%%%

%%%%%%%%%%%%%%%%%%%%%%%%%%%%%%%%%%%%%%%%%%%%%%%%%%%%

\section{Hard-wall benchmark}
\label{sec:hardwall_benchmark}

In this section, we study the AdS$_3$ hard-wall model as an analytic benchmark for the multipartite junction problem. 
Our purpose is twofold. 
First, this model provides a fully tractable setting in which the relevant multipartite candidate cuts can be written explicitly and compared analytically. 
Second, it gives a clean reference point for the later analysis in smooth confining geometries, where the same qualitative competition among saddles persists while the detailed calculations become more involved.

The hard-wall model is also the natural place to introduce the competing multipartite configurations themselves. 
Unlike in the smooth backgrounds, here they can be visualized directly as geodesic networks on a constant-time hyperbolic slice. 
For that reason, we first focus on the $BC$-symmetric tripartition, which is the direct hard-wall analogue of the later confining setups, and only afterward briefly discuss the more general asymmetric case.

\subsection{Geometry and basic formulas}

We work on a constant-time slice of AdS$_3$ with a hard IR cutoff at $z=z_0$:
\begin{equation}
ds^2=\frac{dx^2+dz^2}{z^2},
\qquad
z\in[\varepsilon,z_0],
\label{eq:hw_metric_main}
\end{equation}
where $\varepsilon$ is the UV cutoff and $z_0$ is the hard-wall scale. 
The AdS radius is set to one. 
The spatial slice is the upper half-plane model of $H^2$.

For two bulk points
\begin{equation}
P=(x,z),
\qquad
Q=(x',z'),
\end{equation}
the hyperbolic distance is
\begin{equation}
d(P,Q)
=
\arcosh\!\left(
1+\frac{(x-x')^2+(z-z')^2}{2zz'}
\right) = 
\arcosh\!\left(
\frac{(x-x')^2+z^2+z'^2}{2zz'}
\right),
\label{eq:hw_distance_formula_main}
\end{equation}
and in particular, for a vertical segment with $x=x'$,
\begin{equation}
d(P,Q)=\left|\log\frac{z'}{z}\right|.
\label{eq:hw_vertical_distance_main}
\end{equation}
We also recall that geodesics in the upper half-plane are semicircles orthogonal to the boundary $z=0$, with vertical lines as a limiting case.

\subsection{Warm-up: bipartite geodesics and the role of the hard wall}

Before turning to multipartite cuts, let us briefly review the corresponding bipartite structure. 
Consider a boundary interval $[a,b]$ with endpoints regulated at
\begin{equation}
P=(a,\varepsilon),
\qquad
Q=(b,\varepsilon),
\qquad a<b.
\end{equation}
In pure AdS$_3$, the geodesic connecting them is the semicircle
\begin{equation}
(x-x_0)^2+z^2=R^2,
\qquad
x_0=\frac{a+b}{2},
\qquad
R=\frac{b-a}{2},
\label{eq:hw_semicircle_main}
\end{equation}
whose maximal depth is
\begin{equation}
z_{\max}=\frac{b-a}{2}.
\end{equation}
Its regulated length is
\begin{equation}
\ell_{\rm AdS}(a,b)
=
2\log\frac{b-a}{\varepsilon}.
\label{eq:hw_ads_length_main}
\end{equation}

In the hard-wall geometry, this connected geodesic is admissible only when it remains within the allowed region, namely
\begin{equation}
b-a<2z_0.
\end{equation}
Thus in the shallow regime,
\begin{equation}
\ell_{\rm shallow}(a,b)=2\log\frac{b-a}{\varepsilon}.
\label{eq:hw_shallow_main}
\end{equation}

When
\begin{equation}
b-a\ge 2z_0,
\end{equation}
the unconstrained semicircle would penetrate beyond the wall. 
A natural competing configuration is then the wall-assisted one consisting of two vertical segments down to the wall, with total length
\begin{equation}
\ell_{\rm wall}(a,b)=2\log\frac{z_0}{\varepsilon}.
\label{eq:hw_deep_main}
\end{equation}
Accordingly,
\begin{equation}
S([a,b])
=
\frac{1}{4G_N}\min\{\ell_{\rm shallow}(a,b),\ell_{\rm wall}(a,b)\}
=
\min\left\{
\frac{c}{3}\log\frac{b-a}{\varepsilon},
\frac{c}{3}\log\frac{z_0}{\varepsilon}
\right\},
\label{eq:hw_interval_entropy_main}
\end{equation}
where
\begin{equation}
c=\frac{3}{2G_N}.
\label{eq:Brown-Henneaux_relation}
\end{equation}

This simple bipartite example already shows the essential role of the hard wall: beyond a certain scale, the RT surface can no longer probe deeper into the bulk, and the entropy saturates at a wall-controlled value.

\subsection{$BC$-symmetric tripartition: benchmark case}
\label{subsec:hw_BC_symmetric}

We now turn to the multipartite problem in the configuration that will serve as the prototype for the later confining backgrounds:
\begin{equation}
A=\left[-\frac{L}{2},\frac{L}{2}\right],
\qquad
B=\left(-\infty,-\frac{L}{2}\right),
\qquad
C=\left(\frac{L}{2},\infty\right).
\label{eq:hw_BC_regions_main}
\end{equation}
Here $A$ is a finite interval, while $B$ and $C$ are semi-infinite. 
This is the direct hard-wall analogue of the $BC$-symmetric tripartition that we will later study in the D4-soliton, D3-soliton, and Klebanov--Strassler backgrounds.

\subsubsection{Candidate \texorpdfstring{$Y$}{Y}: connected bulk junction}

\begin{figure}[tbp]
\centering
\begin{tikzpicture}[scale=1.05, >=latex]

% axes
\draw[->] (-3.8,0) -- (3.8,0) node[below] {$x$};
\draw[->] (-3.4,0) -- (-3.4,3.6) node[left] {$z$};

% boundary and IR wall
\draw[line width=1.1pt] (-3.5,0) -- (3.5,0);
\node[below] at (3.1,0.4) {$z=\varepsilon$};

\draw[dashed, line width=1.1pt] (-3.5,2.8) -- (3.5,2.8);
\node[left] at (-3.5,2.8) {$z_0$};
\node[right] at (3.5,2.8) {$z=z_0$};

% terminals
\coordinate (x1) at (-1.2,0);
\coordinate (x2) at (1.2,0);

% ticks
\draw[thick] (x1) -- ++(0,-0.15);
\draw[thick] (x2) -- ++(0,-0.15);

\node[below=4pt] at (x1) {$-L/2$};
\node[below=4pt] at (x2) {$L/2$};

% subsystems
\node[below=18pt] at (-2.5, 0) {\large $B$};
\node[below=18pt] at (0, 0) {\large $A$};
\node[below=18pt] at (2.5, 0) {\large $C$};

% junction and endpoint on wall
\coordinate (J) at (0, 2.0784);
\coordinate (EA) at (0, 2.8);

% geodesics
\draw[thick] (J) arc[start angle=120, end angle=180, radius=2.4];
\draw[thick] (J) arc[start angle=60, end angle=0, radius=2.4];
\draw[thick] (J) -- (EA);

% 120-degree markers
\def\r{0.32}
\draw ($(J)+(90:\r)$)  arc[start angle=90, end angle=210, radius=\r];
\draw ($(J)+(210:\r)$) arc[start angle=210, end angle=330, radius=\r];
\draw ($(J)+(330:\r)$) arc[start angle=330, end angle=450, radius=\r];
\node at ($(J)+(150:0.52)$) {\scriptsize $120^\circ$};

% dots and labels
\fill (EA) circle (1.7pt) node[above=2pt] {$E_A$};
\fill (J) circle (1.7pt) node[below=15pt, right=-8pt] {$Y$};

%\node[align=left] at (1,3.8) {\small Candidate multi-way cut for $BC$-symmetric configuration};
\node[align=left] at (0,3.8) {\small 
Candidate Y: connected bulk junction
};

\end{tikzpicture}
\caption{
Connected bulk junction for the $BC$-symmetric configuration. 
%Multi-way cut for the $BC$-symmetric configuration.
The Y-junction forms over subsystem $A$ and connects to the hard wall at $E_A$. Because subsystems $B$ and $C$ are semi-infinite, they do not possess disconnected vertical geodesics ending on the wall in the finite region.}
\label{fig:bulkY_BC_symmetric}
\end{figure}
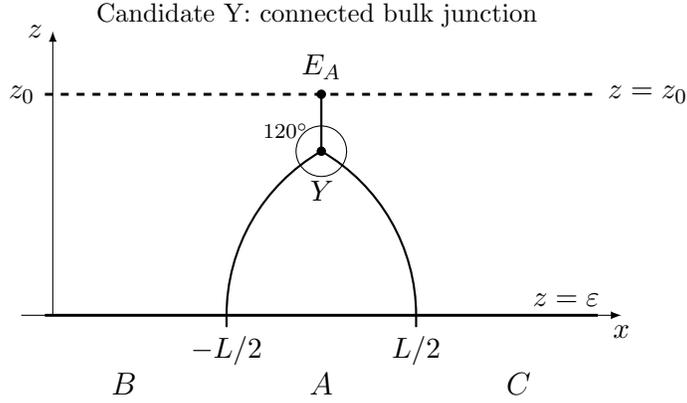

The first candidate is the connected $Y$-type network shown in Fig.~\ref{fig:bulkY_BC_symmetric}. 
By symmetry, the junction lies at $x=0$, and the total cost reduces to
\begin{equation}
L_Y(z_\ast)
=
2\,\operatorname{arcosh}\!\left(\frac{(L/2)^2+z_\ast^2+\varepsilon^2}{2z_\ast\varepsilon}\right)
+
\log\frac{z_0}{z_\ast},
\qquad
0<z_\ast\le z_0,
\label{eq:hw_BC_LY_main}
\end{equation}
where $z_\ast$ is the vertical position of the junction, to be determined by extremizing the cost function $L_Y(z_\ast)$. 
Extremization gives
\begin{equation}
z_\ast^{(0)}=\frac{\sqrt{3}}{2}L  \, \approx \, 0.866 L,
\label{eq:hw_BC_zstar_main}
\end{equation}
and substituting this back yields
\begin{equation}
L_Y^{\rm (Ste)}
=
\log\left(
\frac{8}{3\sqrt{3}}
\frac{Lz_0}{\varepsilon^2}
\right).
\label{eq:hw_BC_Y_main}
\end{equation}

This extremal configuration satisfies the $120^\circ$ condition~\cite{Gadde:2022cqi}, which plays a key role in determining the extremal Y-type network configuration later. To see that the condition indeed is satisfied, let us describe the two side branches explicitly. 
In the UV limit $\varepsilon\to 0$, geodesics on the constant-time slice are semicircles orthogonal to the boundary $z=0$. 
For the symmetric endpoints
\begin{equation}
x_1=-\frac{L}{2},
\qquad
x_2=\frac{L}{2},
\end{equation}
the two side branches are semicircles of radius $L$ centered at
\begin{equation}
\left(\frac{L}{2},0\right),
\qquad
\left(-\frac{L}{2},0\right),
\end{equation}
respectively. 
Hence their equations are
\begin{equation}
\left(x-\frac{L}{2}\right)^2+z^2=L^2,
\qquad
\left(x+\frac{L}{2}\right)^2+z^2=L^2.
\label{eq:BC_symmetric_side_arcs}
\end{equation}
These two semicircles intersect the vertical branch at the symmetric junction point
\begin{equation}
Y=(0,z_\ast),
\qquad
z_\ast=\frac{\sqrt{3}}{2}L.
\label{eq:BC_symmetric_junction_point}
\end{equation}

To compute the angle, differentiate the right arc in \eqref{eq:BC_symmetric_side_arcs}:
\begin{equation}
2\left(x-\frac{L}{2}\right)+2zz'=0
\qquad\Rightarrow\qquad
z'=-\frac{x-\frac{L}{2}}{z}.
\end{equation}
Evaluating this at the junction point \eqref{eq:BC_symmetric_junction_point}, we obtain
\begin{equation}
z'\Big|_{Y}
=
-\frac{-L/2}{(\sqrt{3}/2)L}
=
\frac{1}{\sqrt{3}}.
\end{equation}
Thus the tangent to the side branch makes an angle $\theta$ with the horizontal satisfying
\begin{equation}
\tan\theta=\frac{1}{\sqrt{3}},
\qquad\Rightarrow\qquad
\theta=\frac{\pi}{6}.
\end{equation}
Therefore the angle between the side branch and the vertical branch is
\begin{equation}
\frac{\pi}{2}-\frac{\pi}{6}
=
\frac{\pi}{3}
=
60^\circ.
\end{equation}
Since the two side branches are symmetric, each of them meets the vertical stem at $60^\circ$, and hence the pairwise angles at the junction are all
\begin{equation}
120^\circ.
\end{equation}
This confirms that the extremal configuration is indeed the symmetric Steiner junction.

\subsubsection{Candidate \texorpdfstring{$W$}{W}: wall-assisted disconnected configuration}

\begin{figure}[htb]
\centering
\begin{tikzpicture}[scale=1.05, >=latex]

% axes
\draw[->] (-3.8,0) -- (3.8,0) node[below] {$x$};
\draw[->] (-3.4,0) -- (-3.4,3.6) node[left] {$z$};

% boundary and hard wall
\draw[line width=1.1pt] (-3.5,0) -- (3.5,0);
\node[below] at (3.1,0.4) {$z=\varepsilon$};

\def\zwall{2.4}
\draw[dashed, line width=1.1pt] (-3.5,\zwall) -- (3.5,\zwall);
\node[right] at (3.5,\zwall) {$z=z_0$};

% endpoints of A
\def\Lhalf{1.2}
\coordinate (xL) at (-\Lhalf,0);
\coordinate (xR) at (\Lhalf,0);

% ticks and labels
\draw[thick] (xL) -- ++(0,-0.15);
\draw[thick] (xR) -- ++(0,-0.15);

\node[below=4pt] at (xL) {$-L/2$};
\node[below=4pt] at (xR) {$L/2$};

% subsystem labels
\node[below=18pt] at (-2.5, 0) {\large $B$};
\node[below=18pt] at (0, 0) {\large $A$};
\node[below=18pt] at (2.5, 0) {\large $C$};

% vertical disconnected branches
\draw[thick] (xL) -- (-\Lhalf,\zwall);
\draw[thick] (xR) -- (\Lhalf,\zwall);

% endpoints on wall
\fill (-\Lhalf,\zwall) circle (1.6pt);
\fill (\Lhalf,\zwall) circle (1.6pt);

% title inside figure
%\node[align=left] at (0,3.35) {\small Candidate $W$: wall-assisted disconnected configuration};

\node[align=left] at (0,4) {\small
 Candidate $W$: wall-assisted disconnected configuration
 };

% brace for width L
\draw[decorate,decoration={brace,amplitude=4pt},yshift=-2pt]
(-\Lhalf,0) -- (\Lhalf,0) node[midway,below=6pt] {$L$};

\end{tikzpicture}
\caption{Wall-assisted disconnected candidate $W$ for the $BC$-symmetric configuration. Since $B$ and $C$ are semi-infinite, the multipartite cut is realized simply by two vertical segments descending from the two endpoints of the finite interval $A$ to the hard wall.}
\label{fig:wall_BC_symmetric}
\end{figure}
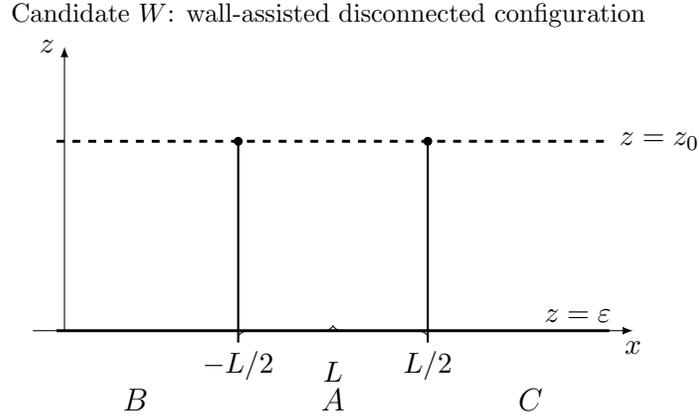

The second candidate is the wall-assisted disconnected configuration shown in Fig.~\ref{fig:wall_BC_symmetric}. 
This configuration has two independent vertical segments descending from the two endpoints of the finite interval $A$ down to the hard wall.

The total cost is therefore just the sum of the two vertical lengths,
\begin{equation}
L_{\rm wall}
=
2\log\frac{z_0}{\varepsilon}.
\label{eq:hw_BC_W_main}
\end{equation}
This configuration is independent of the interval size $L$. 
It becomes favorable when the connected Y-type configuration is pushed sufficiently deep into the bulk, where the Steiner network ceases to be energetically preferred, possibly even before the junction reaches the hard wall.

\subsubsection{Multi-entropy and entanglement entropies}

Using the candidate multi-way cut lengths obtained above, the tripartite multi-entropy is given by the minimum of \eqref{eq:hw_BC_Y_main} and \eqref{eq:hw_BC_W_main},
\begin{equation}
	\begin{aligned}
		S^{(3)}(A:B:C)
		&=\min\left\{ \frac{1}{4G_{N}}L_Y^{(\rm Ste)}, \, \frac{1}{4G_{N}} L_{\rm wall} \right\}\\
		&=\min\left\{ \frac{c}{6} \log\!\left( \frac{8}{3\sqrt{3}} \frac{L z_0}{\varepsilon^2} \right), \,  \frac{c}{3}\log \left(  \frac{z_0}{\varepsilon} \right)  \right\}\\
		&= \begin{dcases}
			\frac{c}{6} \log\!\left( \frac{8}{3\sqrt{3}} \frac{L z_0}{\varepsilon^2} \right) & \text{for } \frac{8}{3\sqrt{3}}L<z_0,\\
			\frac{c}{3}\log \left( \frac{z_0}{\varepsilon} \right) & \text{for } z_0<\frac{8}{3\sqrt{3}}L.
		\end{dcases}
	\end{aligned}
\label{eq:multi-entropy_hardwall-BC-symmetric}
\end{equation}
where we again used \eqref{eq:Brown-Henneaux_relation}.
Note that the transition occurs at 
\begin{equation}
z_0=\frac{8}{3\sqrt{3}}L \, \approx \, 1.54 L ,
\end{equation}
before the junction reaches the hard wall at $z_0=\frac{\sqrt{3}}{2}L \approx 0.866L$, \eqref{eq:hw_BC_zstar_main}.

For the $BC$-symmetric configuration given by \eqref{eq:hw_BC_regions_main}, the ordinary bipartite entanglement entropies are
\begin{equation}
	\begin{aligned}
		S(A)&=\min\left\{\frac{c}{3}\log \frac{L}{\varepsilon},\, \frac{c}{3}\log \frac{z_0}{\varepsilon} \right\}\\
		&=\begin{dcases}
			\frac{c}{3}\log\frac{L}{\varepsilon} & \text{for } L<z_0, \\
			\frac{c}{3}\log\frac{z_0}{\varepsilon} & \text{for } z_0<L,
		\end{dcases}
	\end{aligned}
\label{eq:EE-A_hardwall-BC-symmetric}
\end{equation}
while for the two semi-infinite regions,
\begin{equation}
		S(B)=S(C)= \frac{c}{6}\log \frac{z_0}{\varepsilon}.
\label{eq:EE-B_C_hardwall-BC-symmetric}
\end{equation}

\subsubsection{Genuine multi-entropy}

Combining these ingredients, the genuine multi-entropy is
\begin{equation}
\GM^{(3)}(A:B:C)
=
S^{(3)}(A:B:C)-\frac12\bigl(S(A)+S(B)+S(C)\bigr).
\label{eq:hw_BC_GM_def_main}
\end{equation}
Evaluating this piecewise gives
\begin{equation}
\GM^{(3)}(A:B:C)
=
\begin{dcases}
\frac{c}{6}\log\left(\frac{8}{3\sqrt{3}}\right),
& z_0>\frac{8}{3\sqrt{3}}L,\\[6pt]
\frac{c}{6}\log\left(\frac{z_0}{L}\right),
& \frac{8}{3\sqrt{3}}L>z_0>L,\\[6pt]
0,
& z_0<L.
\end{dcases}
\label{eq:hw_BC_GM_main}
\end{equation}

This is the main hard-wall benchmark result. 
In particular, once the wall lies below the finite interval scale,
\begin{equation}
z_0<L,
\end{equation}
the genuine multipartite contribution vanishes:
\begin{equation}
\GM^{(3)}(A:B:C)=0.
\label{eq:hw_BC_junction_law_main}
\end{equation}
This is the hard-wall manifestation of the effective junction law, and it is precisely this structure that we will later test in smooth confining geometries.

\begin{figure}[tbp]
	\centering
	\includegraphics[width=0.65\textwidth]{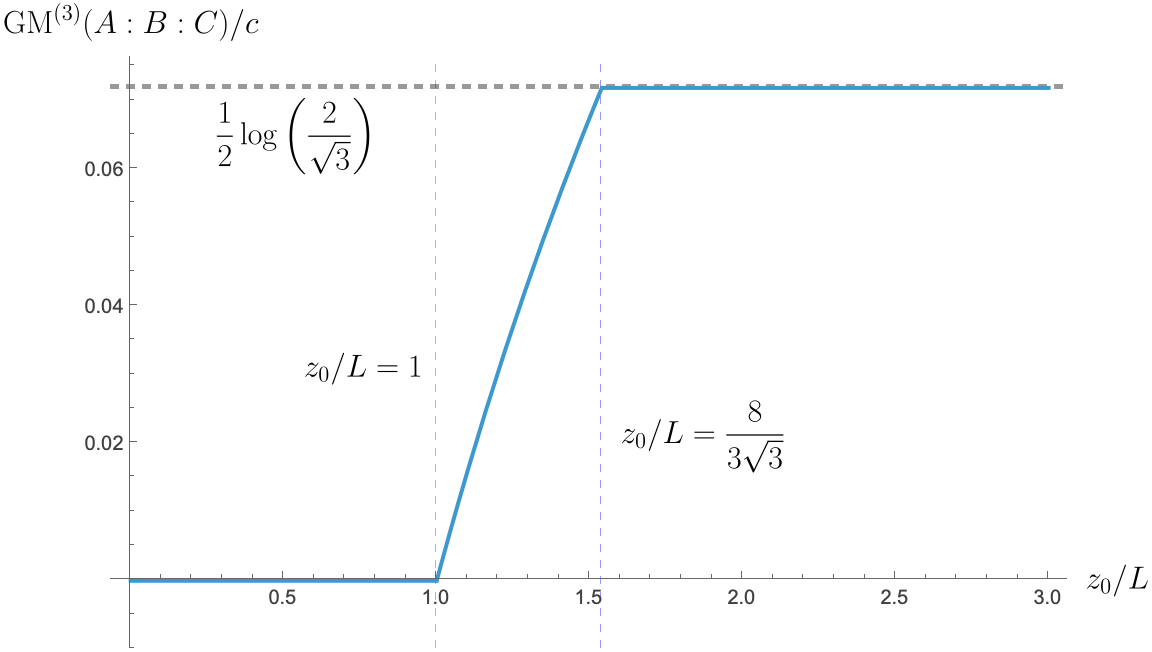}\\[1em]
	\includegraphics[width=0.65\textwidth]{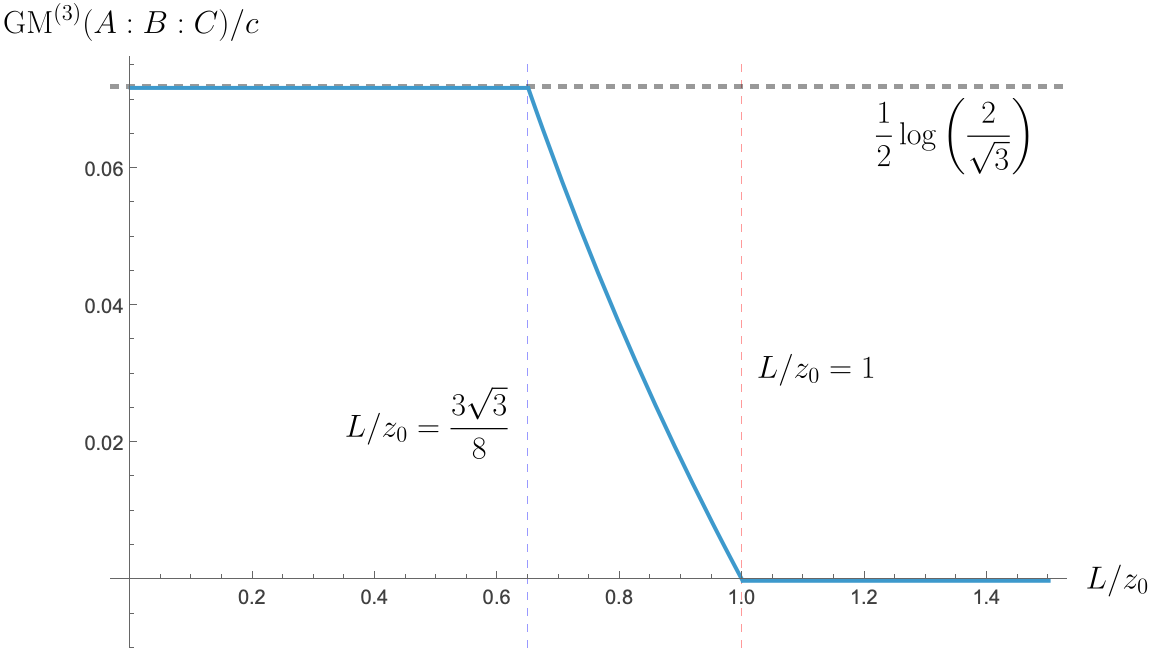}
    \caption{Plots of the rescaled genuine multi-entropy $\GM^{(3)}(A:B:C)/c$ for the $BC$-symmetric configuration. (Top) The entropy as a function of the hard-wall depth $z_0$ with a fixed subsystem size $L$. (Bottom) The entropy as a function of the subsystem size $L$ with a fixed hard-wall depth $z_0$.}
	\label{fig:GM_plots-BC_symmetric-BAC}
\end{figure}

We illustrate this behavior in Fig.~\ref{fig:GM_plots-BC_symmetric-BAC}, which shows the phase structure of the genuine multi-entropy as a function of both the hard-wall depth $z_0$ and the subsystem size $L$.

One key structural feature visible in the present setup is the ordering of the two transition scales. 
For a fixed hard-wall scale $z_0$, let $L_{\rm Crit.(3)}$ denote the transition point at which the multi-entropy changes phase, and let $L_{\rm Crit.}$ denote the corresponding transition point for the bipartite entanglement entropy. 
In the present example, we find
\begin{equation}
L_{\rm Crit.(3)} = \frac{3 \sqrt{3}}{8} z_0 \,, \quad  L_{\rm Crit.} = z_0 \,, \quad L_{\rm Crit.(3)} < L_{\rm Crit.}  
\end{equation}
This means that, as we increase $L$, the  multipartite phase is lost before the bipartite connected phase disappears.
In other words, the junction-supported multipartite structure breaks down at a shorter scale than the bipartite connectivity itself.

\subsection{General asymmetric tripartition}
\label{subsec:hw_asymmetric_summary}

Having established the benchmark $BC$-symmetric case, we now briefly summarize the more general asymmetric tripartition,
\begin{equation}
A=\{-L_A<x<0\},
\qquad
B=\{0<x<L_B\},
\qquad
C=\{x<-L_A,\;x>L_B\},
\label{eq:hw_asymmetric_regions_main}
\end{equation}
and define
\begin{equation}
L_{\min}=\min\{L_A,L_B\},
\qquad
L_{\max}=\max\{L_A,L_B\}.
\label{eq:hw_Lminmax_main}
\end{equation}
This is the most general hard-wall configuration needed in the present paper, and it makes the full saddle structure visible. 
Since the analysis is more involved and will not play a central role in the later D4-, D3-, and Klebanov--Strassler sections, readers primarily interested in those smooth confining backgrounds may skip this subsection on a first reading.

The tripartite multi-entropy
\begin{equation}
S^{(3)}(A:B:C),
\end{equation}
can be obtained holographically by minimizing the total length of a multi-way cut separating the three boundary subsystems.

The hard-wall model is especially useful because the relevant candidates can be identified explicitly as concrete geodesic networks. 
In the setups studied below, three basic types arise:
\begin{enumerate}
\item a fully connected bulk $Y$-type Steiner network,
\item a wall-assisted disconnected configuration, denoted by $W$,
\item and, in asymmetric cases, a partially disconnected (mixed) configuration, denoted by $M$.
\end{enumerate}
The physical tripartite entropy is obtained by minimizing among these candidates.

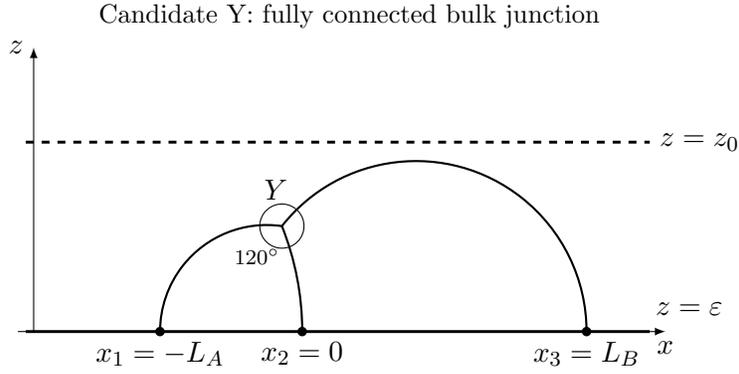
\begin{figure}[htb]
\centering
\begin{tikzpicture}[scale=1.05, >=latex]

% axes (extended for asymmetric LB)
\draw[->] (-3.6,0) -- (4.6,0) node[below] {$x$};
\draw[->] (-3.4,0) -- (-3.4,3.6) node[left] {$z$};

% boundary and (far) hard wall
\draw[line width=1.1pt] (-3.5,0) -- (4.4,0);
\node[below] at (4.9,0.5) {$z=\varepsilon$};

\draw[dashed, line width=1.1pt] (-3.5,2.4) -- (4.4,2.4);
\node[right] at (4.4,2.4) {$z=z_0$};

% terminals: x1=-LA, x2=0, x3=+LB
\coordinate (x1) at (-1.8,0);
\coordinate (x2) at (0,0);
\coordinate (x3) at (3.6,0);

\fill (x1) circle (1.7pt) node[below] {$x_1=-L_A$};
\fill (x2) circle (1.7pt) node[below] {$x_2=0$};
\fill (x3) circle (1.7pt) node[below] {$x_3=L_B$};

% Exact Steiner point
\coordinate (J) at (-0.257, 1.336);

% draw the three geodesic branches using exact arcs:
\draw[thick] (J) arc[start angle=81.79, end angle=180, radius=1.35];
\draw[thick] (J) arc[start angle=21.79, end angle=0, radius=3.6];
\draw[thick] (J) arc[start angle=141.79, end angle=0, radius=2.16];

\node[above left, xshift=6pt, yshift=6pt] at (J) {$Y$};

% 120-degree markers at J
\def\r{0.28}
\draw ($(J)+(51.79:\r)$)  arc[start angle=51.79, end angle=171.79, radius=\r];
\draw ($(J)+(171.79:\r)$) arc[start angle=171.79, end angle=291.79, radius=\r];
\draw ($(J)+(291.79:\r)$) arc[start angle=291.79, end angle=411.79, radius=\r];
\node at ($(J)+(230.79:0.5)$) {\scriptsize $120^\circ$};

%\node[align=left] at (0.6,3.3) {\small Fully connected candidate};
\node[align=left] at (0.6,4) {\small Candidate Y: fully connected bulk junction};

\end{tikzpicture}
\caption{Connected $Y$-type multipartite candidate in the asymmetric hard-wall setup. The three geodesic branches meet at a bulk Steiner point with $120^\circ$ angles.}
\label{fig:hw_asymmetric_Y}
\end{figure}

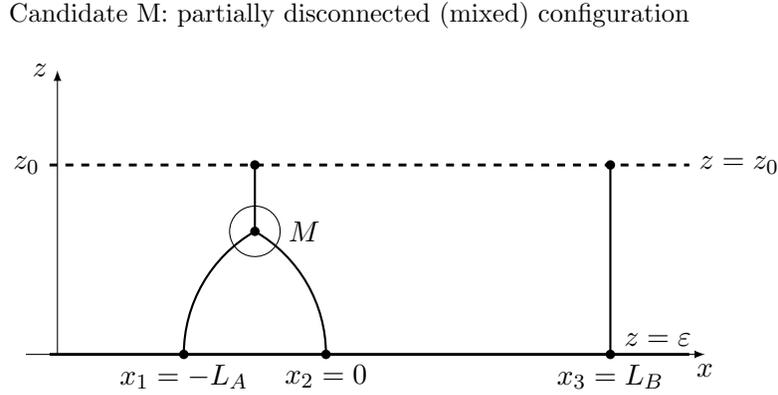
\begin{figure}[htb]
\centering
\begin{tikzpicture}[scale=1.05, >=latex]

% axes
\draw[->] (-3.8,0) -- (4.8,0) node[below] {$x$};
\draw[->] (-3.4,0) -- (-3.4,3.6) node[left] {$z$};

% boundary and wall
\draw[line width=1.1pt] (-3.5,0) -- (4.6,0);
\node[below] at (4.2,0.4) {$z=\varepsilon$};

\draw[dashed, line width=1.1pt] (-3.5,2.4) -- (4.6,2.4);
\node[left] at (-3.5,2.4) {$z_0$};
\node[right] at (4.6,2.4) {$z=z_0$};

% terminals
\coordinate (x1) at (-1.8,0);
\coordinate (x2) at (0,0);
\coordinate (x3) at (3.6,0);

\fill (x1) circle (1.7pt) node[below] {$x_1=-L_A$};
\fill (x2) circle (1.7pt) node[below] {$x_2=0$};
\fill (x3) circle (1.7pt) node[below] {$x_3=L_B$};

% exact junction etc.
\coordinate (J) at (-0.9,1.5588);
\coordinate (EA) at (-0.9,2.4);
\coordinate (EBC) at (3.6,2.4);

\draw[thick] (J) arc[start angle=120, end angle=180, radius=1.8];
\draw[thick] (J) arc[start angle=60, end angle=0, radius=1.8];
\draw[thick] (J) -- (EA);
\draw[thick] (x3) -- (EBC);

\def\r{0.32}
\draw ($(J)+(90:\r)$)  arc[start angle=90, end angle=210, radius=\r];
\draw ($(J)+(210:\r)$) arc[start angle=210, end angle=330, radius=\r];
\draw ($(J)+(330:\r)$) arc[start angle=330, end angle=450, radius=\r];

\fill (EA) circle (1.7pt);
\fill (EBC) circle (1.7pt);
\fill (J) circle (1.7pt);
%\node[below right] at (J) {$M$};
\node[right] at ($(J)+(0.3,0)$) {$M$};

%\node[align=left] at (0.8,3.3) {\small Partially disconnected candidate};
\node[align=left] at (0.3,4.3) {\small Candidate M: partially disconnected (mixed) configuration};

\end{tikzpicture}
\caption{Partially disconnected (mixed) candidate $M$ in the asymmetric hard-wall setup. A $Y$-junction forms only over the smaller side, while the larger side is already disconnected and reaches the wall directly.}
\label{fig:hw_asymmetric_M}
\end{figure}

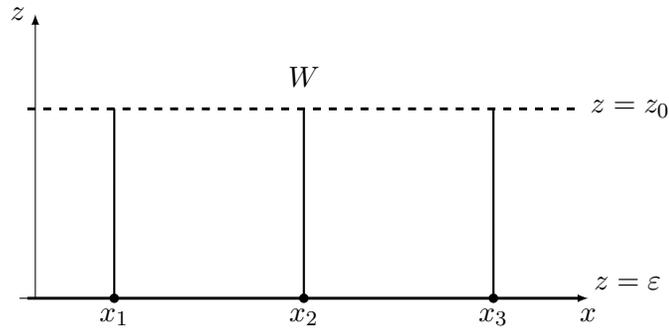
\begin{figure}[htb]
\centering
\begin{tikzpicture}[scale=1.05, >=latex]

% axes
\draw[->] (-3.6,0) -- (3.6,0) node[below] {$x$};
\draw[->] (-3.4,0) -- (-3.4,3.6) node[left] {$z$};

% boundary and hard wall
\draw[line width=1.1pt] (-3.5,0) -- (3.5,0);
\node[below] at (4.1,0.4) {$z=\varepsilon$};

\def\zwall{2.4}
\draw[dashed, line width=1.1pt] (-3.5,\zwall) -- (3.5,\zwall);
\node[right] at (3.5,\zwall) {$z=z_0$};

\def\Len{2.4}
\coordinate (x1) at (-\Len,0);
\coordinate (x2) at (0,0);
\coordinate (x3) at (\Len,0);

\fill (x1) circle (1.7pt) node[below] {$x_1$};
\fill (x2) circle (1.7pt) node[below] {$x_2$};
\fill (x3) circle (1.7pt) node[below] {$x_3$};

\draw[thick] (x1) -- (-\Len,\zwall);
\draw[thick] (x2) -- (0,\zwall);
\draw[thick] (x3) -- (\Len,\zwall);

\node[above] at (0,2.55) {$W$};
%\node[align=left] at (0,3.3) {\small Fully wall-assisted disconnected candidate};
\node[align=left] at (0.2,4.3) {\small Candidate W: fully wall-assisted disconnected configuration};

\end{tikzpicture}
\caption{Wall-assisted disconnected candidate $W$ in the hard-wall geometry. Once all branches terminate on the wall, the cost becomes independent of the horizontal separations.}
\label{fig:hw_wall_candidate}
\end{figure}

\FloatBarrier

These three configurations already capture the main lesson of the hard-wall model. 
The multipartite problem is not controlled by a single connected saddle, but by a competition among geometrically distinct candidates. 
Moreover, asymmetry introduces a genuinely new possibility, namely the partially disconnected saddle $M$, which has no analogue in the symmetric case.

\subsubsection{Candidate \texorpdfstring{$Y$}{Y}: fully connected bulk Steiner network}

The first candidate is the fully connected bulk $Y$ network shown in Fig.~\ref{fig:hw_asymmetric_Y}. 
Its total cost is
\begin{align}
L_Y(z_\ast,x_\ast)
&=
\operatorname{arcosh}\!\left(\frac{(x_\ast + L_A)^2+z_\ast^2+\varepsilon^2}{2z_\ast\varepsilon}\right)
+
\operatorname{arcosh}\!\left(\frac{(L_B-x_\ast)^2+z_\ast^2+\varepsilon^2}{2z_\ast\varepsilon}\right) \nonumber \\
&\qquad  +
\operatorname{arcosh}\!\left(\frac{x_\ast^2+z_\ast^2+\varepsilon^2}{2z_\ast\varepsilon}\right)
\label{eq:hw_asymmetric_LY_full}
\end{align}
with $0<z_\ast\le z_0$. Here, $(z_\ast,x_\ast)$ is the position of the junction, to be determined by extremizing the cost function.

In the UV limit, extremization reproduces the usual hyperbolic Steiner condition, namely that the three branches meet pairwise at $120^\circ$. 
Solving the extremality conditions gives the junction location
\begin{equation}
x_\ast^{(0)}
=
\frac{(L_A-L_B)L_A L_B}{2(L_A^2+L_A L_B+L_B^2)},
\qquad
z_\ast^{(0)}
=
\frac{\sqrt{3}\,L_A L_B(L_A+L_B)}{2(L_A^2+L_A L_B+L_B^2)}.
\label{eq:hw_asymmetric_extremal_point_main}
\end{equation}
Substituting this back into \eqref{eq:hw_asymmetric_LY_full}, one obtains
\begin{equation}
L_Y^{\rm (Ste)}
=
\log\left(
\frac{8}{3\sqrt{3}}
\frac{L_A L_B(L_A+L_B)}{\varepsilon^3}
\right).
\label{eq:hw_asymmetric_Y_main}
\end{equation}
For $L_A=L_B=L$, this reduces to the familiar symmetric result. 
A detailed verification of the $120^\circ$ condition is given in Appendix~\ref{app:hardwall_technical_details}.
\subsubsection{Candidate \texorpdfstring{$M$}{M}: partially disconnected (mixed) configuration}

A second class of saddles appears only in the asymmetric problem. 
Here the multi-way cut forms a $Y$-junction over the smaller finite interval, while the larger side is already disconnected and reaches the wall directly, as shown in Fig.~\ref{fig:hw_asymmetric_M}. 
Denoting by $L_{\min}$ the smaller of $L_A$ and $L_B$, the corresponding extremal length is obtained by combining the symmetric $Y$-junction contribution for the interval of size $L_{\min}$ with one additional vertical wall segment:
\begin{equation}
L_M
=
\log\!\left( \frac{8}{3\sqrt{3}} \frac{L_{\min} z_0}{\varepsilon^2} \right)
+
\log\!\left( \frac{z_0}{\varepsilon} \right)
=
\log\!\left(
\frac{8}{3\sqrt{3}}
\frac{L_{\min} z_0^2}{\varepsilon^3}
\right).
\label{eq:hw_asymmetric_M_main}
\end{equation}
This saddle is the crucial new ingredient of the asymmetric hard-wall phase diagram.

\subsubsection{Candidate \texorpdfstring{$W$}{W}: wall-assisted disconnected configuration}

Finally, there is the fully wall-assisted disconnected candidate of Fig.~\ref{fig:hw_wall_candidate}, whose cost is simply
\begin{equation}
L_W
=
3\log\frac{z_0}{\varepsilon}.
\label{eq:hw_asymmetric_W_main}
\end{equation}
This is independent of the interval sizes, reflecting the fact that once all branches terminate on the hard wall, the geometry no longer probes the detailed horizontal separations.

Thus the tripartite multi-entropy is
\begin{align}
&S^{(3)}(A:B:C)
=
\frac{1}{4G_N}\min\{L_Y^{\rm (Ste)},L_M,L_W\}
\label{eq:hw_asymmetric_S3_main} \\
&=
\begin{cases}
\displaystyle
\frac{c}{6}\log\!\left(
\frac{8}{3\sqrt{3}}
\frac{L_{\min}L_{\max}(L_{\min}+L_{\max})}{\varepsilon^3}
\right),
& \mbox{if} \,
\begin{cases}
\begin{array}{l}
L_{\max}(L_{\min}+L_{\max})\le z_0^2,\\[2pt]
L_{\min}L_{\max}(L_{\min}+L_{\max})\le \dfrac{3\sqrt{3}}{8}z_0^3,
\end{array}
\end{cases}
\\[5mm]
\displaystyle
\frac{c}{6}\log\!\left(
\frac{8}{3\sqrt{3}}
\frac{L_{\min}z_0^2}{\varepsilon^3}
\right),
& \mbox{if} \,
\begin{cases}
\begin{array}{l}
L_{\max}(L_{\min}+L_{\max})\ge z_0^2,\\[2pt]
L_{\min}\le \dfrac{3\sqrt{3}}{8}z_0,
\end{array}
\end{cases}
\\[5mm]
\displaystyle
\frac{c}{6}\log\!\left(
\frac{z_0^3}{\varepsilon^3}
\right),
& \mbox{if} \,
\begin{cases}
\begin{array}{l}
L_{\min}L_{\max}(L_{\min}+L_{\max})\ge \dfrac{3\sqrt{3}}{8}z_0^3,\\[2pt]
L_{\min}\ge \dfrac{3\sqrt{3}}{8}z_0.
\end{array}
\end{cases}
\end{cases}
\end{align}
Thus the behavior of the tripartite multi-entropy depends nontrivially on the values of $L_{\max}$ and $L_{\min}$, and the full phase structure is correspondingly more complicated. 
For completeness, we classify all cases and present the corresponding plots of the multi-entropy and genuine multi-entropy in Appendix~\ref{app:hardwall_asymmetric_details}.

The qualitative lesson is that asymmetry introduces an additional intermediate phase, controlled by the partially disconnected saddle $M$, which has no analogue in the symmetric benchmark case. 
Nevertheless, the main conclusion remains the same: once the hard wall lies above the shortest relevant boundary scale, the genuine multipartite contribution disappears. 
Detailed derivations of the asymmetric phase structure are collected in Appendix~\ref{app:hardwall_asymmetric_details}.

\subsection{Summary of the hard-wall benchmark}

The hard-wall model already exhibits the essential structural features that will reappear in smooth confining backgrounds:
\begin{enumerate}
\item competition between connected and wall-assisted multipartite saddles,
\item a critical scale set by the IR cutoff,
\item and the disappearance of the genuine multipartite contribution once the wall lies above the shortest relevant interval scale.
\end{enumerate}
The $BC$-symmetric case will serve as the main benchmark for the later sections, while the asymmetric hard-wall analysis provides a useful guide to the richer phase structure that can arise away from symmetry.

%%%%%%%%%%%%%%%%%%%%%%%%%%%%%%%%%%%%%%%%%%%%%%%%%%%%%%%%%%%%%%%%%%%%%%%

\section{General framework for smooth confining geometries}
\label{sec:smooth_general_framework}

In this section, we formulate the general framework that will be used in the subsequent analyses of the D4-soliton, D3-soliton, and Klebanov--Strassler backgrounds. Our goal is not to solve a specific model here, but rather to isolate the common structure of the multipartite problem in smooth confining geometries.

The key contrast with the hard-wall benchmark of Sec.~\ref{sec:hardwall_benchmark} is that the IR region now ends smoothly, rather than at a sharp cutoff. As a result, while the competition among connected and cap-assisted saddles remains, the geometric realization of that competition is different. In particular, the role played in the hard-wall model by explicit piecewise geodesic networks is replaced by an effective variational problem for codimension-two surfaces in a smooth cigar-like geometry.

\subsection{Common setup and reduced one-dimensional functional}

In all smooth backgrounds considered in this paper, we will focus on the $BC$-symmetric tripartition
\begin{equation}
A=\left[-\frac{L}{2},\frac{L}{2}\right],
\qquad
B=\left(-\infty,-\frac{L}{2}\right),
\qquad
C=\left(\frac{L}{2},\infty\right).
\label{eq:smooth_BCsymmetric_tripartition}
\end{equation}
Here $A$ is a finite interval (or strip), while $B$ and $C$ are semi-infinite. This is the direct analogue of the benchmark configuration studied in Sec.~\ref{subsec:hw_BC_symmetric}, and it is the natural setup in which a symmetric multipartite junction can be compared across different confining geometries.

The main reason for focusing on this configuration is that it cleanly separates the two basic ingredients of the problem:
\begin{enumerate}
\item the tripartite saddle competition entering $S^{(3)}(A:B:C)$,
\item the ordinary bipartite saddle competition entering $S(A)$.
\end{enumerate}
In the later sections, this will allow us to distinguish the transition scale associated with the tripartite multi-entropy from the one associated with the ordinary entanglement entropy.

The holographic entanglement surfaces relevant for strip-like regions can be reduced to an effective one-dimensional variational problem. In all cases of interest, the on-shell area functional can be written in the form
\begin{equation}
S_{\rm branch}
=
\mathcal{N}\int dx\,F(u)\sqrt{1+\beta(u)\,u'(x)^2},
\label{eq:smooth_branch_functional_general}
\end{equation}
where $u$ denotes the radial coordinate appropriate to the background, $\mathcal{N}$ is an overall positive constant, and the functions $F(u)$ and $\beta(u)$ depend on the metric and dilaton of the given geometry.

This form is general enough to cover the D4-soliton, D3-soliton, and KS backgrounds discussed later. The details of $F(u)$ and $\beta(u)$ differ from one background to another, but the basic variational structure is always the same.

The radial coordinate $u$ runs from the asymptotic boundary at $u\to\infty$ down to a smooth IR cap at $u=u_{\rm IR}$, where the geometry ends without a singularity. In D4 and D3 this is the cigar tip at $U=U_0$, while in the KS case it is the deformed-conifold tip at $\tau=0$.

\subsection{Connected and disconnected tripartite candidates}

It is convenient to rewrite \eqref{eq:smooth_branch_functional_general} in terms of the effective two-dimensional metric
\begin{equation}
ds_{\rm eff}^2
=
F(u)^2\left(dx^2+\beta(u)\,du^2\right).
\label{eq:smooth_effective_metric}
\end{equation}
Then the branch functional becomes simply proportional to the proper length in this effective metric,
\begin{equation}
S_{\rm branch}=\mathcal{N}\int ds_{\rm eff}.
\end{equation}

This observation is important because it makes the multipartite problem geometrically parallel to the hard-wall case. In particular, once written in the form \eqref{eq:smooth_effective_metric}, the connected tripartite saddle can again be viewed as an effective $Y$-network. Since all three branches carry the same local tension in this metric, the force-balance condition at a symmetric junction is the usual Steiner condition:
\begin{equation}
\text{the three branches meet pairwise at }120^\circ.
\end{equation}
Thus the $120^\circ$ rule is not a special feature of AdS$_3$, but a general property of the equal-tension effective network problem.

The connected candidate for the tripartite multi-entropy consists of
\begin{itemize}
\item a left side branch from $x=-L/2$ at the boundary down to the junction,
\item a right side branch from $x=+L/2$ at the boundary down to the same junction,
\item and a vertical stem from the IR cap up to the junction.
\end{itemize}
Because of reflection symmetry, it is enough to study the right side branch on
\begin{equation}
0\le x\le \frac{L}{2},
\qquad
u(0)=u_J,
\qquad
u(L/2)=\infty,
\label{eq:smooth_side_branch_bc}
\end{equation}
where $u_J$ is the radial position of the junction.

Since the reduced Lagrangian in \eqref{eq:smooth_branch_functional_general} does not depend explicitly on $x$, there is a conserved quantity
\begin{equation}
-c_J
=
u'\frac{\partial\mathcal{L}}{\partial u'}-\mathcal{L}
=
-\frac{F(u)}{\sqrt{1+\beta(u)\,u'(x)^2}}.
\label{eq:smooth_first_integral}
\end{equation}
This conserved quantity has a simple geometric interpretation in the effective metric \eqref{eq:smooth_effective_metric}. Since the physical slope of the side branch is measured by
\begin{equation}
\tan\theta=\sqrt{\beta(u)}\,u'(x),
\end{equation}
where $\theta$ is the angle measured from the $x$-direction in the effective metric. Thus, one has
\begin{equation}
\cos\theta=\frac{1}{\sqrt{1+\beta(u)\,u'(x)^2}}.
\end{equation}
Therefore from \eqref{eq:smooth_first_integral}, the conserved quantity $c_J$ is simply the horizontal projection of the local tension,
\begin{equation}
c_J=F(u)\cos\theta.
\end{equation}
At the symmetric Steiner junction, the side branch meets the vertical stem at angle $\pi/3$. Since $\theta$ is measured from the $x$-direction, this implies $\theta=\pi/6$, and hence
\begin{equation}
c_J=F(u_J)\cos\frac{\pi}{6}
=\frac{\sqrt{3}}{2}F(u_J).
\label{eq:smooth_cJ_general}
\end{equation}
This relation will appear repeatedly in the later sections.

Solving \eqref{eq:smooth_first_integral} for $dx/du$, the half-width condition becomes
\begin{equation}
\frac{L}{2}
=
\int_{u_J}^{\infty}
du\,
\frac{\sqrt{\beta(u)}\,c_J}{\sqrt{F(u)^2-c_J^2}},
\qquad
c_J=\frac{\sqrt{3}}{2}F(u_J).
\label{eq:smooth_L_uJ_general}
\end{equation}
This equation determines the junction position $u_J$ as a function of the strip width $L$.

The corresponding connected tripartite candidate is then
\begin{equation}
S_Y^{(3)}(L)
=
2S_{\rm side}(u_J)+S_{\rm stem}(u_J),
\label{eq:smooth_SY_general}
\end{equation}
where
\begin{equation}
S_{\rm side}(u_J)
=
\mathcal{N}\int_{u_J}^{u_\infty}
du\,
\frac{\sqrt{\beta(u)}\,F(u)^2}{\sqrt{F(u)^2-c_J^2}},
\label{eq:smooth_Sside_general}
\end{equation}
and
\begin{equation}
S_{\rm stem}(u_J)
=
\mathcal{N}\int_{u_{\rm IR}}^{u_J}du\,\sqrt{\beta(u)}\,F(u).
\label{eq:smooth_Sstem_general}
\end{equation}
Here $u_\infty$ is the UV cutoff.

The second universal candidate is the cap-assisted disconnected configuration. In the $BC$-symmetric setup, it consists of two independent vertical pieces from the endpoints of $A$ down to the IR cap. Its cost is
\begin{equation}
S_W^{(3)}
=
2\mathcal{N}\int_{u_{\rm IR}}^{u_\infty}du\,\sqrt{\beta(u)}\,F(u).
\label{eq:smooth_SW_general}
\end{equation}
This is the smooth-cap analogue of the wall-assisted $W$ saddle in the hard-wall model.

Although both $S_Y^{(3)}$ and $S_W^{(3)}$ are UV divergent, their difference is finite:
\begin{align}
\frac{S_Y^{(3)}-S_W^{(3)}}{\mathcal{N}}
&=
2\int_{u_J}^{u_\infty}du\,\sqrt{\beta(u)}
\left[
\frac{F(u)^2}{\sqrt{F(u)^2-c_J^2}}-F(u)
\right]
-
\int_{u_{\rm IR}}^{u_J}du\,\sqrt{\beta(u)}\,F(u).
\label{eq:smooth_SY_minus_SW_general}
\end{align}
This finite quantity is the natural object to study when locating the tripartite phase transition.

\subsection{Ordinary entanglement entropy and two transition scales}

To construct the genuine multi-entropy, we also need the ordinary entanglement entropy of the finite interval $A$.

The connected bipartite surface is again described by \eqref{eq:smooth_branch_functional_general}, but now the appropriate boundary condition is not the Steiner condition at a junction. Instead, the connected RT surface has a turning point $u_R$ where
\begin{equation}
u'(0)=0.
\end{equation}
In the geometric language introduced above, this means that at the turning point the surface is locally horizontal, namely $\theta=0$. Thus, unlike the tripartite junction where the Steiner condition gives $\theta=\pi/6$ (equivalently, the side branch meets the vertical stem at angle $\pi/3$), the connected RT surface satisfies the special case $\cos\theta=1$ at $u=u_R$.
Using \eqref{eq:smooth_first_integral}, this fixes the conserved quantity to be
\begin{equation}
c_R=F(u_R).
\label{eq:smooth_cR_general}
\end{equation}
Accordingly, the half-width relation becomes
\begin{equation}
\frac{L}{2}
=
\int_{u_R}^{\infty}
du\,
\frac{\sqrt{\beta(u)}\,c_R}{\sqrt{F(u)^2-c_R^2}},
\qquad
c_R=F(u_R),
\label{eq:smooth_L_uR_general}
\end{equation}
which determines the turning point $u_R$ as a function of $L$.

The connected and disconnected candidates for $S(A)$ are then
\begin{equation}
S_{\rm conn}(A)
=
2\mathcal{N}\int_{u_R}^{u_\infty}
du\,
\frac{\sqrt{\beta(u)}\,F(u)^2}{\sqrt{F(u)^2-c_R^2}},
\label{eq:smooth_SA_conn_general}
\end{equation}
and
\begin{equation}
S_{\rm disc}(A)
=
2\mathcal{N}\int_{u_{\rm IR}}^{u_\infty}du\,\sqrt{\beta(u)}\,F(u),
\label{eq:smooth_SA_disc_general}
\end{equation}
so that
\begin{equation}
S(A)=\min\{S_{\rm conn}(A),S_{\rm disc}(A)\}.
\label{eq:smooth_SA_general}
\end{equation}

For the semi-infinite regions $B$ and $C$, the entanglement surfaces are simply the disconnected cap-reaching ones. Thus,
\begin{equation}
S(B)=S(C)=\mathcal{N}\int_{u_{\rm IR}}^{u_\infty}du\,\sqrt{\beta(u)}\,F(u).
\label{eq:smooth_SB_SC_general}
\end{equation}

A central point, already anticipated by the hard-wall analysis, is that the tripartite transition scale and the bipartite transition scale need not coincide.

Indeed, the physical tripartite entropy is determined by
\begin{equation}
S^{(3)}(A:B:C)=\min\{S_Y^{(3)},S_W^{(3)}\},
\label{eq:smooth_S3_general}
\end{equation}
while the physical entanglement entropy of $A$ is determined by
\begin{equation}
S(A)=\min\{S_{\rm conn}(A),S_{\rm disc}(A)\}.
\end{equation}
Therefore, the critical scale at which the multipartite saddle changes can differ from the one at which the ordinary RT surface changes.

This distinction will be important in the later sections, because the genuine multi-entropy
\begin{equation}
\GM^{(3)}(A:B:C)
=
S^{(3)}(A:B:C)-\frac{1}{2}\Bigl(S(A)+S(B)+S(C)\Bigr)
\label{eq:smooth_GM_general}
\end{equation}
can continue to change even after $S^{(3)}$ itself has already switched from the connected to the disconnected branch.
We note that $\mathrm{GM}^{(3)} \geq 0$, as guaranteed by the general
argument of \cite{Iizuka:2025ioc}, which applies directly to the present
smooth-cap setting.

\subsection{Expected comparison with the hard-wall benchmark}

The hard-wall benchmark of Sec.~\ref{sec:hardwall_benchmark} suggests a natural set of questions for smooth confining backgrounds.

In the hard-wall model, the sharp IR boundary supports a wall-assisted configuration of finite cost, and this leads to plateau behavior in $\GM^{(3)}$. By contrast, in smooth cigar-like geometries, the IR region ends continuously. The cap-assisted disconnected saddle still exists, but the geometric mechanism responsible for the hard-wall plateau is no longer obviously present.

Thus the main questions to be addressed in the following sections are:
\begin{enumerate}
\item Does the connected $Y$ saddle still dominate at small $L$ and the cap-assisted $W$ saddle at large $L$?
\item Does $\GM^{(3)}$ still vanish beyond a critical scale?
\item Most importantly, does the hard-wall plateau survive in smooth confining geometries, or is it a special feature of the sharp IR wall?
\end{enumerate}

The remaining sections answer these questions in the D4-soliton, D3-soliton, and Klebanov--Strassler backgrounds. Although the explicit forms of $F(u)$ and $\beta(u)$ differ from case to case, the general framework developed here applies to all of them.

\section{The D4-soliton background}
\label{sec:D4_soliton}

In this section, we study the tripartite multi-entropy and the genuine multi-entropy for the $BC$-symmetric tripartition in the confining D4-soliton background. This provides the first smooth-cap analogue of the hard-wall benchmark analyzed in Sec.~\ref{sec:hardwall_benchmark}.

Our goal is to determine how the hard-wall phase structure is modified when the sharp IR cutoff is replaced by a smooth cigar geometry. As we will see, the connected $Y$-type multipartite saddle and the cap-assisted disconnected saddle still compete, but the resulting behavior of $\GM^{(3)}$ differs qualitatively from that in the hard-wall case: in particular, the hard-wall plateau disappears, and the small-$L$ behavior is replaced by a nontrivial power-law decay.

\subsection{Geometry, tripartition, and reduced functional}

We consider the near-horizon geometry of $N_c$ D4-branes compactified on a spatial circle with anti-periodic boundary conditions for fermions. In string frame, the metric and dilaton are
\begin{align}
\frac{ds^2}{\alpha'}
&=
\left(\frac{U}{R}\right)^{3/2}
\left[
-dt^2+dx_1^2+dx_2^2+dx_3^2+f(U)\,dx_4^2
\right]
+
\left(\frac{R}{U}\right)^{3/2}\frac{dU^2}{f(U)}
+
R^{3/2}U^{1/2}d\Omega_4^2,
\label{eq:D4_metric_main}
\\
e^{-2\phi}
&=
\left(\frac{2\pi}{g_{\rm YM}}\right)^4\left(\frac{R}{U}\right)^{3/2},
\qquad
f(U)=1-\left(\frac{U_0}{U}\right)^3 .
\label{eq:D4_dilaton_main}
\end{align}
The compact coordinate satisfies $x_4\sim x_4+2\pi R_4$, and smoothness at the tip requires
\begin{equation}
R_4=\frac{2}{3}\frac{R^{3/2}}{U_0^{1/2}},
\qquad\Longleftrightarrow\qquad
U_0=\left(\frac{2}{3}\right)^2\frac{R^3}{R_4^2}.
\label{eq:D4_periodicity_main}
\end{equation}
Thus the $(U,x_4)$ directions form a cigar geometry with a smooth IR cap at $U=U_0$.

Throughout this section, we take the entangling direction to be $x\equiv x_1$.

We consider the same $BC$-symmetric tripartition as in the hard-wall benchmark,
\begin{equation}
A=\left[-\frac{L}{2},\frac{L}{2}\right],
\qquad
B=\left(-\infty,-\frac{L}{2}\right),
\qquad
C=\left(\frac{L}{2},\infty\right).
\label{eq:D4_tripartition_main}
\end{equation}
This is the direct smooth-cap analogue of the hard-wall configuration studied in Sec.~\ref{subsec:hw_BC_symmetric}.

The relevant extremal surfaces extend along the transverse directions $x_2,x_3$, the Kaluza–Klein circle $x_4$, and the internal $S^4$.

For a strip-like branch described by a profile $U(x)$, the entropy functional reduces to the universal form introduced in Sec.~\ref{sec:smooth_general_framework},
\begin{equation}
S_{\rm branch}
=
\mathcal{N}\int dx\,F(U)\sqrt{1+\beta(U)\,U'(x)^2}.
\label{eq:D4_branch_general_main}
\end{equation}
For the D4-soliton background, one finds
\begin{equation}
F(U)=U\sqrt{U^3-U_0^3},
\qquad
\beta(U)=\frac{R^3}{U^3-U_0^3}.
\label{eq:D4_F_beta_main}
\end{equation}
The overall constant $\mathcal{N}$, including the (infinite) volume of the transverse directions, plays no role in the phase structure and will therefore be left implicit.

A particularly useful simplification is
\begin{equation}
\sqrt{\beta(U)}\,F(U)=R^{3/2}U,
\label{eq:D4_sqrtbetaF_main}
\end{equation}
which makes the vertical stem contribution analytic.

\subsection{Connected and disconnected tripartite candidates}

As explained in Sec.~\ref{sec:smooth_general_framework}, it is convenient to introduce the effective optical metric
\begin{equation}
ds_{\rm eff}^2
=
F(U)^2\left(dx^2+\beta(U)\,dU^2\right).
\label{eq:D4_effective_metric_main}
\end{equation}
In this metric, the side branches and the vertical stem all carry the same local tension, so the force-balance condition at a symmetric junction is the usual $120^\circ$ Steiner rule.

Accordingly, for the connected $Y$ candidate, the conserved quantity
\begin{equation}
-c_J=-\frac{F(U)}{\sqrt{1+\beta(U)\,U'(x)^2}}
\label{eq:D4_first_integral_main}
\end{equation}
is fixed by the Steiner condition to be
\begin{equation}
c_J=\frac{\sqrt{3}}{2}F(U_J)
=
\frac{\sqrt{3}}{2}U_J\sqrt{U_J^3-U_0^3},
\label{eq:D4_cJ_main}
\end{equation}
where $U_J$ denotes the radial position of the junction.

Because of reflection symmetry, it is sufficient to study one side branch on
\begin{equation}
0\le x\le \frac{L}{2},
\qquad
U(0)=U_J,
\qquad
U(L/2)=\infty.
\end{equation}
The half-width condition then becomes
\begin{equation}
\frac{L}{2}
=
R^{3/2}c_J
\int_{U_J}^{\infty}
\frac{dU}
{\sqrt{U^3-U_0^3}\,
\sqrt{U^2(U^3-U_0^3)-c_J^2}},
\qquad
c_J=\frac{\sqrt{3}}{2}U_J\sqrt{U_J^3-U_0^3}.
\label{eq:D4_L_UJ_main}
\end{equation}
This determines $U_J$ as a function of $L$.

The on-shell action of one side branch is
\begin{equation}
S_{\rm side}(U_J)
=
\mathcal{N}R^{3/2}
\int_{U_J}^{U_\infty}
dU\,
\frac{U^2\sqrt{U^3-U_0^3}}
{\sqrt{U^2(U^3-U_0^3)-c_J^2}},
\label{eq:D4_Sside_main}
\end{equation}
while the stem contribution is
\begin{equation}
S_{\rm stem}(U_J)
=
\mathcal{N}R^{3/2}\int_{U_0}^{U_J}U\,dU
=
\frac{\mathcal{N}R^{3/2}}{2}\left(U_J^2-U_0^2\right).
\label{eq:D4_Sstem_main}
\end{equation}
Hence the connected tripartite candidate is
\begin{equation}
S_Y^{(3)}(L)
=
2S_{\rm side}(U_J)+S_{\rm stem}(U_J).
\label{eq:D4_SY_main}
\end{equation}

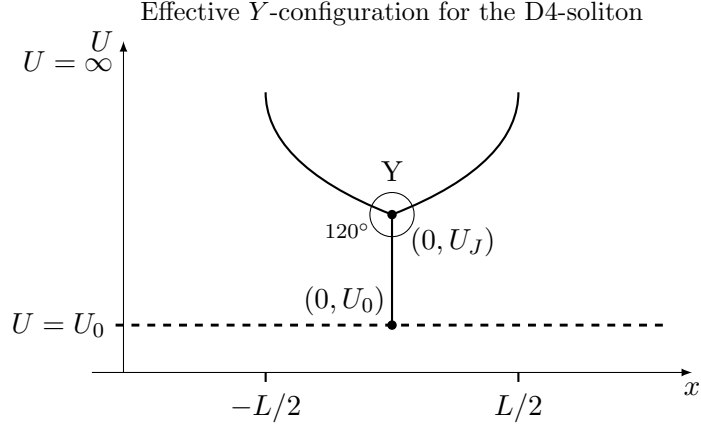
\begin{figure}[htbp]
\centering
\begin{tikzpicture}[scale=1.05, >=latex]

% axes
\draw[->] (-3.8,0) -- (3.8,0) node[below] {$x$};
\draw[->] (-3.4,0) -- (-3.4,4.2) node[left] {$U$};

% IR cap
\draw[dashed, line width=1.1pt] (-3.5,0.6) -- (3.5,0.6);
\node[left] at (-3.5,0.6) {$U=U_0$};

% boundary
\node[left] at (-3.4,3.95) {$U=\infty$};

% x ticks
\draw[thick] (-1.6,0) -- ++(0,-0.12);
\draw[thick] (1.6,0) -- ++(0,-0.12);
\node[below=4pt] at (-1.6,0) {$-L/2$};
\node[below=4pt] at (1.6,0) {$L/2$};

% points
\coordinate (J) at (0,2.0);
\coordinate (E) at (0,0.6);
\coordinate (Lu) at (-1.6,3.55);
\coordinate (Ru) at (1.6,3.55);

% side branches
\draw[thick] (Lu) .. controls (-1.6,3.0) and (-1.15,2.45) .. (J);
\draw[thick] (Ru) .. controls (1.6,3.0) and (1.15,2.45) .. (J);

% stem
\draw[thick] (E) -- (J);

% labels
\fill (J) circle (1.7pt);
\fill (E) circle (1.7pt);
\node[above=9pt] at (J) {Y};
\node[right=23pt,below=1pt] at (J) {$(0,U_J)$};
\node[left=18pt,above] at (E) {$(0,U_0)$};

% 120 degrees
\def\r{0.28}
\draw ($(J)+(90:\r)$)  arc[start angle=90, end angle=210, radius=\r];
\draw ($(J)+(210:\r)$) arc[start angle=210, end angle=330, radius=\r];
\draw ($(J)+(330:\r)$) arc[start angle=330, end angle=450, radius=\r];
\node at ($(J)+(2000:0.6)$) {\scriptsize $120^\circ$};

\node[align=left] at (0,4.55) {\small Effective $Y$-configuration for the D4-soliton};

% % labels
% \fill (J) circle (1.7pt);
% \fill (E) circle (1.7pt);
% \node[right=3pt] at (J) {Y$\,(0,U_J)$};
% \node[right=3pt] at (E) {$(0,U_0)$};

% % 120 degrees
% \def\r{0.28}
% \draw ($(J)+(90:\r)$)  arc[start angle=90, end angle=210, radius=\r];
% \draw ($(J)+(210:\r)$) arc[start angle=210, end angle=330, radius=\r];
% \draw ($(J)+(330:\r)$) arc[start angle=330, end angle=450, radius=\r];
% \node at ($(J)+(150:0.48)$) {\scriptsize $120^\circ$};

% \node[align=left] at (1.1,4.05) {\small Effective $Y$-configuration for the D4-soliton};

\end{tikzpicture}
\caption{Effective connected $Y$-configuration for the $BC$-symmetric tripartition in the D4-soliton background.}
\label{fig:D4_Y_config}
\end{figure}

The disconnected cap-assisted candidate consists of two independent vertical pieces extending from the endpoints of $A$ down to the IR cap. Its cost is
\begin{equation}
S_W^{(3)}
=
2\mathcal{N}R^{3/2}\int_{U_0}^{U_\infty}U\,dU
=
\mathcal{N}R^{3/2}(U_\infty^2-U_0^2).
\label{eq:D4_SW_main}
\end{equation}
Although both $S_Y^{(3)}$ and $S_W^{(3)}$ are UV divergent, their difference is finite:
\begin{align}
\frac{S_Y^{(3)}-S_W^{(3)}}{\mathcal{N}}
&=
2\int_{U_J}^{U_\infty}dU\,\sqrt{\beta(U)}
\left[
\frac{F(U)^2}{\sqrt{F(U)^2-c_J^2}}-F(U)
\right]
-\frac{R^{3/2}}{2}(U_J^2-U_0^2).
\label{eq:D4_SY_minus_SW_main}
\end{align}
This is the quantity used to locate the tripartite phase transition.

\subsection{Ordinary entanglement entropy and numerical phase structure}

For the finite interval $A$, the connected RT surface is characterized by a turning point $U_R$ rather than a Steiner junction. The corresponding conserved quantity is
\begin{equation}
c_R=F(U_R)=U_R\sqrt{U_R^3-U_0^3},
\label{eq:D4_cR_main}
\end{equation}
and the width relation is
\begin{equation}
\frac{L}{2}
=
R^{3/2}c_R
\int_{U_R}^{\infty}
\frac{dU}
{\sqrt{U^3-U_0^3}\,
\sqrt{U^2(U^3-U_0^3)-c_R^2}}.
\label{eq:D4_L_UR_main}
\end{equation}
The connected and disconnected candidates for $S(A)$ are
\begin{equation}
S_{\rm conn}(A)
=
2\mathcal{N}R^{3/2}
\int_{U_R}^{U_\infty}
dU\,
\frac{U^2\sqrt{U^3-U_0^3}}
{\sqrt{U^2(U^3-U_0^3)-c_R^2}},
\label{eq:D4_SA_conn_main}
\end{equation}
and
\begin{equation}
S_{\rm disc}(A)
=
\mathcal{N}R^{3/2}(U_\infty^2-U_0^2),
\label{eq:D4_SA_disc_main}
\end{equation}
so that
\begin{equation}
S(A)=\min\{S_{\rm conn}(A),S_{\rm disc}(A)\}.
\label{eq:D4_SA_main}
\end{equation}
For the semi-infinite regions,
\begin{equation}
S(B)=S(C)=\frac{1}{2}\mathcal{N}R^{3/2}(U_\infty^2-U_0^2).
\label{eq:D4_SB_SC_main}
\end{equation}

\begin{figure}[htbp]
\centering
\begin{minipage}[b]{0.45\textwidth}
\centering
\includegraphics[width=\textwidth]{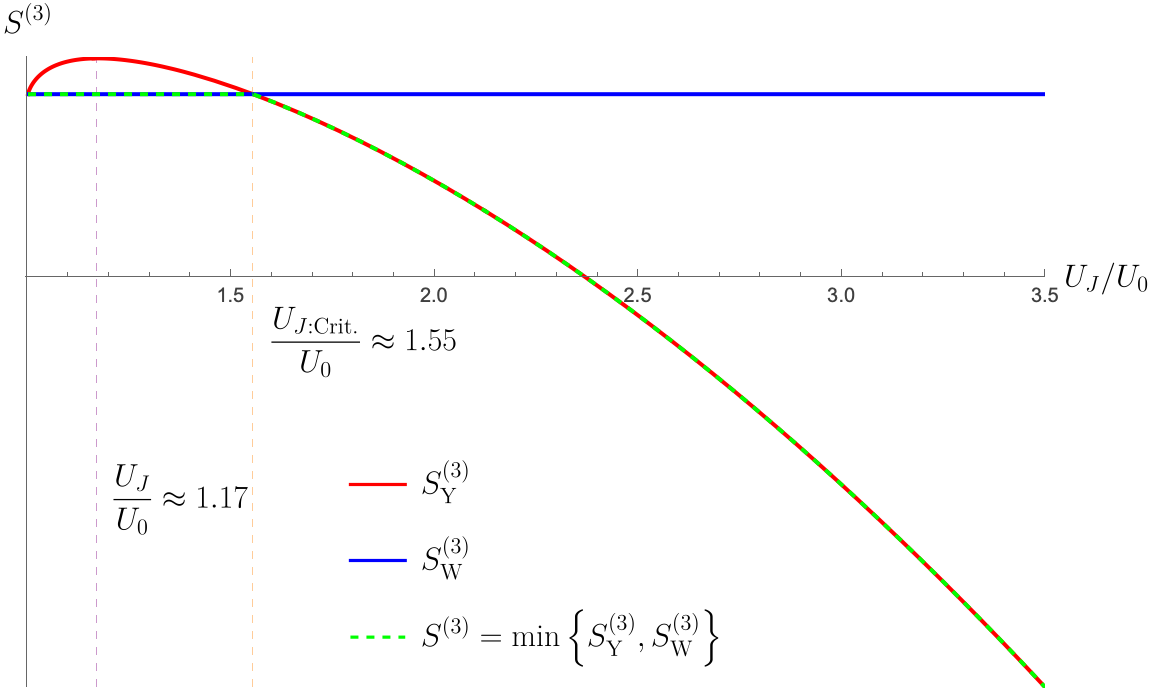}
\\ (a) $S^{(3)}$
\end{minipage}
\hfill
\begin{minipage}[b]{0.45\textwidth}
\centering
\includegraphics[width=\textwidth]{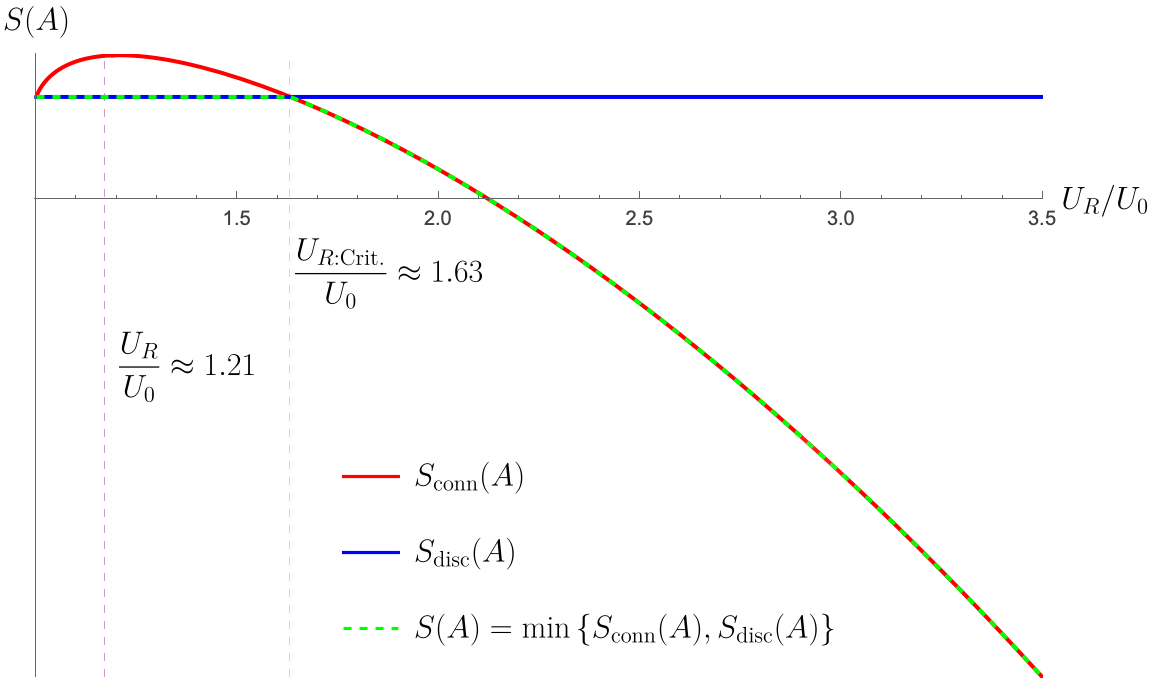}
\\ (b) $S(A)$
\end{minipage}
\caption{Phase transitions in the D4-soliton background. The physical entropy is obtained by minimizing between the connected and cap-assisted disconnected candidates. Note that the horizontal axis is not at zero; both functions remain positive throughout.}
\label{fig:D4_entropy_transitions}
\end{figure}

The relations \eqref{eq:D4_L_UJ_main} and \eqref{eq:D4_L_UR_main} determine the junction position $U_J$ and the turning point $U_R$ as functions of $L$. Numerically, both relations exhibit the familiar two-branch structure, as shown in Fig.~\ref{fig:D4_entropy_transitions}, with maximal lengths
\begin{equation}
L_{\max(3)}\approx 0.938\,R_4,
\qquad
L_{\max}\approx 1.42\,R_4. 
\label{eq:D4_Lmax_main}
\end{equation}
The critical transition scales are
\begin{equation}
L_{\rm Crit.(3)}\approx 0.854\,R_4,
\qquad
L_{\rm Crit.}\approx 1.29\,R_4, 
\label{eq:D4_Lcrit_main}
\end{equation}
where these values are attained at $U_J/U_0\approx 1.55$  and $U_R/U_0\approx 1.63$  respectively. 
As in the bipartite case, the physical transition occurs before the curve turns around, so the small branch is never dominant. 
The phase structure is shown in Fig.~\ref{fig:D4_entropy_transitions}.

\begin{figure}[htbp]
 \centering
 \includegraphics[width=0.78\textwidth]{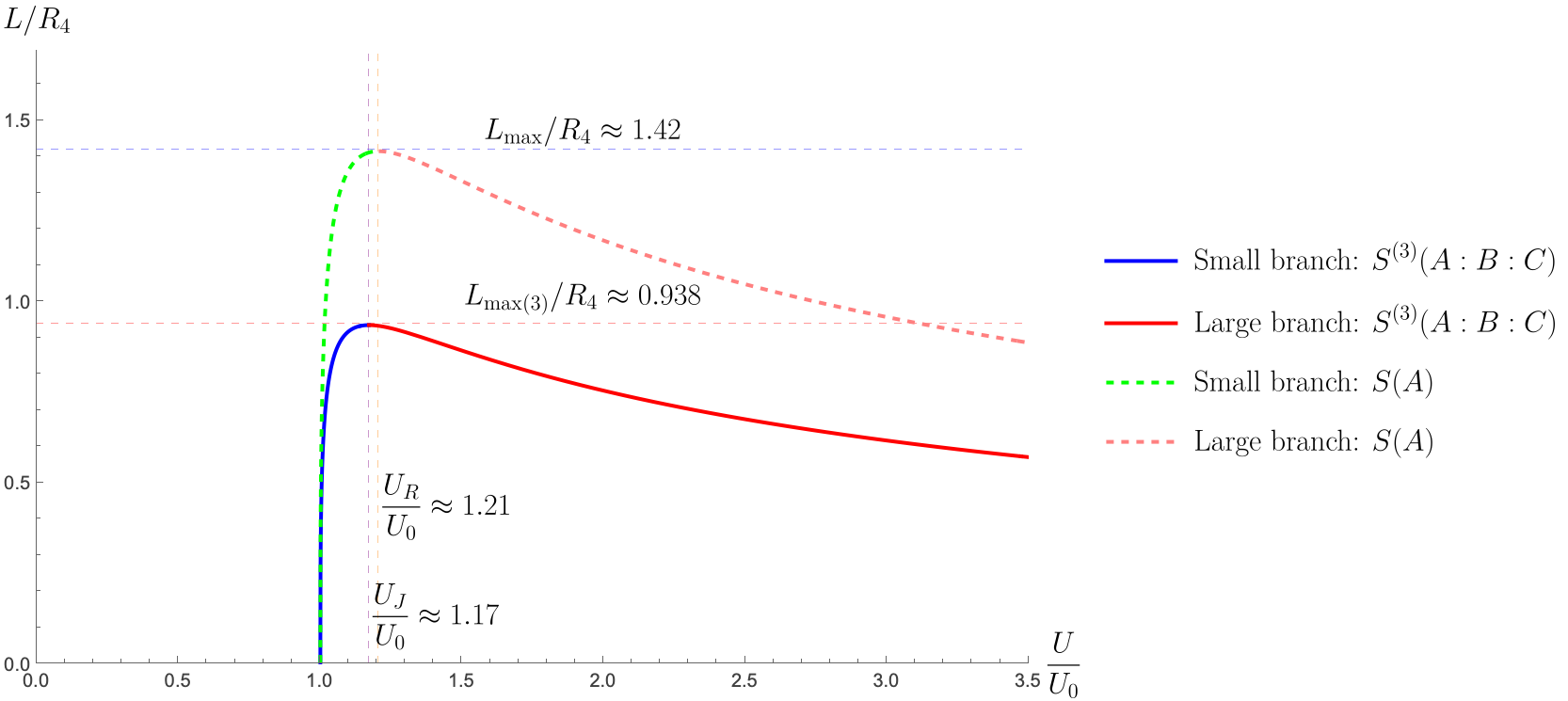}
 \caption{The relations between $L$ and $U_J/U_0$ and between $L$ and $U_R/U_0$ in the D4-soliton background. The solid curves correspond to the tripartite problem, while the dashed curves correspond to the bipartite entanglement entropy.}
 \label{fig:D4_L_vs_U}
\end{figure}

\subsection{Genuine multi-entropy}

\begin{figure}[htbp]
  \centering
  \includegraphics[width=0.82\textwidth]{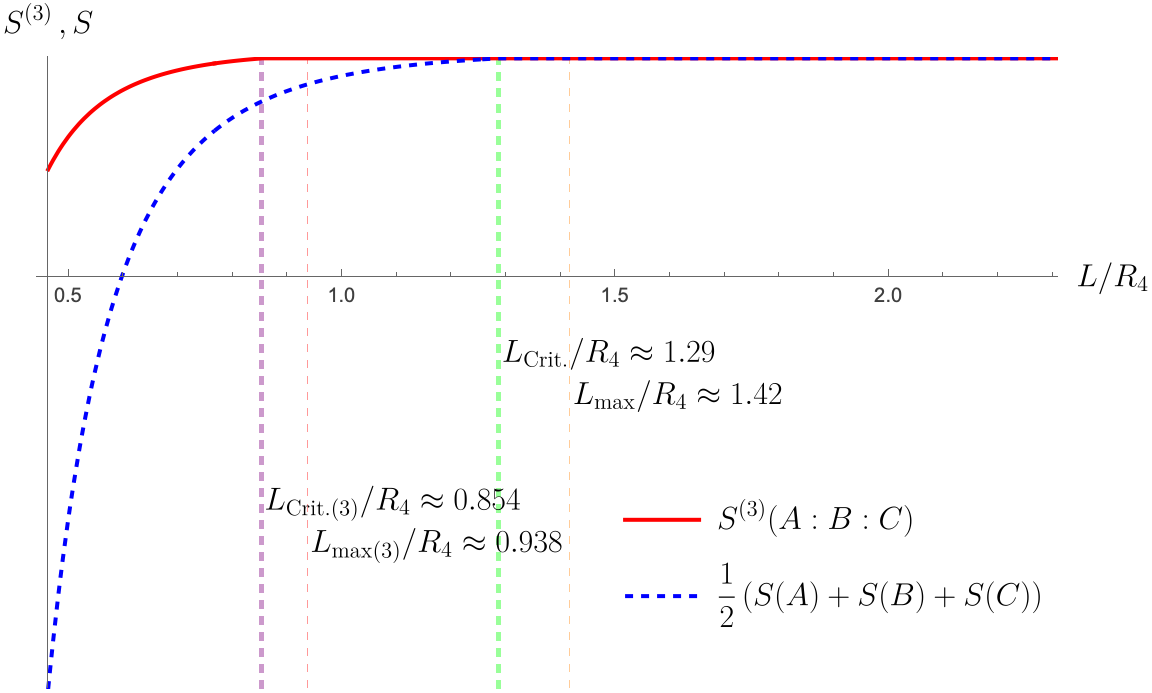}
  \caption{The tripartite multi-entropy $S^{(3)}(A:B:C)$ and the bipartite combination $\frac12(S(A)+S(B)+S(C))$ in the D4-soliton background. Again, note that the horizontal axis is not at zero; both functions remain positive throughout.}
  \label{fig:D4_S3_and_S}
\end{figure}

\begin{figure}[htbp]
  \centering
  \includegraphics[width=0.72\textwidth]{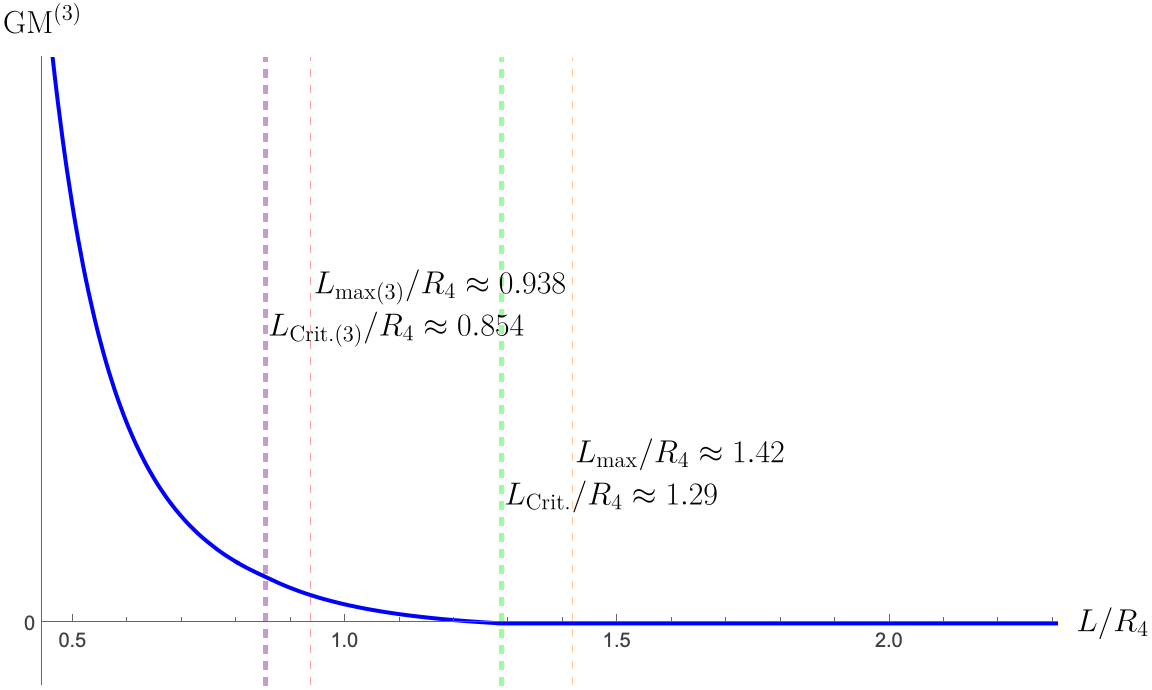}
  \caption{The genuine multi-entropy $\GM^{(3)}(A:B:C)$ in the D4-soliton background. It decreases monotonically and vanishes at $L=L_{\rm Crit.}$.}
  \label{fig:D4_GM}
\end{figure}

The genuine multi-entropy is
\begin{equation}
\GM^{(3)}(A:B:C)
=
S^{(3)}(A:B:C)-\frac12\Bigl(S(A)+S(B)+S(C)\Bigr).
\label{eq:D4_GM_main}
\end{equation}
In Fig.~\ref{fig:D4_S3_and_S}, the first and second terms of \eqref{eq:D4_GM_main} are plotted as a function of $L$, and Fig.~\ref{fig:D4_GM} shows the resulting genuine multi-entropy. 
Numerically, one finds that $\GM^{(3)}$ decreases monotonically with $L$ and vanishes exactly at
\begin{equation}
L=L_{\rm Crit.}\approx 1.29\,R_4.
\label{eq:D4_GM_zero_main}
\end{equation}
The mechanism behind this vanishing is straightforward.
For $L>L_{\rm Crit.(3)}$, the tripartite multi-entropy has already switched to the
cap-assisted disconnected branch,
\begin{equation}
S^{(3)}(A:B:C)=S_W^{(3)}=2S(B).
\end{equation}
Therefore, in the intermediate regime
\begin{equation}
L_{\rm Crit.(3)}<L<L_{\rm Crit.},
\end{equation}
the genuine multi-entropy reduces to
\begin{equation}
\GM^{(3)}(A:B:C)=S(B)-\frac12 S_{\rm conn}(A).
\end{equation}
As $L$ increases, $S_{\rm conn}(A)$ grows monotonically and reaches
\begin{equation}
S_{\rm disc}(A)=2S(B)
\end{equation}
at $L=L_{\rm Crit.}$, so $\GM^{(3)}$ vanishes exactly there.
For $L>L_{\rm Crit.}$, both the tripartite and bipartite sectors are on disconnected branches, and the cancellation remains exact.
Thus the disappearance of the genuinely multipartite contribution is controlled by the bipartite transition, not by the tripartite one.

\subsection{Small-\texorpdfstring{$L$}{L} asymptotics and comparison with the hard-wall benchmark}

We now analyze this decreasing behavior for small $L$.
In the small-$L$ regime, both $U_J$ and $U_R$ lie deep in the UV, so the leading asymptotics can be extracted by taking the limit $U_0\to 0$. In this limit,
\begin{equation}
F(U)\sim U^{5/2},
\qquad
\beta(U)\sim \frac{R^3}{U^3},
\qquad
\sqrt{\beta(U)}\,F(U)\sim R^{3/2}U .
\label{eq:D4_smallL_UVdata_main}
\end{equation}
The width relations for both the tripartite and bipartite connected branches then reduce to
\begin{equation}
L \sim \frac{R^{3/2}}{U_J^{1/2}},
\qquad
L \sim \frac{R^{3/2}}{U_R^{1/2}},
\label{eq:D4_smallL_scaling_U_main}
\end{equation}
so that
\begin{equation}
U_J\sim \frac{R^3}{L^2},
\qquad
U_R\sim \frac{R^3}{L^2}.
\label{eq:D4_U_scaling_main}
\end{equation}

\begin{figure}[htbp]
    \centering
    \includegraphics[width=0.82\textwidth]{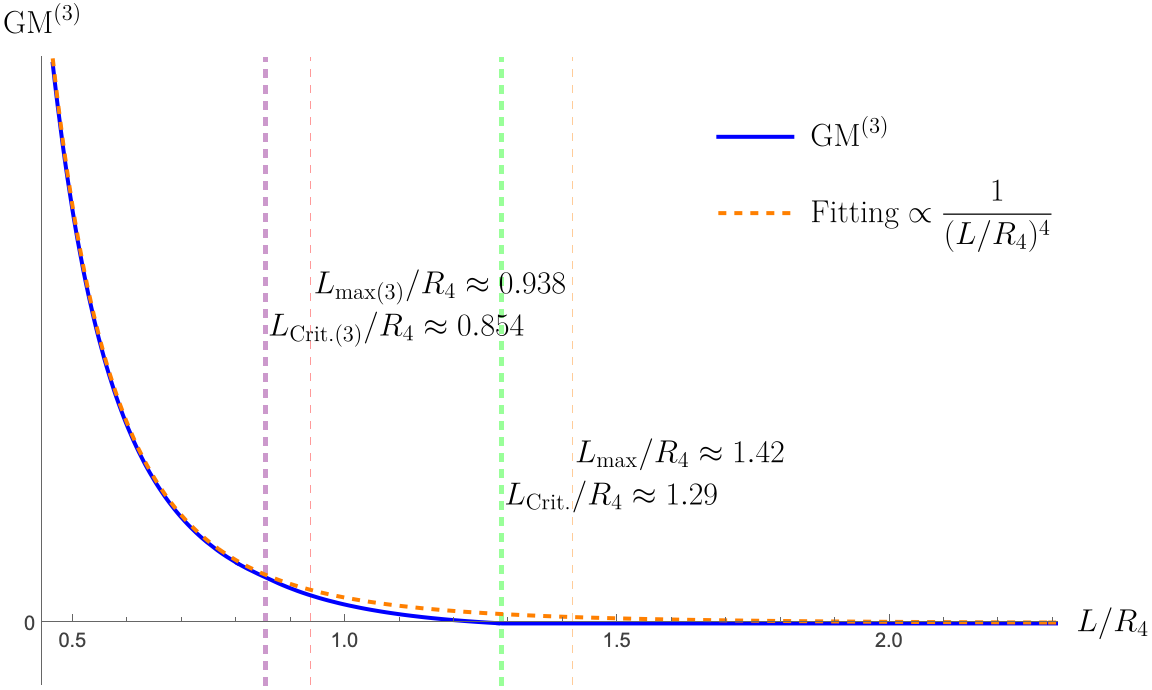}
    \caption{Small-$L$ behavior of the genuine multi-entropy in the D4-soliton background. The numerical data are well fitted by a $1/L^4$ fall-off.}
    \label{fig:D4_GM_fit}
\end{figure}

It is also straightforward to see the scaling of the finite entropy differences entering $\GM^{(3)}$.  
For the connected tripartite branch, the finite quantity
\begin{equation}
S_Y^{(3)}-S_W^{(3)}
\end{equation}
contains the stem contribution
\begin{equation}
S_{\rm stem}(U_J)\sim R^{3/2}U_J^2,
\end{equation}
and the subtracted side-branch contribution scales in the same way. Indeed, after writing
\begin{equation}
c_J^2\sim U_J^5
\end{equation}
and rescaling $U=U_J y$, the side-branch difference becomes
\begin{equation}
\int_{U_J}^{\infty} dU\,\sqrt{\beta(U)}
\left[
\frac{F(U)^2}{\sqrt{F(U)^2-c_J^2}}-F(U)
\right]
\sim
R^{3/2}U_J^2
\int_1^\infty dy
\left[
\frac{y^{7/2}}{\sqrt{y^5-\frac34}}-y
\right],
\end{equation}
so its $U_J$-dependence is also $R^{3/2}U_J^2$.  
Exactly the same argument for the bipartite connected/disconnected difference gives
\begin{equation}
S_{\rm conn}(A)-S_{\rm disc}(A)\sim R^{3/2}U_R^2 .
\end{equation}
Therefore, after the leading UV-divergent pieces cancel in the combination defining $\GM^{(3)}$, the remaining finite contribution scales as
\begin{equation}
\GM^{(3)}(A:B:C)\sim U_J^2\sim U_R^2\sim \frac{R^6}{L^4}.
\end{equation}
Hence
\begin{equation}
\GM^{(3)}(A:B:C)\propto \frac{1}{L^4},
\qquad
L\to 0.
\label{eq:D4_smallL_GM_main}
\end{equation}
This agrees with the numerical fit as in Fig.~\ref{fig:D4_GM_fit}.

The main qualitative difference from the hard-wall benchmark is now clear. In the hard-wall model, $\GM^{(3)}$ exhibits a plateau at small $L/z_0$,
reflecting the existence of a sharp IR boundary that supports a wall-assisted configuration of finite cost. In the D4-soliton background, by contrast, the IR region ends smoothly, and the genuine multi-entropy decreases monotonically without developing such a plateau.

At the same time, the D4-soliton background preserves the basic tripartite logic of the hard-wall analysis: the connected $Y$ saddle dominates at small $L$, the cap-assisted disconnected saddle dominates at large $L$, and the competition between these saddles determines the behavior of $\GM^{(3)}$. Thus the hard-wall benchmark remains qualitatively useful, but the smooth cap changes the detailed shape of the answer in an essential way.

\section{The D3-soliton background}
\label{sec:D3_soliton}

We now turn to the D3-soliton background, which provides a second smooth confining geometry in which the same tripartite construction can be implemented. This section is parallel to Sec.~\ref{sec:D4_soliton}, but it will also be important for isolating which features are robust across smooth caps and which depend on the detailed UV/IR structure of the background.

As we will see, the qualitative phase structure closely parallels that of the D4 case: the hard-wall plateau again disappears, and $\GM^{(3)}$ decreases monotonically to zero. However, the short-distance fall-off is different, scaling as $1/L^2$ rather than $1/L^4$.

\subsection{Geometry, tripartition, and reduced functional}

We consider the near-horizon geometry of $N_c$ D3-branes compactified on a circle with anti-periodic boundary conditions for fermions. In string frame,
\begin{equation}
ds_{10}^2
=
\left(\frac{U}{R}\right)^2
\left[
-dt^2+dx_1^2+dx_2^2+h(U)\,dx_3^2
\right]
+
\left(\frac{R}{U}\right)^2\frac{dU^2}{h(U)}
+
R^2 d\Omega_5^2,
\label{eq:D3_metric_main}
\end{equation}
where
\begin{equation}
h(U)=1-\left(\frac{U_0}{U}\right)^4.
\label{eq:D3_h_main}
\end{equation}
The compact coordinate satisfies $x_3\sim x_3+2\pi R_3$, and the geometry ends smoothly at the cigar tip $U=U_0$.
Note that smoothness at the tip requires
\begin{equation}
R_3=\frac{1}{2}\frac{R^{2}}{U_0},
\qquad\Longleftrightarrow\qquad
U_0=\frac{1}{2}\frac{R^2}{R_3}.
\label{eq:D3_periodicity_main}
\end{equation}

As in the D4 case, we take the entangling direction to be $x\equiv x_1$.

We again consider
\begin{equation}
A=\left[-\frac{L}{2},\frac{L}{2}\right],
\qquad
B=\left(-\infty,-\frac{L}{2}\right),
\qquad
C=\left(\frac{L}{2},\infty\right),
\label{eq:D3_tripartition_main}
\end{equation}
so that the comparison with both the hard-wall and D4 analyses is direct.

The reduced strip functional again takes the general form
\begin{equation}
S_{\rm branch}
=
\mathcal N_{D3}\int dx\,F(U)\sqrt{1+\beta(U)\,U'(x)^2},
\label{eq:D3_branch_general_main}
\end{equation}
with
\begin{equation}
F(U)=U\sqrt{U^4-U_0^4},
\qquad
\beta(U)=\frac{R^4}{U^4-U_0^4}.
\label{eq:D3_F_beta_main}
\end{equation}
As in the D4 case, the effective optical metric
\begin{equation}
ds_{\rm eff}^2
=
F(U)^2\left(dx^2+\beta(U)\,dU^2\right)
\label{eq:D3_effective_metric_main}
\end{equation}
makes the equal-tension nature of the problem manifest, and the Steiner condition at the junction gives
\begin{equation}
c_J=\frac{\sqrt{3}}{2}F(U_J)
=
\frac{\sqrt{3}}{2}U_J\sqrt{U_J^4-U_0^4}.
\label{eq:D3_cJ_main}
\end{equation}

A useful simplification, analogous to the D4 case, is
\begin{equation}
\sqrt{\beta(U)}\,F(U)=R^2U.
\label{eq:D3_sqrtbetaF_main}
\end{equation}

\subsection{Connected and disconnected tripartite candidates}

The half-width relation for the connected $Y$ candidate is
\begin{equation}
\frac{L}{2}
=
R^2 c_J
\int_{U_J}^{\infty}
\frac{dU}
{\sqrt{U^4-U_0^4}\,
\sqrt{U^2(U^4-U_0^4)-c_J^2}},
\qquad
c_J=\frac{\sqrt{3}}{2}U_J\sqrt{U_J^4-U_0^4}.
\label{eq:D3_L_UJ_main}
\end{equation}
The connected side-branch and stem contributions are
\begin{equation}
S_{\rm side}(U_J)
=
\mathcal N_{D3}R^2
\int_{U_J}^{U_\infty}
dU\,
\frac{U^2\sqrt{U^4-U_0^4}}
{\sqrt{U^2(U^4-U_0^4)-c_J^2}},
\label{eq:D3_Sside_main}
\end{equation}
and
\begin{equation}
S_{\rm stem}(U_J)
=
\frac{\mathcal N_{D3}R^2}{2}(U_J^2-U_0^2),
\label{eq:D3_Sstem_main}
\end{equation}
so that
\begin{equation}
S_Y^{(3)}(L)=2S_{\rm side}(U_J)+S_{\rm stem}(U_J).
\label{eq:D3_SY_main}
\end{equation}

The cap-assisted disconnected candidate is
\begin{equation}
S_W^{(3)}
=
\mathcal N_{D3}R^2(U_\infty^2-U_0^2),
\label{eq:D3_SW_main}
\end{equation}
and the finite difference is
\begin{align}
\frac{S_Y^{(3)}-S_W^{(3)}}{\mathcal N_{D3}}
&=
2\int_{U_J}^{U_\infty}dU\,\sqrt{\beta(U)}
\left[
\frac{F(U)^2}{\sqrt{F(U)^2-c_J^2}}-F(U)
\right]
-\frac{R^2}{2}(U_J^2-U_0^2).
\label{eq:D3_SY_minus_SW_main}
\end{align}

\subsection{Ordinary entanglement entropy and numerical phase structure}

For the finite interval $A$, the turning-point condition gives
\begin{equation}
c_R=F(U_R)=U_R\sqrt{U_R^4-U_0^4},
\label{eq:D3_cR_main}
\end{equation}
and the width relation is
\begin{equation}
\frac{L}{2}
=
R^2 c_R
\int_{U_R}^{\infty}
\frac{dU}
{\sqrt{U^4-U_0^4}\,
\sqrt{U^2(U^4-U_0^4)-c_R^2}}.
\label{eq:D3_L_UR_main}
\end{equation}
The connected and disconnected candidates are
\begin{equation}
S_{\rm conn}(A)
=
2\mathcal N_{D3}R^2
\int_{U_R}^{U_\infty}
dU\,
\frac{U^2\sqrt{U^4-U_0^4}}
{\sqrt{U^2(U^4-U_0^4)-c_R^2}},
\label{eq:D3_SA_conn_main}
\end{equation}
and
\begin{equation}
S_{\rm disc}(A)
=
\mathcal N_{D3}R^2(U_\infty^2-U_0^2),
\label{eq:D3_SA_disc_main}
\end{equation}
so that
\begin{equation}
S(A)=\min\{S_{\rm conn}(A),S_{\rm disc}(A)\}.
\label{eq:D3_SA_main}
\end{equation}
Meanwhile,
\begin{equation}
S(B)=S(C)=\frac12 \mathcal N_{D3}R^2(U_\infty^2-U_0^2).
\label{eq:D3_SB_SC_main}
\end{equation}

\begin{figure}[htbp]
    \centering
    \includegraphics[width=0.78\textwidth]{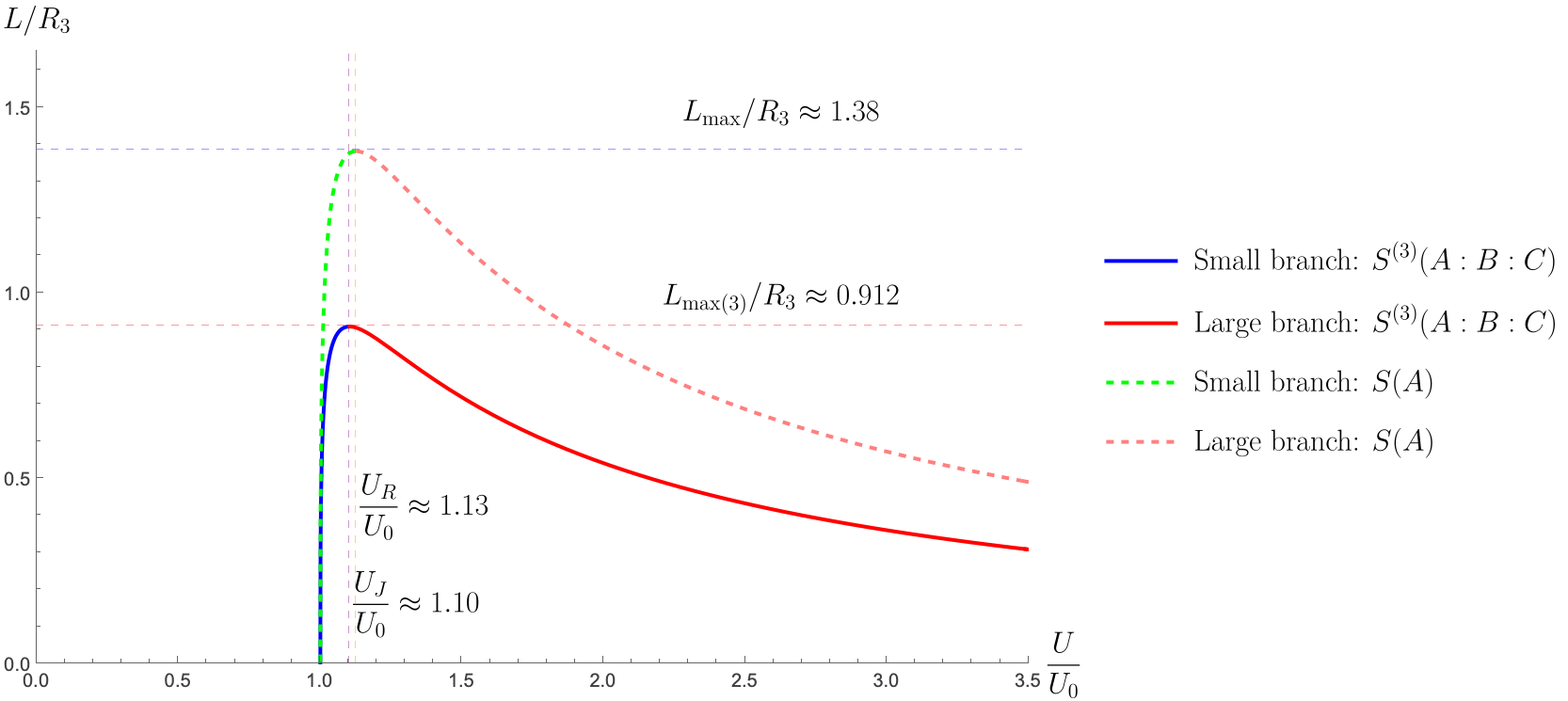}
    \caption{The relations between $L$ and $U_J/U_0$ and between $L$ and $U_R/U_0$ in the D3-soliton background.}
    \label{fig:D3_L_vs_U}
\end{figure}

\begin{figure}[htbp]
    \centering
    \begin{minipage}[b]{0.46\textwidth}
        \centering
        \includegraphics[width=\textwidth]{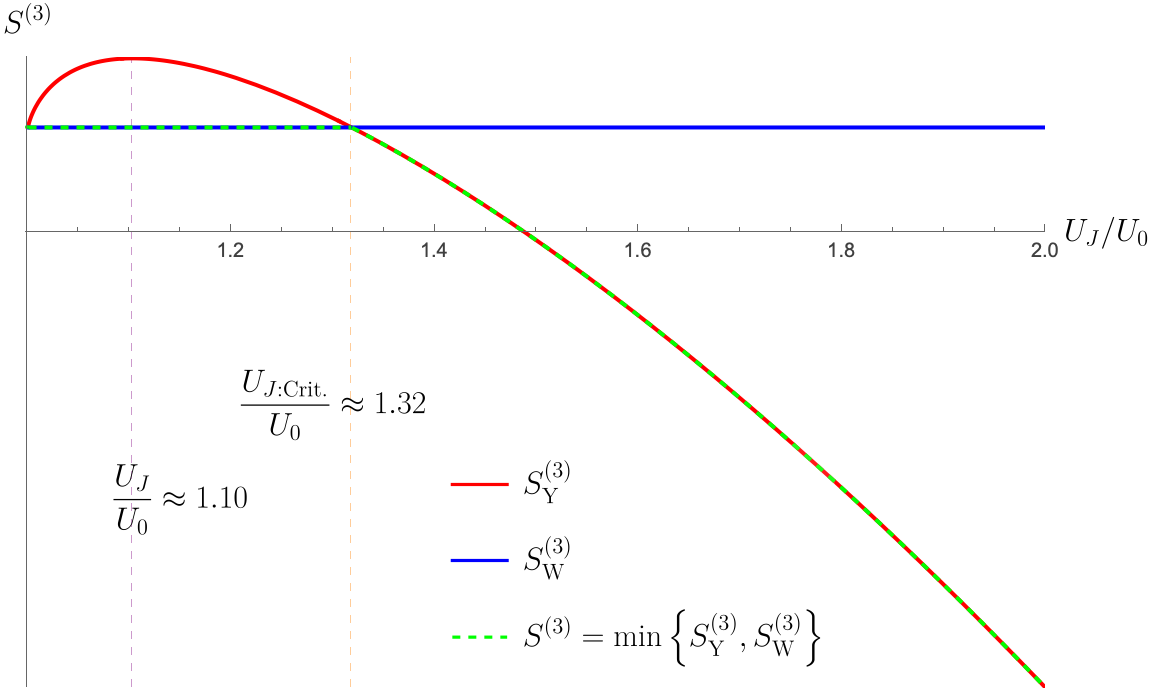}
        \\ (a) $S^{(3)}$
    \end{minipage}
    \hfill
    \begin{minipage}[b]{0.46\textwidth}
        \centering
        \includegraphics[width=\textwidth]{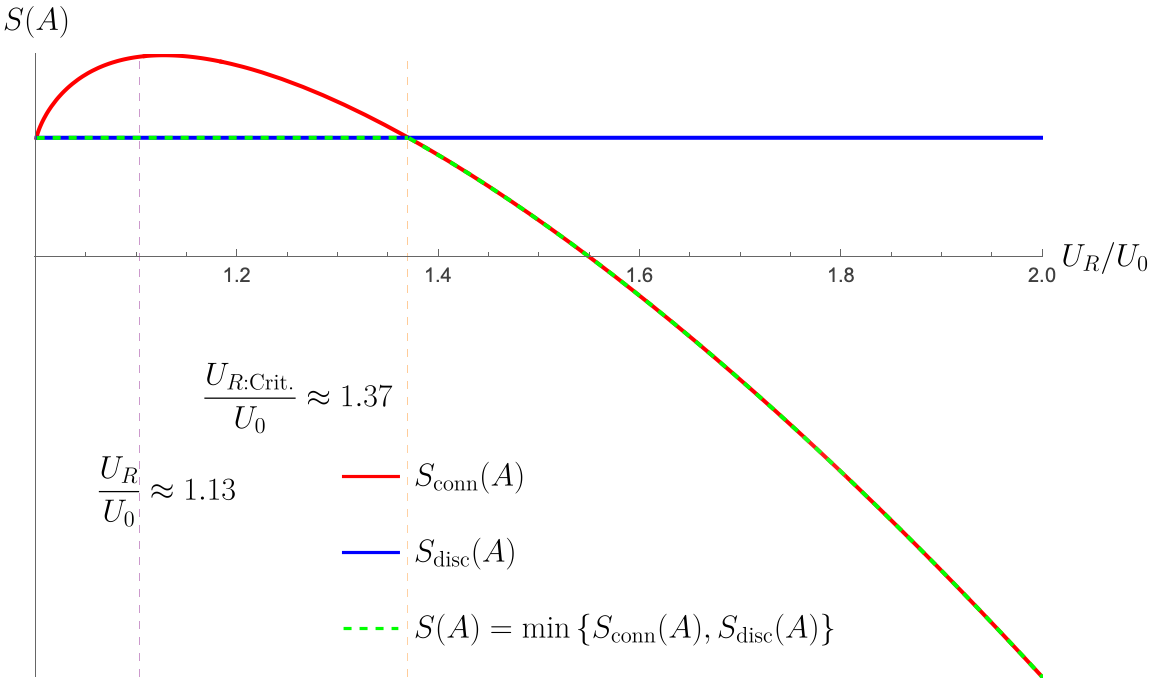}
        \\ (b) $S(A)$
    \end{minipage}
    \caption{Phase transitions in the D3-soliton background. In each case, the physical entropy is obtained by minimizing between the connected and disconnected candidates. Note that the horizontal axis is not at zero; both functions remain positive throughout.}
    \label{fig:D3_entropy_transitions}
\end{figure}

The numerical analysis parallels the D4 case.
As in Fig.~\ref{fig:D3_L_vs_U}, the relations between $L$ and $U_J/U_0$ and between $L$ and $U_R/U_0$ exhibit the familiar two-branch structure, with maximal lengths
\begin{equation}
L_{\max(3)}\approx 0.912\,R_3,
\qquad
L_{\max}\approx 1.38\,R_3. 
\label{eq:D3_Lmax_main}
\end{equation}
The critical transition scales are
\begin{equation}
L_{\rm Crit.(3)}\approx 0.813\,R_3,
\qquad
L_{\rm Crit.}\approx 1.23\,R_3, 
\label{eq:D3_Lcrit_main}
\end{equation} 
where these values are attained at $U_J/U_0\approx 1.32$ and $U_R/U_0\approx1.37$ respectively.
As in the bipartite and D4 cases, the physical transition occurs before the curve turns around, so the small branch is never dominant. See Fig.~\ref{fig:D3_entropy_transitions} for the phase structure.

\subsection{Genuine multi-entropy}

\begin{figure}[htbp]
    \centering
    \includegraphics[width=0.72\textwidth]{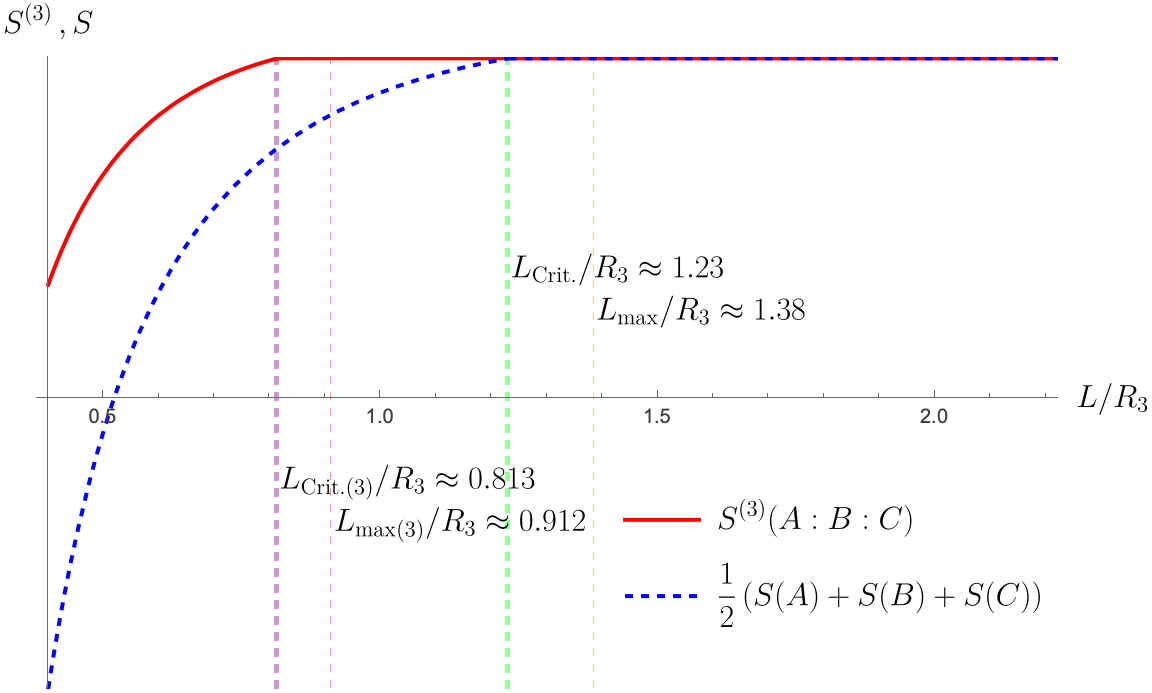}
    \caption{The tripartite multi-entropy $S^{(3)}(A:B:C)$ and the bipartite combination $\frac12(S(A)+S(B)+S(C))$ in the D3-soliton background. Note that the horizontal axis is not at zero; both functions remain positive throughout.}
    \label{fig:D3_S3_and_S}
\end{figure}

\begin{figure}[htbp]
    \centering
    \includegraphics[width=0.72\textwidth]{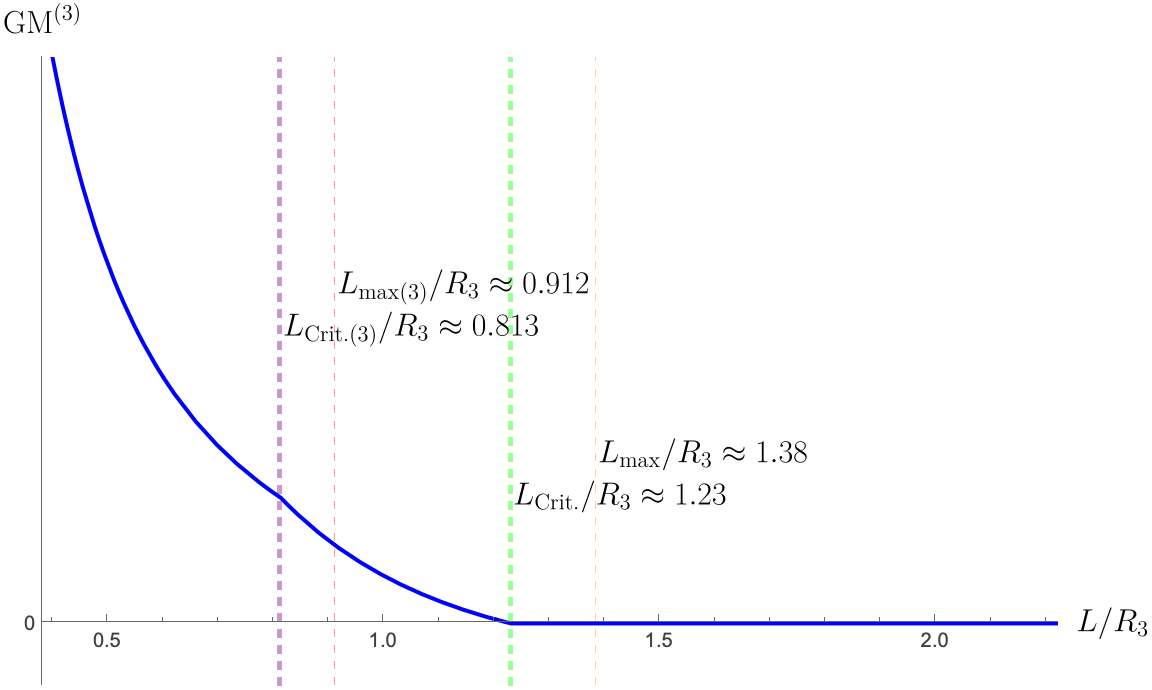}
    \caption{The genuine multi-entropy $\GM^{(3)}(A:B:C)$ in the D3-soliton background.}
    \label{fig:D3_GM}
\end{figure}

The genuine multi-entropy is again
\begin{equation}
\GM^{(3)}(A:B:C)
=
S^{(3)}(A:B:C)-\frac12\Bigl(S(A)+S(B)+S(C)\Bigr).
\label{eq:D3_GM_main}
\end{equation}
Fig.~\ref{fig:D3_S3_and_S} shows the first and second terms of \eqref{eq:D3_GM_main} as functions of $L$, and Fig.~\ref{fig:D3_GM} shows the resulting genuine multi-entropy. 
Numerically, $\GM^{(3)}$ behaves qualitatively in the same way as in the D4 case: it decreases monotonically with $L$, exhibits no hard-wall-like plateau, and vanishes exactly at the entanglement transition scale. Thus,
\begin{equation}
\GM^{(3)}(A:B:C)=0
\qquad
\text{for }L\ge L_{\rm Crit.}.
\label{eq:D3_GM_zero_main}
\end{equation}

\subsection{Small-\texorpdfstring{$L$}{L} asymptotics and comparison with D4 and the hard wall}

As in the D4 case, we can analyze this decreasing behavior for small $L$.
In the small-$L$ regime, we again take the UV limit $U_0\to 0$. In this limit,
\begin{equation}
F(U)\sim U^3,
\qquad
\beta(U)\sim \frac{R^4}{U^4},
\qquad
\sqrt{\beta(U)}\,F(U)\sim R^2 U .
\label{eq:D3_smallL_UVdata_main}
\end{equation}
The width relations for both the tripartite and bipartite connected branches then reduce to
\begin{equation}
L\sim \frac{R^2}{U_J},
\qquad
L\sim \frac{R^2}{U_R},
\label{eq:D3_smallL_scaling_main}
\end{equation}
so that
\begin{equation}
U_J\sim \frac{R^2}{L},
\qquad
U_R\sim \frac{R^2}{L}.
\label{eq:D3_U_scaling_main}
\end{equation}

\begin{figure}[htbp]
    \centering
    \includegraphics[width=0.82\textwidth]{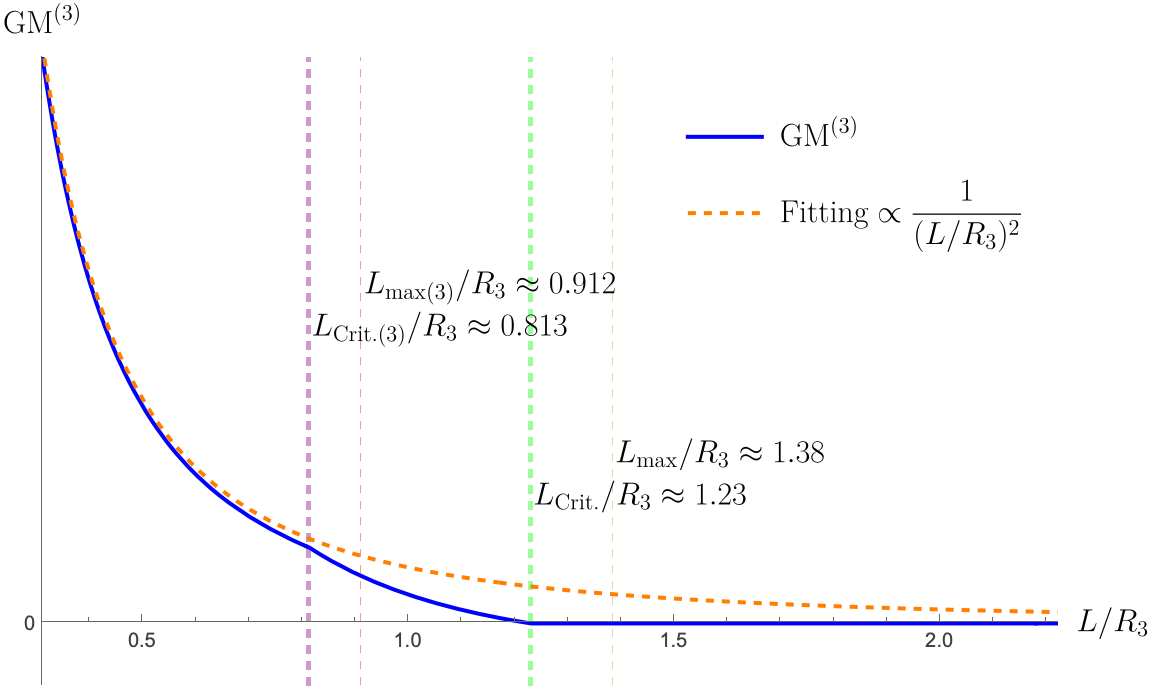}
    \caption{Small-$L$ behavior of the genuine multi-entropy in the D3-soliton background. The numerical data are well fitted by a $1/L^2$ fall-off.}
    \label{fig:D3_GM_fit}
\end{figure}

The scaling of the finite entropy differences can again be seen directly. 
For the connected tripartite branch, one has
\begin{equation}
c_J^2\sim U_J^6 .
\end{equation}
Writing $U=U_J y$, the side-branch difference behaves as
\begin{equation}
\int_{U_J}^{\infty} dU\,\sqrt{\beta(U)}
\left[
\frac{F(U)^2}{\sqrt{F(U)^2-c_J^2}}-F(U)
\right]
\sim
R^2 U_J^2
\int_1^\infty dy
\left[
\frac{y^4}{\sqrt{y^6-\frac34}}-y
\right],
\end{equation}
so its leading $U_J$-dependence is $R^2 U_J^2$.
The stem contribution scales as
\begin{equation}
S_{\rm stem}(U_J)\sim R^2 U_J^2 ,
\end{equation}
and is therefore of the same parametric order.
Likewise, for the bipartite sector one finds
\begin{equation}
S_{\rm conn}(A)-S_{\rm disc}(A)\sim R^2 U_R^2 .
\end{equation}
Therefore, after the leading UV-divergent pieces cancel in the combination defining $\GM^{(3)}$, the remaining finite contribution scales as
\begin{equation}
\GM^{(3)}(A:B:C)\sim U_J^2\sim U_R^2\sim \frac{R^4}{L^2}.
\end{equation}
Since the background-dependent prefactors are fixed, this implies
\begin{equation}
\GM^{(3)}(A:B:C)\propto \frac{1}{L^2},
\qquad
L\to 0.
\label{eq:D3_smallL_GM_main}
\end{equation}
The comparison with the numerical result is shown in Fig.~\ref{fig:D3_GM_fit}, and it shows good agreement. 
Thus the D3-soliton shares the qualitative phase structure of the D4-soliton, but differs in its UV asymptotics.

The comparison is now sharp.

Relative to the hard-wall benchmark, both the D4- and D3-soliton backgrounds eliminate the plateau and replace it with a monotonic decrease. Thus the hard-wall plateau is not a generic feature of confining geometries, but instead appears to be tied to the sharpness of the IR cutoff.

At the same time, the D4 and D3 results are not identical. Their small-$L$ asymptotics differ:
\begin{equation}
\GM^{(3)}_{\rm D4}\sim \frac{1}{L^4},
\qquad
\GM^{(3)}_{\rm D3}\sim \frac{1}{L^2}.
\end{equation}
This suggests a useful separation between two kinds of properties:
\begin{itemize}
\item qualitative phase-structure features that are robust across smooth caps,
\item short-distance exponents that are background-dependent.
\end{itemize}

\section{The Klebanov--Strassler background}
\label{sec:KS_background}

We now turn to the Klebanov--Strassler (KS) warped deformed conifold background. This section is the KS analogue of Secs.~\ref{sec:D4_soliton} and \ref{sec:D3_soliton}. At the present stage, our main goal is to formulate the tripartite problem in a way directly parallel to the D4 and D3 analyses, so that the corresponding numerical results can later be incorporated without changing the overall structure of the paper.

The key new feature of the KS geometry is that the IR cap is provided not by a cigar tip in a wrapped D-brane background, but by the smooth finite-size $S^3$ at the tip of the deformed conifold. This makes KS an important test of how universal the smooth-cap behavior of $\GM^{(3)}$ really is.

\subsection{Geometry, tripartition, and reduced functional}

The Klebanov--Strassler solution is a warped product of $\mathbb{R}^{1,3}$ with the deformed conifold,
\begin{equation}
ds^2_{10}
=
h(\tau)^{-1/2}\eta_{\mu\nu}\,dx^\mu dx^\nu
+
h(\tau)^{1/2}ds_6^2,
\label{eq:KS_metric_main}
\end{equation}
where $\tau$ is the radial coordinate on the deformed conifold,
and $h(\tau)$ is the warp factor given by
\begin{equation}
h(\tau) = (g_s M \alpha')^2\, 2^{2/3}\, \epsilon^{-8/3}
\int_\tau^\infty d\tau'\;
\frac{\tau'\coth\tau' - 1}{\sinh^2\tau'}\,(\sinh 2\tau' - 2\tau')^{1/3}\,.
\label{eq:warp_factor}
\end{equation}
Here, $g_s M$ is the ‘t Hooft coupling. The internal metric is
\begin{equation}
ds_6^2
=
\frac{\epsilon^{4/3}}{2}K(\tau)\Bigg[
\frac{1}{3K(\tau)^3}\big(d\tau^2+(g^5)^2\big)
+\cosh^2\!\left(\frac{\tau}{2}\right)\big((g^3)^2+(g^4)^2\big)
+\sinh^2\!\left(\frac{\tau}{2}\right)\big((g^1)^2+(g^2)^2\big)
\Bigg],
\label{eq:KS_internal_main}
\end{equation}
with
\begin{equation}
K(\tau)=\frac{(\sinh 2\tau-2\tau)^{1/3}}{2^{1/3}\sinh\tau}.
\label{eq:KS_Ktau_main}
\end{equation}
Here, $g_i$ is the angular forms \cite{Klebanov:2000hb}.
The warp factor $h(\tau)$ is finite at $\tau=0$, and the geometry ends smoothly at the deformed-conifold tip.
Note that this background has a natural physical scale, the glueball mass scale $m_{\text{glueball}}$, defined by
\begin{equation}
m_{\text{glueball}} = \frac{\epsilon^{2/3}}{g_s M \alpha'}.\label{eq:gluball-mass}
\end{equation}

As in the previous sections, we take the entangling direction to be $x\equiv x_1$.

As before, we consider
\begin{equation}
A=\left[-\frac{L}{2},\frac{L}{2}\right],
\qquad
B=\left(-\infty,-\frac{L}{2}\right),
\qquad
C=\left(\frac{L}{2},\infty\right).
\label{eq:KS_tripartition_main}
\end{equation}
This provides the direct KS analogue of the D4, D3, and hard-wall benchmark setups.

For a strip-like branch described by a profile $\tau(x)$, the entropy functional again reduces to the universal one-dimensional form
\begin{equation}
S_{\rm branch}
=
\mathcal N_{\rm KS}\int dx\,F(\tau)\sqrt{1+\beta(\tau)\,\tau'(x)^2}.
\label{eq:KS_branch_general_main}
\end{equation}
In the KS background,
\begin{equation}
\beta(\tau)=\frac{h(\tau)\,\epsilon^{4/3}}{6K(\tau)^2},
\label{eq:KS_beta_main}
\end{equation}
and
\begin{equation}
H(\tau)=\frac{8\pi^6}{3}\,\epsilon^{20/3}h(\tau)K(\tau)^2\sinh^4\tau,
\qquad
F(\tau)=\sqrt{H(\tau)}.
\label{eq:KS_H_F_main}
\end{equation}
Unlike in the D4 and D3 cases, the combination $\sqrt{\beta}\,F$ does not reduce to a simple monomial in the radial variable, so the stem and disconnected contributions must in general be evaluated numerically.

\subsection{Connected and disconnected tripartite candidates}

The effective optical metric is
\begin{equation}
ds_{\rm eff}^2
=
F(\tau)^2\left(dx^2+\beta(\tau)\,d\tau^2\right).
\label{eq:KS_effective_metric_main}
\end{equation}
The connected tripartite saddle again consists of two side branches and a vertical stem, and the equal-tension Steiner condition gives the same universal relation
\begin{equation}
c_J=\frac{\sqrt{3}}{2}F(\tau_J),
\label{eq:KS_cJ_main}
\end{equation}
where $\tau_J$ is the radial position of the junction.

The half-width condition becomes
\begin{equation}
\frac{L}{2}
=
\int_{\tau_J}^{\infty}d\tau\,
\frac{\sqrt{\beta(\tau)}\,c_J}{\sqrt{F(\tau)^2-c_J^2}},
\qquad
c_J=\frac{\sqrt{3}}{2}F(\tau_J).
\label{eq:KS_L_tauJ_main}
\end{equation}
Thus the connected tripartite candidate is
\begin{equation}
S_Y^{(3)}(L)
=
2S_{\rm side}(\tau_J)+S_{\rm stem}(\tau_J),
\label{eq:KS_SY_main}
\end{equation}
with
\begin{equation}
S_{\rm side}(\tau_J)
=
\mathcal N_{\rm KS}
\int_{\tau_J}^{\infty}d\tau\,
\frac{\sqrt{\beta(\tau)}\,F(\tau)^2}{\sqrt{F(\tau)^2-c_J^2}},
\label{eq:KS_Sside_main}
\end{equation}
and
\begin{equation}
S_{\rm stem}(\tau_J)
=
\mathcal N_{\rm KS}\int_{0}^{\tau_J}d\tau\,\sqrt{\beta(\tau)}\,F(\tau).
\label{eq:KS_Sstem_main}
\end{equation}

The cap-assisted disconnected tripartite candidate is
\begin{equation}
S_W^{(3)}
=
2\mathcal N_{\rm KS}\int_{0}^{\infty}d\tau\,\sqrt{\beta(\tau)}\,F(\tau).
\label{eq:KS_SW_main}
\end{equation}
As in the D4 and D3 cases, the difference
\begin{equation}
S_Y^{(3)}-S_W^{(3)}
\label{eq:KS_SY_minus_SW_main}
\end{equation}
is UV finite and is the natural quantity to use in locating the tripartite phase transition.

\subsection{Ordinary entanglement entropy and genuine multi-entropy}

For the finite interval $A$, the connected RT surface is characterized by a turning point $\tau_R$, with conserved quantity
\begin{equation}
c_R=F(\tau_R).
\label{eq:KS_cR_main}
\end{equation}
The corresponding width relation is
\begin{equation}
\frac{L}{2}
=
\int_{\tau_R}^{\infty}d\tau\,
\frac{\sqrt{\beta(\tau)}\,c_R}{\sqrt{F(\tau)^2-c_R^2}},
\qquad
c_R=F(\tau_R),
\label{eq:KS_L_tauR_main}
\end{equation}
and the connected and disconnected candidates are
\begin{equation}
S_{\rm conn}(A)
=
2\mathcal N_{\rm KS}
\int_{\tau_R}^{\infty}d\tau\,
\frac{\sqrt{\beta(\tau)}\,F(\tau)^2}{\sqrt{F(\tau)^2-c_R^2}},
\label{eq:KS_SA_conn_main}
\end{equation}
and
\begin{equation}
S_{\rm disc}(A)
=
2\mathcal N_{\rm KS}\int_{0}^{\infty}d\tau\,\sqrt{\beta(\tau)}\,F(\tau).
\label{eq:KS_SA_disc_main}
\end{equation}
Thus
\begin{equation}
S(A)=\min\{S_{\rm conn}(A),S_{\rm disc}(A)\},
\label{eq:KS_SA_main}
\end{equation}
while for the semi-infinite regions
\begin{equation}
S(B)=S(C)=\mathcal N_{\rm KS}\int_{0}^{\infty}d\tau\,\sqrt{\beta(\tau)}\,F(\tau).
\label{eq:KS_SB_SC_main}
\end{equation}

The genuine multi-entropy is therefore
\begin{equation}
\GM^{(3)}(A:B:C)
=
S^{(3)}(A:B:C)-\frac12\Bigl(S(A)+S(B)+S(C)\Bigr),
\label{eq:KS_GM_main}
\end{equation}
with
\begin{equation}
S^{(3)}(A:B:C)=\min\{S_Y^{(3)},S_W^{(3)}\}.
\label{eq:KS_S3_main}
\end{equation}
This is the direct KS analogue of the D4 and D3 constructions.

\subsection{Numerical analysis for genuine multi-entropy}

The numerical analysis in the KS background proceeds in direct parallel with the D4 and D3 cases. One first tabulates $h(\tau)$ and hence $F(\tau)$ and $\beta(\tau)$, then solves \eqref{eq:KS_L_tauJ_main} and \eqref{eq:KS_L_tauR_main} for $\tau_J(L)$ and $\tau_R(L)$, evaluates the finite differences $S_Y^{(3)}-S_W^{(3)}$ and $S_{\rm conn}(A)-S_{\rm disc}(A)$, determines the physical branches by minimization, and finally constructs $\GM^{(3)}$ from \eqref{eq:KS_GM_main}.

Below we show that the KS background follows qualitatively the same monotonic smooth-cap pattern as the D4 and D3 cases. The connected $Y$ saddle dominates at small $L$, while the cap-assisted disconnected saddle dominates at large $L$, and $\GM^{(3)}$ vanishes beyond the entanglement transition scale.

It is useful to summarize the comparison at the level of formalism. The reduced strip functional has the same universal form, the connected $Y$ saddle is again determined by the universal Steiner relation $c_J=\frac{\sqrt{3}}{2}F$, and the bipartite connected surface is again determined by $c_R=F$. The main new feature is that, unlike in D4 and D3, the KS stem integral is not analytically simple.

Thus the KS case is not conceptually different from the D4 and D3 analyses, but it is technically less explicit. This makes it the natural final example in the sequence: once the framework has been established in Secs.~\ref{sec:D4_soliton} and \ref{sec:D3_soliton}, KS can be treated as the most general smooth-cap test of the same multipartite mechanism.

\begin{figure}[htbp]
    \centering
    \includegraphics[width=0.78\textwidth]{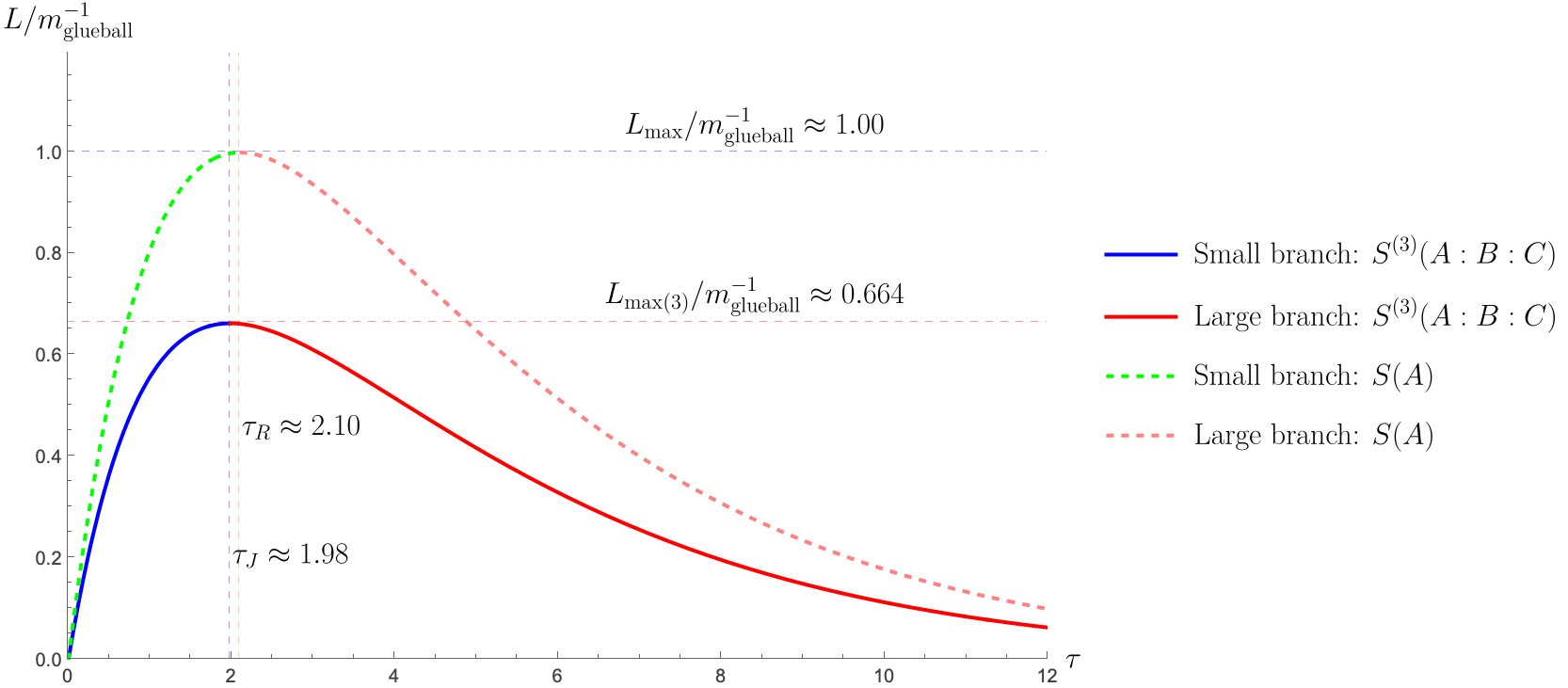}
    \caption{The relations between $L/m_{\text{glueball}}^{-1}$ and $\tau_J$, and between $L/m_{\text{glueball}}^{-1}$ and $\tau_R$ in the KS background.}
    \label{fig:KS_L_vs_tau}
\end{figure}

\begin{figure}[htbp]
    \centering
    \begin{minipage}[b]{0.46\textwidth}
        \centering
        \includegraphics[width=\textwidth]{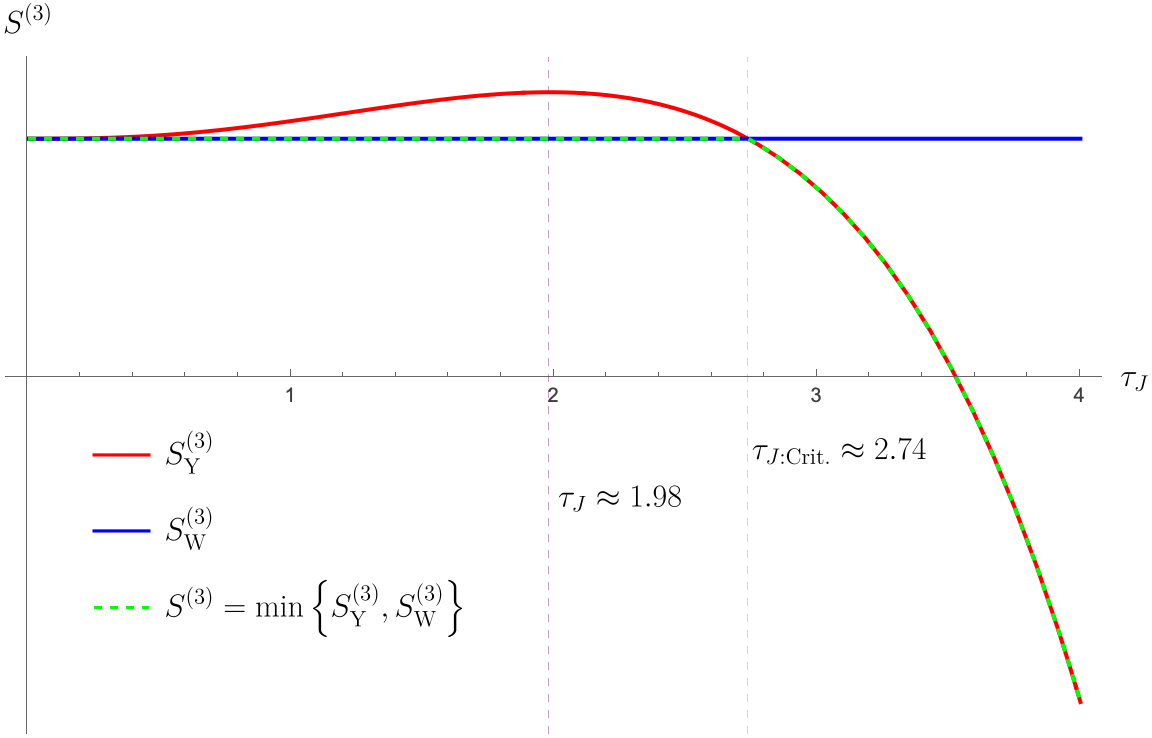}
        \\ (a) $S^{(3)}$
    \end{minipage}
    \hfill
    \begin{minipage}[b]{0.46\textwidth}
        \centering
        \includegraphics[width=\textwidth]{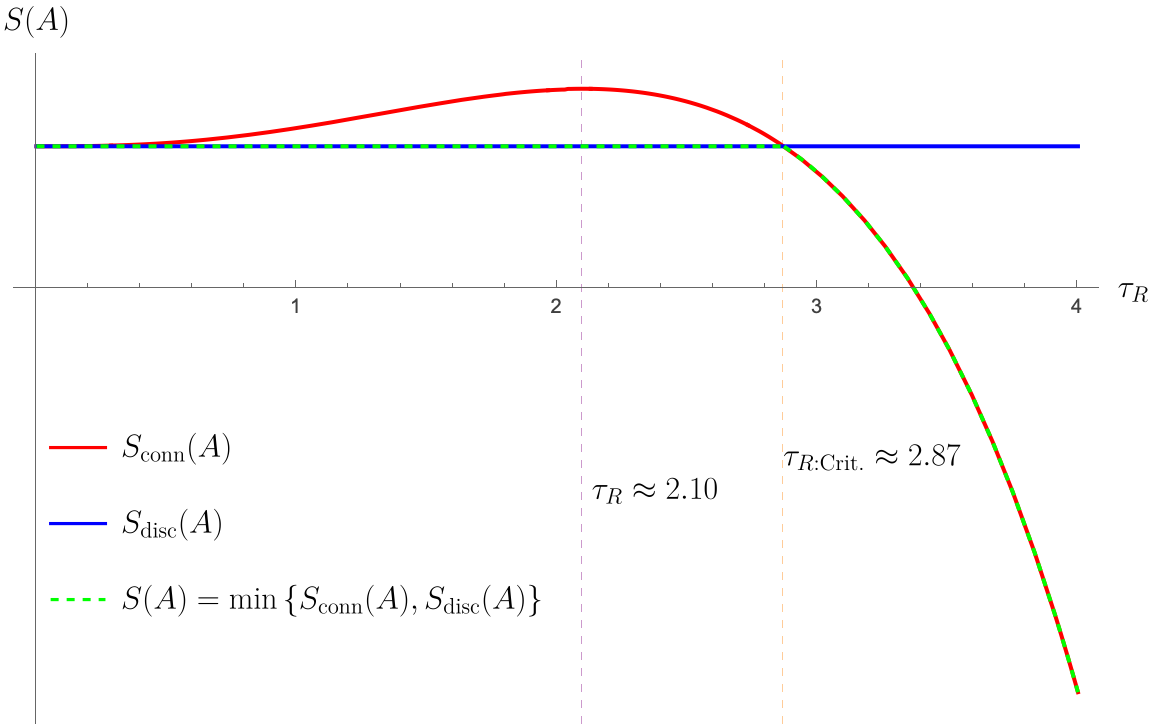}
        \\ (b) $S(A)$
    \end{minipage}
    \caption{Phase transitions in the KS background. In each case, the physical entropy is obtained by minimizing between the connected and disconnected candidates. Note that the horizontal axis is not at zero; both functions remain positive throughout.}
    \label{fig:KS_entropy_transitions}
\end{figure}

First, in Fig.~\ref{fig:KS_L_vs_tau}, the relations between $L/m_{\text{glueball}}^{-1}$ and $\tau_{J}$ \eqref{eq:KS_L_tauJ_main}, and between $L/m_{\text{glueball}}^{-1}$ and $\tau_{R}$ \eqref{eq:KS_L_tauR_main}, show the familiar large-branch/small-branch structure, with maximal lengths
\begin{equation}
L_{\max(3)}/m_{\text{glueball}}^{-1} \approx 0.664,
\qquad 
L_{\max}/m_{\text{glueball}}^{-1} \approx 1.00.
\label{eq:KS_Lmax_main}
\end{equation}
The critical transition scales for the multi-entropy \eqref{eq:KS_S3_main} and the entanglement entropy \eqref{eq:KS_SA_main} are
\begin{equation}
L_{\text{Crit.}(3)}/m_{\text{glueball}}^{-1} \approx 0.632,
\qquad
L_{\text{Crit.}}/m_{\text{glueball}}^{-1} \approx 0.952,
\label{eq:KS_Lcrit_main}
\end{equation}
where these values are attained at $\tau_J\approx 2.74$ and $\tau_R\approx 2.87$ respectively.
As in the D4 and D3 cases, the physical phase transition occurs before the curve turns around, confirming that the small branch is never dominant. See Fig.~\ref{fig:KS_entropy_transitions} for the phase structure.

\begin{figure}[htbp]
    \centering
    \includegraphics[width=0.82\textwidth]{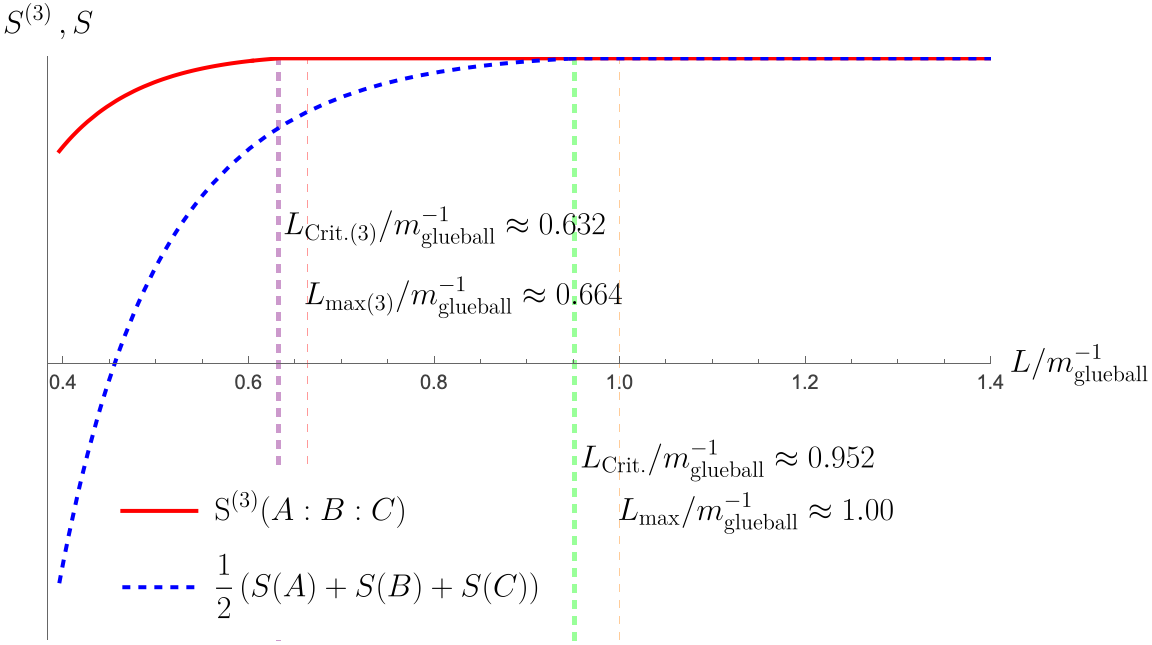}
    \caption{The tripartite multi-entropy $S^{(3)}(A:B:C)$ and the bipartite combination $\frac12(S(A)+S(B)+S(C))$ in the KS background. Note that the horizontal axis is not at zero; both functions remain positive throughout.}
    \label{fig:KS_S3_and_S}
\end{figure}

\begin{figure}[htbp]
    \centering
    \includegraphics[width=0.62\textwidth]{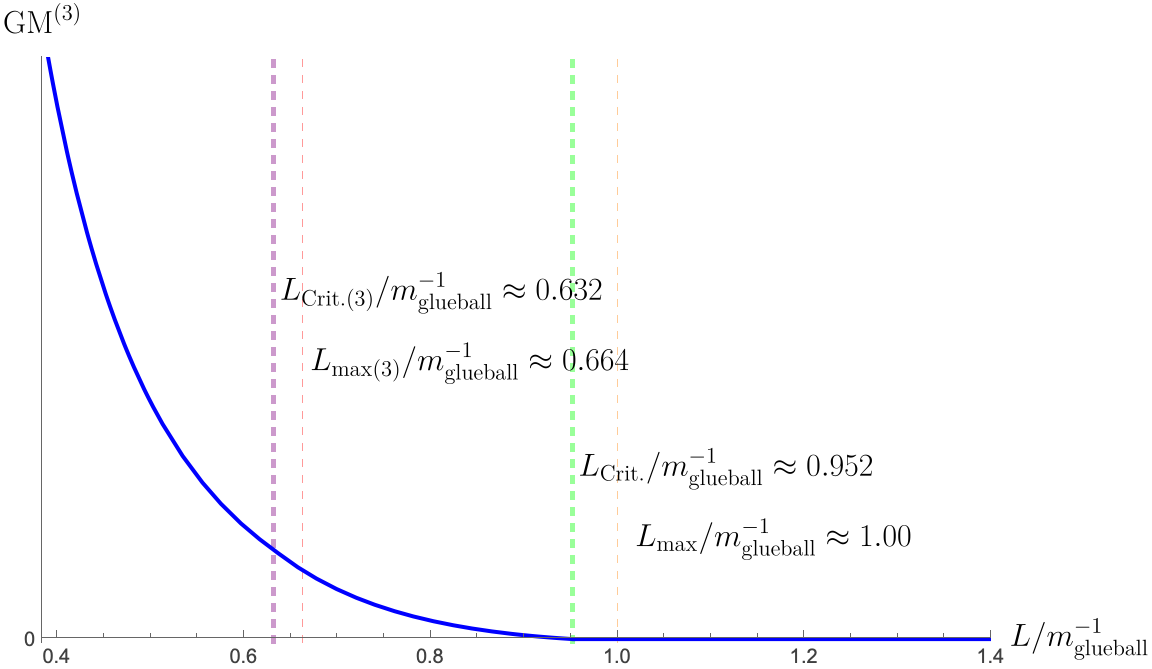}
    \caption{The genuine multi-entropy $\GM^{(3)}(A:B:C)$ in the KS background.}
    \label{fig:KS_GM}
\end{figure}

In Fig.~\ref{fig:KS_S3_and_S}, the first and second terms of \eqref{eq:KS_GM_main} are plotted, while Fig.~\ref{fig:KS_GM} shows the resulting genuine multi-entropy. Numerically, $\GM^{(3)}$ behaves qualitatively in the same way as in the D4 and D3 cases: it decreases monotonically with $L$, exhibits no hard-wall-like plateau, and vanishes exactly at the entanglement transition scale. Thus,
\begin{equation}
\GM^{(3)}(A:B:C)=0
\qquad
\text{for }L\ge L_{\rm Crit.}.
\label{eq:KS_GM_zero_main}
\end{equation}

\subsection{Small-$L$ asymptotics and comparison with D4 and D3}
Regarding the small-$L$ asymptotics, the KS geometry also demonstrates qualitatively the same monotonic smooth-cap pattern as the D4 and D3 geometries, but with an additional logarithmic correction. We can systematically analyze these asymptotics by taking $\tau_J, \tau_R \gg 1$. In this limit, the background functions simplify to
\begin{equation}
	F(\tau)\sim  \sqrt{\tau} e^{\tau} , \qquad \beta(\tau)\sim  \tau e^{-2\tau/3}  , \qquad \sqrt{\beta(\tau)}F(\tau)\sim \tau e^{2\tau/3}.
	\label{eq:KS-asymp-Functions}
\end{equation}
Using these forms, the relations \eqref{eq:KS_L_tauJ_main} and \eqref{eq:KS_L_tauR_main} reduce to 
\begin{equation}
L/m_{\text{glueball}}^{-1} \sim \mathcal{C}_{J} \sqrt{\tau_J} e^{-\tau_J/3}, \qquad L/m_{\text{glueball}}^{-1} \sim \mathcal{C}_{R} \sqrt{\tau_R} e^{-\tau_R/3},
	\label{eq:KS-asymp-L-and-tauJR}
\end{equation}
where $\mathcal{C}_{J}$ and $\mathcal{C}_{R}$ are constants.

Before proceeding, it is illuminating to note the structural relation between
the KS asymptotics and the D3 case.  Introducing the radial coordinate
$U \propto e^{\tau/3}$, or equivalently $\tau \sim 3\log U$, the asymptotic
forms \eqref{eq:KS-asymp-Functions} become
\begin{equation}
	F(\tau)\sim \sqrt{\log U} \cdot U^{3}, \quad
	d\tau \cdot \sqrt{\beta(\tau)}\sim dU\cdot \sqrt{\log U}\cdot\frac{1}{U^{2}}, \quad
	d\tau \cdot \sqrt{\beta(\tau)}F(\tau)\sim dU\cdot \log U \cdot U.
	\label{eq:KS-asymp-Functions-U}
\end{equation}
The right-hand sides have exactly the same $U$-dependence as the D3
asymptotics \eqref{eq:D3_smallL_UVdata_main}, up to overall logarithmic
factors.  This already anticipates the result derived below: the KS
small-$L$ behavior should match that of the D3-soliton, dressed by
logarithmic corrections originating from these $\sqrt{\log U}$ and $\log U$
factors.  The Lambert $W$ analysis that follows will make this expectation
quantitative.

The relations \eqref{eq:KS-asymp-L-and-tauJR} cannot be solved algebraically for $\tau_{J}$ and $\tau_{R}$. However, by utilizing the lower branch of the Lambert W function, $W_{-1}(z)$, which is defined by the relation $W_{-1}(z)e^{W_{-1}(z)}=z \quad (-1/e\leq z<0)$, and applying its small-$z$ asymptotic expansion $W_{-1}(z) = \log(-z) - \log(-\log(-z)) + \mathcal{O}(1)$, we can express $\tau_{J}$ and $\tau_{R}$ as
\begin{equation}
	\begin{gathered}
		\tau_{J}\approx    -3 \log (L/m_{\text{glueball}}^{-1}) + \frac{3}{2} \log(-\log (L/m_{\text{glueball}}^{-1})) + \dots, \\
		\tau_{R}\approx -3 \log (L/m_{\text{glueball}}^{-1}) + \frac{3}{2} \log(-\log (L/m_{\text{glueball}}^{-1})) + \dots,
	\end{gathered}
	\label{eq:tau-J_R-asymptotic}
\end{equation}
where the ellipses denote $\mathcal{O}(1)$ terms. 

The finite entropy differences can be evaluated similarly to the D4 and D3 cases. With $c_{J}^{2}\sim \tau_J e^{2\tau_{J}}$, the side-branch difference is approximated by 
\begin{equation}
	\int_{\tau_J}^{\infty} d\tau \sqrt{\beta(\tau)} \left( \frac{F(\tau)^{2} }{\sqrt{F(\tau)^2 - c_J^2}} - F(\tau) \right)\sim \tau_{J} e^{2\tau_{J}/3} \int_1^{\infty} dy \,  \left[ \frac{ y^4 }{ \sqrt{ y^6 - \frac{3}{4} } } - y \right],
\end{equation}
where we approximated the slowly varying polynomial factors as constants (e.g., evaluating $\tau \approx \tau_{J}$), kept only the rapidly varying exponential factors, and changed the variable to $y = e^{(\tau - \tau_{J})/3}$. Consequently, the $\tau_J$-dependence of the side-branch difference scales as $\tau_{J} e^{2\tau_{J}/3}$. The stem contributions also scale with the same parametric order: $\tau_{J} e^{2\tau_{J}/3}$. Similarly, the bipartite part has an analogous parametric dependence: $\tau_{R} e^{2\tau_{R}/3}$.

\begin{figure}[htbp]
    \centering
    \includegraphics[width=1\textwidth]{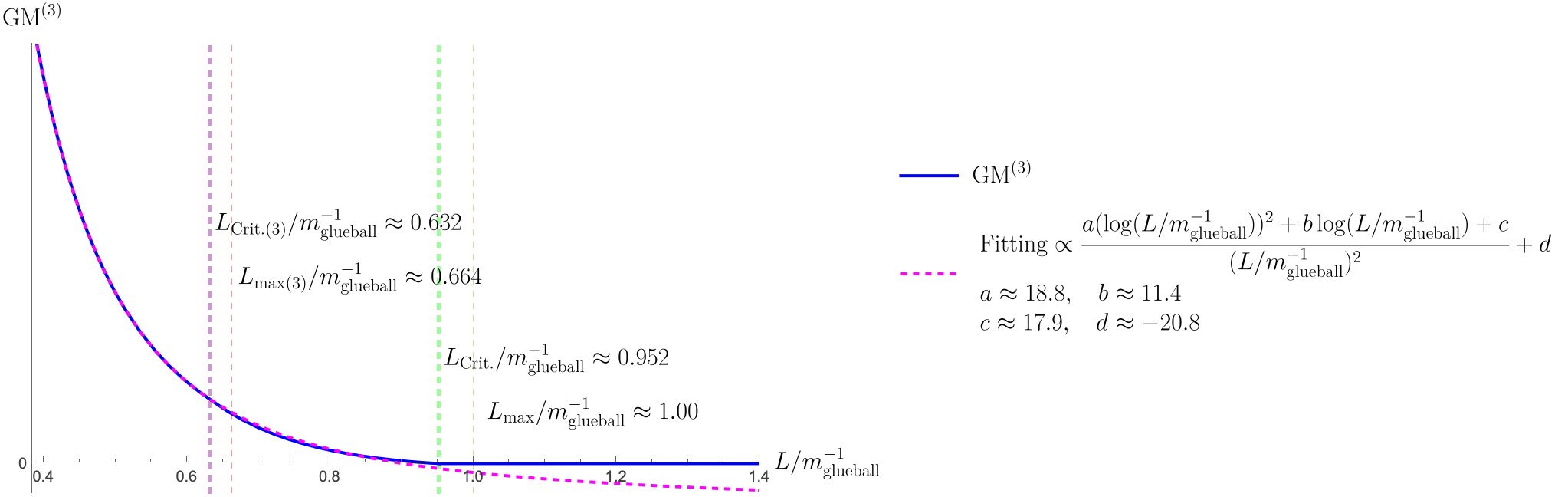}
    \caption{Small-$L$ behavior of the genuine multi-entropy in the KS background. The numerical data are excellently fitted by a $\frac{\left(\log (L/m_{\text{glueball}}^{-1})\right)^2}{(L/m_{\text{glueball}}^{-1})^2}$ fall-off, including sub-leading corrections.}
    \label{fig:KS_GM_fit}
\end{figure}

After the leading UV-divergent pieces strictly cancel in $\GM^{(3)}$, the remaining finite contribution behaves as
\begin{equation}
	\GM^{(3)}(A:B:C) \sim \tau_{J} e^{2\tau_{J}/3} \sim \tau_{R} e^{2\tau_{R}/3} \sim \frac{\left(\log (L/m_{\text{glueball}}^{-1})\right)^2}{(L/m_{\text{glueball}}^{-1})^2},
\end{equation}
where we used the expansions in \eqref{eq:tau-J_R-asymptotic}.
This analytical behavior is consistent with the numerical data, as shown in Fig.~\ref{fig:KS_GM_fit}.

This confirms the structural expectation anticipated below
\eqref{eq:KS-asymp-Functions-U}: the KS small-$L$ behavior indeed reproduces
the D3 power law $1/L^{2}$, dressed by the $(\log L)^{2}$ factor whose
origin can be traced directly to the $\sqrt{\log U}$ and $\log U$ factors
in $F$ and $\sqrt{\beta}\,F$.  Thus the KS geometry shares the qualitative
phase structure of the D4- and D3-solitons, and its UV asymptotics sit in
the same universality class as D3 up to logarithmic dressing.

In conclusion, the numerical evaluations confirm that the genuine
multi-entropy in the KS background follows the same monotonic smooth-cap
pattern observed in the D4 and D3 cases, with the only background-dependent
feature being the logarithmic correction to the short-distance scaling.

\section{Discussion}
\label{sec:universal_vs_background_dependent}

In this paper, we studied the junction law for multipartite entanglement in a sequence of confining holographic backgrounds, using the genuine multi-entropy $\GM^{(3)}$ as our main diagnostic. Our strategy was to begin with an analytically tractable hard-wall benchmark and then move to smooth confining geometries, namely the D4-soliton, D3-soliton, and Klebanov--Strassler backgrounds.

The main motivation was to distinguish features that are genuinely tied to multipartite junction structure from features that depend more sensitively on the detailed infrared realization of confinement. From this point of view, the hard-wall model is useful not because it is realistic, but because it makes the relevant saddle competition completely explicit. By contrast, the smooth confining geometries allow us to test which aspects of that picture survive once the sharp IR cutoff is replaced by a cigar-like cap or a deformed-conifold tip.

The first robust lesson is that the multipartite problem continues to be governed by a competition between a connected $Y$-type configuration and a cap-assisted disconnected configuration. In the hard-wall model this competition can be described directly in terms of multi-way cuts, while in the smooth backgrounds it is reformulated as an effective one-dimensional variational problem together with a Steiner-type junction condition. Despite this technical difference, the underlying multipartite logic is the same.

A second robust lesson is that the connected multipartite saddle is naturally associated with a junction-localized genuinely multipartite contribution. In all cases analyzed so far, the quantity $\GM^{(3)}$ is nonzero precisely when the multipartite entropy retains a genuinely connected contribution that cannot be reduced to the ordinary bipartite entropies alone. Conversely, once the disconnected phase dominates strongly enough, $\GM^{(3)}$ disappears.

A third robust lesson is that there are, in general, two distinct transition scales in the problem: $L_{\rm Crit.(3)}$, at which the $Y$-shaped bulk saddle of the tripartite entropy $S^{(3)}$ ceases to dominate, and $L_{\rm Crit.}$, at which the ordinary bipartite entropy $S(A)$ undergoes its own saddle transition. In all backgrounds studied here we find the ordering
\begin{equation}
L_{\rm Crit.(3)} < L_{\rm Crit.},
\label{eq:orderingall}
\end{equation}
so that the $Y$-shaped geometry characteristic of the tripartite sector disappears earlier than the bipartite connected saddle; in this sense, the tripartite sector is more IR-sensitive than the bipartite one, consistent with the general picture that multipartite structure is more fragile against IR effects~\cite{Iizuka:2025bcc}.
The fate of $\GM^{(3)}(A:B:C) = S^{(3)}(A:B:C) - \tfrac{1}{2}(S(A)+S(B)+S(C))$ is, however, more subtle: once $L$ exceeds $L_{\rm Crit.(3)}$, $S^{(3)}$ is taken over by a disconnected saddle, yet the subtraction of the bipartite pieces generically leaves a nonzero residue, so that $\GM^{(3)} \neq 0$ in the intermediate window $L_{\rm Crit.(3)} < L < L_{\rm Crit.}$, and only at the bipartite transition does genuine multipartite entanglement finally vanish, $\GM^{(3)}=0$ for $L > L_{\rm Crit.}$. Thus, the vanishing of $\GM^{(3)}$ is controlled not by the tripartite saddle transition but by the bipartite one. In particular, the disappearance of the $Y$-shaped saddle at $L_{\rm Crit.(3)}$ does not by itself imply $\GM^{(3)}=0$.

These features strongly suggest that the basic junction-law interpretation of $\GM^{(3)}$ is not an artifact of the hard-wall toy model, but instead reflects a broader structural property of confining holographic geometries.

At the same time, our results make it equally clear that the detailed shape of $\GM^{(3)}$ is not universal.

The clearest example is the plateau structure found in the hard-wall benchmark. Its origin is not merely that the IR wall supports a wall-assisted configuration, but that in the AdS$_3$ hard-wall model the relevant $L$-dependence of the multipartite and bipartite sectors becomes identical over the plateau regime. As a result, when forming $\GM^{(3)}$, the $L$-dependent pieces cancel, leaving an approximately constant contribution. By contrast, in the D4-soliton, D3-soliton and Klebanov--Strassler backgrounds, where the IR geometry caps off smoothly, no analogous cancellation occurs. Instead, $\GM^{(3)}$ decreases monotonically and vanishes at a finite critical scale. This indicates that the plateau is not a generic consequence of confinement itself, but rather a special feature of the AdS$_3$ hard-wall setup.

A second background-dependent feature is the short-distance scaling of $\GM^{(3)}$. Even among smooth confining geometries, the ultraviolet asymptotics are not the same. 
In the D4-soliton case, we find
\begin{equation}
\GM^{(3)}(A:B:C)\sim \frac{1}{L^4},
\qquad
L\to 0,
\end{equation}
while in the D3-soliton case,
\begin{equation}
\GM^{(3)}(A:B:C)\sim \frac{1}{L^2},
\qquad
L\to 0,
\end{equation}
and in the Klebanov--Strassler case,
\begin{equation}
\GM^{(3)}(A:B:C)\sim \frac{(\log L)^{2}}{L^2},
\qquad
L\to 0.
\end{equation}
Thus, although these examples share the same qualitative phase structure, they differ quantitatively in a way that reflects the detailed background dependence of the effective functions $F(u)$ and $\beta(u)$ entering the reduced entropy functional.

This suggests a useful conceptual separation. The existence of a connected multipartite saddle, the relevance of the junction condition, and the eventual disappearance of $\GM^{(3)}$ beyond a critical scale appear to be robust qualitative features. By contrast, plateau behavior and the precise small-$L$ falloff encode finer information about the detailed IR and UV structure of the geometry.

At the same time, the comparison with the D4-soliton, D3-soliton and Klebanov--Strassler backgrounds shows that one must be careful not to overinterpret hard-wall-specific phenomena. In particular, the plateau in $\GM^{(3)}$ is best understood as a sharp-wall phenomenon rather than as a universal signature of confinement. In this sense, the hard-wall benchmark is useful precisely because it allows one to separate genuinely structural features from artifacts of a particularly simple IR model.

There are several natural directions for further work.

First, it would be interesting to move beyond the $BC$-symmetric configuration and analyze more general asymmetric tripartitions in smooth confining geometries. In the hard-wall benchmark, asymmetry led to partially disconnected saddles and to a richer phase structure. It would be important to understand to what extent analogous phenomena survive in smooth caps.

Second, our results suggest that genuine multi-entropy may provide a useful new probe of confinement, complementary to ordinary bipartite entanglement entropy. The bipartite entropy already detects confinement through the standard connected/disconnected transition, but $\GM^{(3)}$ appears to encode additional information about whether a genuinely multipartite junction can still be supported by the geometry. It would be interesting to understand whether this viewpoint extends to broader classes of holographic models, including finite-temperature transitions, more general strip networks, or time-dependent settings.

Finally, one would like to sharpen the relation between the junction-law interpretation of $\GM^{(3)}$ and the broader program of characterizing multipartite entanglement in holography. The present paper suggests that there is a meaningful sense in which genuinely multipartite entanglement is localized near an effective junction, but also that the detailed realization of that localization depends on the global infrared structure of the bulk geometry. Understanding this interplay more systematically may help clarify which multipartite observables are truly universal in holography and which are best interpreted as refined probes of particular geometric backgrounds.

To summarize, the results obtained here support the following picture. The junction-law interpretation of genuine multi-entropy survives beyond the simplest toy models and continues to organize the multipartite problem in confining holographic backgrounds. However, the detailed behavior of $\GM^{(3)}$ is not fully universal. In particular, the hard-wall plateau appears to be tied to the sharpness of the IR cutoff, whereas smooth confining caps lead instead to monotonic decay and background-dependent short-distance scaling.

In this sense, genuine multi-entropy plays a double role. It is universal enough to diagnose the existence or disappearance of a genuinely multipartite connected contribution, but sensitive enough to distinguish sharp-wall confinement from smooth-cap confinement. This makes it a potentially useful probe of how multipartite entanglement is geometrized in holographic theories with nontrivial infrared structure.

\acknowledgments
The work of N.I. and A.M. was supported in part by MEXT KAKENHI Grant-in-Aid for Transformative Research Areas A “Extreme Universe” No. 21H05184. The work of N.I. was also supported in part by NSTC of Taiwan Grant Number 114-2112-M-007-025-MY3. The work of A.M. was also supported by JSPS KAKENHI Grant Number JP26KJ0186. The authors used generative AI tools for limited language polishing.

%%%%%%%%%%%%%%%%%%%%%%%%%%%%%%%%%%%%%%%%%%%%%%%%%%%%%%%%%%%%%%%%%%%%%%%%
%%%%%%%%%%%%%%%%%%%%%%%%%%%%%%%%%%%%%%%%%%%%%%%%%%%%%%%%%%%%%%%%%%%%%%%%
\appendix
\section{AdS$_3$ geometry and hyperbolic distance formulas}
\label{app:AdS3_distance_geometry}

In this appendix, we collect several standard geometric facts about the constant-time slice of AdS$_3$ used in the main text.

\subsection{AdS$_3$ as an embedded hyperboloid}

Consider the flat embedding space $\mathbb{R}^{2,2}$ with coordinates
\begin{equation}
X^A=(X^0,X^1,X^2,X^3)
\end{equation}
and metric
\begin{equation}
ds^2_{\rm emb}
=
-(dX^0)^2-(dX^3)^2+(dX^1)^2+(dX^2)^2,
\qquad
\eta_{AB}=\mathrm{diag}(-,+,+,-).
\end{equation}
Lorentzian AdS$_3$ with radius $R_{\rm AdS}$ is defined by the hyperboloid
\begin{equation}
(X^0)^2+(X^3)^2-(X^1)^2-(X^2)^2=R_{\rm AdS}^2.
\label{eq:app_ads_hyperboloid}
\end{equation}
Equivalently, introducing the embedding-space inner product
\begin{equation}
X\cdot Y
=
\eta_{AB}X^A Y^B
=
- X^0Y^0+X^1Y^1+X^2Y^2-X^3Y^3,
\end{equation}
the constraint may be written as
\begin{equation}
X\cdot X=-R_{\rm AdS}^2.
\label{eq:app_ads_norm}
\end{equation}

A convenient Poincar\'e parametrization is
\begin{align}
X^0 &= R_{\rm AdS}\,\frac{t}{z}, \\
X^1 &= R_{\rm AdS}\,\frac{x}{z}, \\
X^2 &= \frac{R_{\rm AdS}}{2z}\left(-t^2+x^2+z^2-1\right), \\
X^3 &= \frac{R_{\rm AdS}}{2z}\left(-t^2+x^2+z^2+1\right),
\end{align}
with $z>0$. This induces the standard AdS$_3$ metric
\begin{equation}
ds^2_{\rm AdS_3}
=
\frac{R_{\rm AdS}^2}{z^2}\left(-dt^2+dx^2+dz^2\right).
\label{eq:app_ads_poincare}
\end{equation}

\subsection{Constant-time slice and $H^2$}

Restricting to a constant-time slice, say $t=0$, the induced spatial metric becomes
\begin{equation}
ds^2_{\rm slice}
=
\frac{R_{\rm AdS}^2}{z^2}\left(dx^2+dz^2\right),
\qquad z>0,
\label{eq:app_H2_metric}
\end{equation}
which is the upper half-plane model of the hyperbolic plane $H^2$.

At $t=0$, the embedding coordinates reduce to
\begin{equation}
X^0=0,\qquad
X^1=R_{\rm AdS}\frac{x}{z},\qquad
X^2=\frac{R_{\rm AdS}}{2z}(x^2+z^2-1),\qquad
X^3=\frac{R_{\rm AdS}}{2z}(x^2+z^2+1).
\label{eq:app_t0_embedding}
\end{equation}
Because $X^0=0$, the slice lies inside the $(+,+,-)$ signature subspace spanned by $(X^1,X^2,X^3)$, and \eqref{eq:app_ads_norm} becomes
\begin{equation}
(X^1)^2+(X^2)^2-(X^3)^2=-R_{\rm AdS}^2,
\end{equation}
which is precisely the hyperboloid model of $H^2$.

\subsection{Hyperbolic distance from the embedding inner product}

A basic fact in the hyperboloid model is that for two points $P,Q\in H^2$,
\begin{equation}
\cosh\!\left(\frac{d(P,Q)}{R_{\rm AdS}}\right)
=
-\frac{X(P)\cdot X(Q)}{R_{\rm AdS}^2}.
\label{eq:app_cosh_distance_general}
\end{equation}
We now evaluate the right-hand side explicitly in Poincar\'e coordinates.

Let
\begin{equation}
P=(x,z),\qquad Q=(x',z'),\qquad (t=0 \text{ for both}),
\end{equation}
and define
\begin{equation}
A=x^2+z^2,\qquad A'=x'^2+z'^2.
\end{equation}
From \eqref{eq:app_t0_embedding},
\begin{equation}
X^1(P)=R_{\rm AdS}\frac{x}{z},\quad
X^2(P)=\frac{R_{\rm AdS}}{2z}(A-1),\quad
X^3(P)=\frac{R_{\rm AdS}}{2z}(A+1),
\end{equation}
and similarly for $Q$. Therefore,
\begin{align}
X(P)\cdot X(Q)
&=
X^1(P)X^1(Q)+X^2(P)X^2(Q)-X^3(P)X^3(Q)
\nonumber\\
&=
\frac{R_{\rm AdS}^2}{zz'}
\left[
xx'
+\frac{(A-1)(A'-1)-(A+1)(A'+1)}{4}
\right]
\nonumber\\
&=
\frac{R_{\rm AdS}^2}{zz'}
\left[
xx'-\frac{A+A'}{2}
\right].
\label{eq:app_inner_product_result}
\end{align}
Hence
\begin{equation}
-\frac{X(P)\cdot X(Q)}{R_{\rm AdS}^2}
=
\frac{A+A'-2xx'}{2zz'}
=
\frac{(x-x')^2+z^2+z'^2}{2zz'}.
\end{equation}
Using
\begin{equation}
z^2+z'^2=(z-z')^2+2zz',
\end{equation}
we obtain
\begin{equation}
-\frac{X(P)\cdot X(Q)}{R_{\rm AdS}^2}
=
1+\frac{(x-x')^2+(z-z')^2}{2zz'}.
\end{equation}
Substituting this into \eqref{eq:app_cosh_distance_general} gives
\begin{equation}
\cosh\!\left(\frac{d(P,Q)}{R_{\rm AdS}}\right)
=
1+\frac{(x-x')^2+(z-z')^2}{2zz'}.
\label{eq:app_hyperbolic_distance_final}
\end{equation}
Thus,
\begin{equation}
d(P,Q)
=
R_{\rm AdS}\,
\arcosh\!\left(
1+\frac{(x-x')^2+(z-z')^2}{2zz'}
\right).
\label{eq:app_hyperbolic_distance_boxed}
\end{equation}
In the main text we set $R_{\rm AdS}=1$.

\subsection{Vertical distance}

For $x=x'$, \eqref{eq:app_hyperbolic_distance_final} simplifies to
\begin{equation}
\cosh\!\left(\frac{d(P,Q)}{R_{\rm AdS}}\right)
=
1+\frac{(z-z')^2}{2zz'}
=
\frac12\left(\frac{z}{z'}+\frac{z'}{z}\right)
=
\cosh\!\left(\log\frac{z'}{z}\right),
\end{equation}
so that
\begin{equation}
d(P,Q)=R_{\rm AdS}\left|\log\frac{z'}{z}\right|.
\label{eq:app_vertical_distance}
\end{equation}

The same result may also be obtained directly from the line element. Along a vertical line with $dx=0$,
\begin{equation}
ds=\frac{R_{\rm AdS}}{z}|dz|,
\end{equation}
and hence, for $z'>z$,
\begin{equation}
\ell_{\rm vert}
=
\int_z^{z'}\frac{R_{\rm AdS}}{z}\,dz
=
R_{\rm AdS}\log\frac{z'}{z},
\end{equation}
in agreement with \eqref{eq:app_vertical_distance}.

\subsection{Geodesics as semicircles}

To determine the geodesic connecting two points, one minimizes the length functional
\begin{equation}
\ell
=
R_{\rm AdS}\int dx\,\frac{\sqrt{1+z'(x)^2}}{z}.
\label{eq:app_length_functional}
\end{equation}
Since the Lagrangian
\begin{equation}
\mathcal{L}(z,z')
=
\frac{\sqrt{1+z'^2}}{z}
\end{equation}
does not depend explicitly on $x$, the quantity
\begin{equation}
\mathcal{L}
-
z'\frac{\partial\mathcal{L}}{\partial z'}
=
\frac{1}{R_0}
\end{equation}
is conserved, where $R_0$ is a constant. This gives
\begin{equation}
\frac{1}{z\sqrt{1+z'^2}}=\frac{1}{R_0},
\end{equation}
or equivalently
\begin{equation}
z^2(1+z'^2)=R_0^2.
\end{equation}
Rearranging,
\begin{equation}
\frac{dx}{dz}
=
\pm \frac{z}{\sqrt{R_0^2-z^2}},
\end{equation}
which integrates to
\begin{equation}
(x-x_0)^2+z^2=R_0^2.
\label{eq:app_circle_geodesic}
\end{equation}
Thus geodesics in the upper half-plane are Euclidean semicircles orthogonal to the boundary $z=0$, with vertical lines arising as the limiting case $R_0\to\infty$.

For two boundary endpoints $(a,\varepsilon)$ and $(b,\varepsilon)$, symmetry fixes
\begin{equation}
x_0=\frac{a+b}{2},
\qquad
R_0=\frac{b-a}{2},
\end{equation}
so the maximal depth is
\begin{equation}
z_{\max}=\frac{b-a}{2}.
\end{equation}
This is the result used in the main text.

%%%%%%%%%%%%%%%%%%%

%%%%%%%%%%%%%%%%%%%%%%

\section{Verification of the \texorpdfstring{$120^\circ$}{120 degree} condition for the asymmetric \texorpdfstring{$Y$}{Y}-junction}
\label{app:hardwall_technical_details}

In this appendix, we verify that the extremized asymmetric $Y$-junction in the hard-wall model satisfies the expected $120^\circ$ condition.

For the arc connecting a boundary point $x_i$ (see Fig.~\ref{fig:hw_asymmetric_Y}) to the bulk junction
\begin{equation}
Y=(x_\ast^{(0)},z_\ast^{(0)}),
\end{equation}
let its Euclidean center and radius be $x_{iY}$ and $r_{iY}$.
They satisfy
\begin{equation}
(x_\ast^{(0)}-x_{iY})^2+(z_\ast^{(0)})^2=r_{iY}^2,
\qquad
(x_i-x_{iY})^2=r_{iY}^2.
\end{equation}
Solving these equations gives
\begin{equation}
x_{iY}
=
\frac{(x_\ast^{(0)})^2+(z_\ast^{(0)})^2-x_i^2}{2(x_\ast^{(0)}-x_i)},
\qquad
r_{iY}
=
\frac{(x_\ast^{(0)}-x_i)^2+(z_\ast^{(0)})^2}{2|x_\ast^{(0)}-x_i|}.
\end{equation}
For $x_1=-L_A$, $x_2=0$, and $x_3=L_B$, one obtains
\begin{equation}
\begin{aligned}
x_{1Y} &= -\frac{L_A^2}{2L_A+L_B},
&\qquad
r_{1Y} &= \frac{L_A(L_A+L_B)}{2L_A+L_B},\\
x_{2Y} &= \frac{L_A L_B}{L_A-L_B},
&\qquad
r_{2Y} &= \frac{L_A L_B}{|L_A-L_B|},\\
x_{3Y} &= \frac{L_B^2}{L_A+2L_B},
&\qquad
r_{3Y} &= \frac{L_B(L_A+L_B)}{L_A+2L_B}.
\end{aligned}
\label{eq:app_centers_radii}
\end{equation}

Parameterizing each arc by an angle $\theta_i$ so that $\theta_i=0$ corresponds to the boundary point,
\begin{equation}
x^{(i)}(\theta_i)=x_{iY}+(x_i-x_{iY})\cos\theta_i,
\qquad
z^{(i)}(\theta_i)=r_{iY}\sin\theta_i,
\end{equation}
the junction is reached at $\theta_i=\theta_{iY\ast}$ determined by
\begin{equation}
\cos\theta_{iY\ast}
=
\frac{x_\ast^{(0)}-x_{iY}}{x_i-x_{iY}},
\qquad
\sin\theta_{iY\ast}
=
\frac{z_\ast^{(0)}}{r_{iY}}.
\label{eq:app_theta_conditions}
\end{equation}
Substituting the explicit values gives
\begin{equation}
\begin{aligned}
\cos \theta_{1Y\ast} &= -\frac{2L_A^2+2L_A L_B-L_B^2}{2(L_A^2+L_A L_B+L_B^2)},
&
\sin \theta_{1Y\ast} &= \frac{\sqrt{3}\,L_B(2L_A+L_B)}{2(L_A^2+L_A L_B+L_B^2)},
\\
\cos \theta_{2Y\ast} &= \frac{L_A^2+4L_A L_B+L_B^2}{2(L_A^2+L_A L_B+L_B^2)},
&
\sin \theta_{2Y\ast} &= \frac{\sqrt{3}\,|L_A^2-L_B^2|}{2(L_A^2+L_A L_B+L_B^2)},
\\
\cos \theta_{3Y\ast} &= \frac{L_A^2-2L_A L_B-2L_B^2}{2(L_A^2+L_A L_B+L_B^2)},
&
\sin \theta_{3Y\ast} &= \frac{\sqrt{3}\,L_A(L_A+2L_B)}{2(L_A^2+L_A L_B+L_B^2)}.
\end{aligned}
\label{eq:app_theta_explicit}
\end{equation}

The unit tangent vector at the junction is
\begin{equation}
\vec v_i
=
\left(
-\mathrm{sgn}(x_i-x_{iY})\sin\theta_{iY\ast},
\;
\cos\theta_{iY\ast}
\right).
\label{eq:app_tangent_vector}
\end{equation}
A direct computation gives
\begin{equation}
\vec v_i\cdot \vec v_j=-\frac12,
\qquad i\neq j,
\end{equation}
which establishes that the three pairwise angles are all $120^\circ$.

\section{Asymmetric hard-wall phase structure of \texorpdfstring{$S^{(3)}$}{S(3)} and \texorpdfstring{$\GM^{(3)}$}{GM(3)}}
\label{app:hardwall_asymmetric_details}

In this appendix, we summarize the phase structure of the tripartite multi-entropy and genuine multi-entropy for the asymmetric hard-wall configuration
\begin{equation}
A=\{-L_A<x<0\},
\qquad
B=\{0<x<L_B\},
\qquad
C=\{x<-L_A,\;x>L_B\}.
\label{eq:app_asymmetric_regions}
\end{equation}
It is convenient to introduce
\begin{equation}
L_{\min}=\min\{L_A,L_B\},
\qquad
L_{\max}=\max\{L_A,L_B\}.
\label{eq:app_Lmin_Lmax}
\end{equation}

\subsection{Tripartite multi-entropy}

From Sec.~\ref{subsec:hw_asymmetric_summary},
%Appendix~\ref{app:hardwall_technical_details},
 the relevant candidate lengths are
\begin{equation}
L_Y^{\rm (Ste)}
=
\log\left(
\frac{8}{3\sqrt{3}}
\frac{L_{\min}L_{\max}(L_{\min}+L_{\max})}{\varepsilon^3}
\right),
\end{equation}
\begin{equation}
L_M
=
\log\left(
\frac{8}{3\sqrt{3}}
\frac{L_{\min}z_0^2}{\varepsilon^3}
\right),
\qquad
L_W
=
3\log\frac{z_0}{\varepsilon}.
\end{equation}
Hence
\begin{equation}
S^{(3)}(A:B:C)
=
\frac{1}{4G_N}\min\{L_Y^{\rm (Ste)},L_M,L_W\}.
\label{eq:app_S3_asym_min}
\end{equation}
Equivalently,
\begin{align}
S^{(3)}(A:B:C)
&=
\begin{cases}
\displaystyle
\frac{c}{6}\log\!\left(
\frac{8}{3\sqrt{3}}
\frac{L_{\min}L_{\max}(L_{\min}+L_{\max})}{\varepsilon^3}
\right),
&
\begin{array}{l}
\sqrt{L_{\max}(L_{\min}+L_{\max})}\le z_0,\\[2pt]
\frac{2}{\sqrt3}\bigl(L_{\min}L_{\max}(L_{\min}+L_{\max})\bigr)^{1/3}\le z_0,
\end{array}
\\[8pt]
\displaystyle
\frac{c}{6}\log\!\left(
\frac{8}{3\sqrt{3}}
\frac{L_{\min}z_0^2}{\varepsilon^3}
\right),
&
\begin{array}{l}
\sqrt{L_{\max}(L_{\min}+L_{\max})}\ge z_0,\\[2pt]
\frac{8}{3\sqrt{3}}L_{\min}\le z_0,
\end{array}
\\[8pt]
\displaystyle
\frac{c}{2}\log\frac{z_0}{\varepsilon},
&
\begin{array}{l}
\frac{2}{\sqrt3}\bigl(L_{\min}L_{\max}(L_{\min}+L_{\max})\bigr)^{1/3}\ge z_0,\\[2pt]
\frac{8}{3\sqrt{3}}L_{\min}\ge z_0.
\end{array}
\end{cases}
\label{eq:app_piecewise_S3}
\end{align}

This implies the following qualitative structure:
\begin{enumerate}
\item For $z_0\to\infty$, the fully connected $Y$ saddle dominates.
\item For $z_0\to 0$, the wall-assisted disconnected saddle $W$ dominates.
\item The mixed saddle $M$ can dominate only in the intermediate regime
\begin{equation}
\frac{8}{3\sqrt{3}}\,L_{\min} \le z_0 \le \sqrt{L_{\max}(L_{\min}+L_{\max})}.
\end{equation}
\end{enumerate}
Therefore, $M$ is realized as an actual phase only when
\begin{equation}
\frac{8}{3\sqrt{3}}\,L_{\min}
<
\sqrt{L_{\max}(L_{\min}+L_{\max})},
\end{equation}
or equivalently,
\begin{equation}
\frac{L_{\max}}{L_{\min}}
>
\frac{\sqrt{849}-9}{18}
\approx 1.12.
\label{eq:app_M_condition}
\end{equation}
If instead
\begin{equation}
\frac{L_{\max}}{L_{\min}}
<
\frac{\sqrt{849}-9}{18},
\end{equation}
the system undergoes a direct transition from $Y$ to $W$.

\begin{figure}[htbp]
	\centering
	\includegraphics[width=0.55\textwidth]{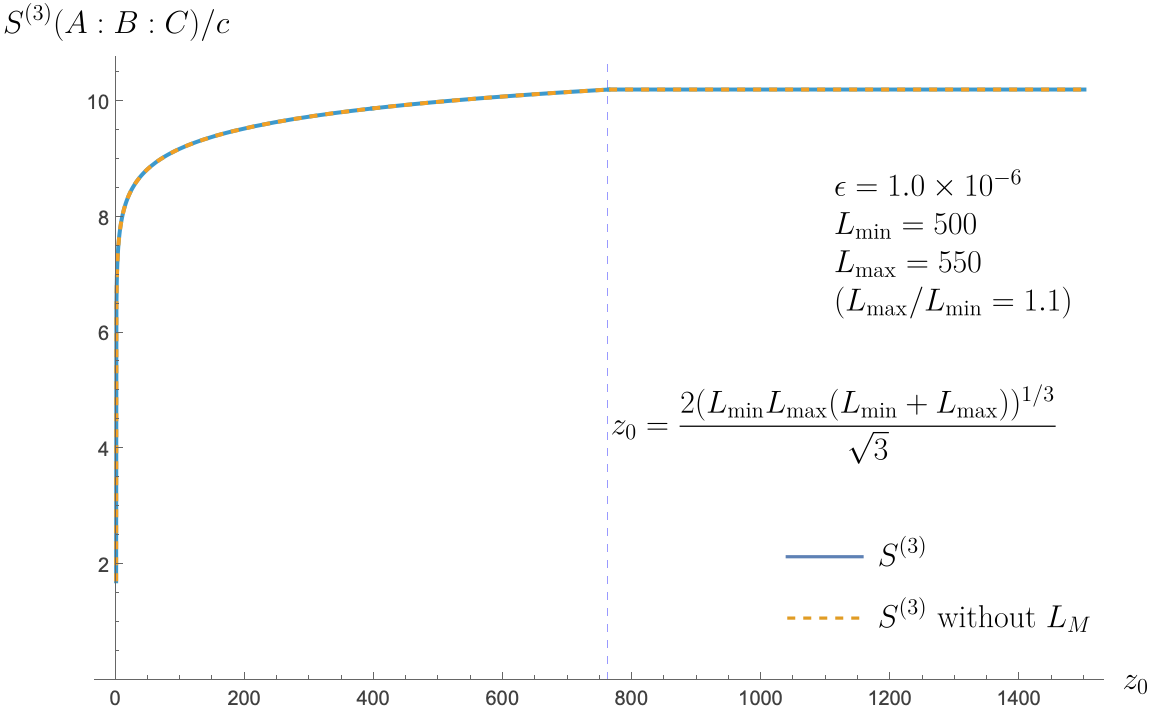}\\[0.5em]
	\includegraphics[width=0.55\textwidth]{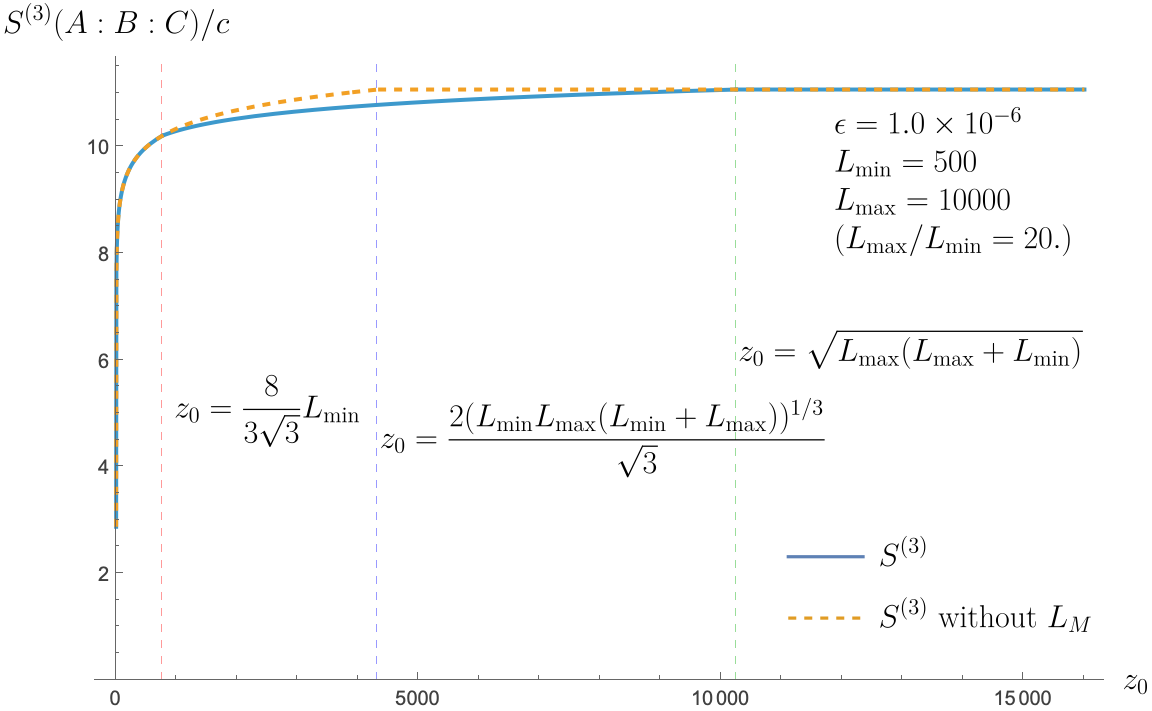}
	\caption{Plots of the rescaled multi-entropy $S^{(3)}(A:B:C)/c$ as a function of the hard-wall depth $z_0$ for asymmetric configurations. The solid blue curve denotes the true minimum including the mixed saddle $M$. The dashed orange curve denotes the result obtained if one artificially restricts the candidates to the fully connected $Y$ saddle and the fully disconnected $W$ saddle.}
	\label{fig:Smulti_plots-Asymmetric-CABC}
\end{figure}

In Fig.~\ref{fig:Smulti_plots-Asymmetric-CABC}, we plot the multi-entropy for the two distinct cases.

\subsection{Ordinary entanglement entropies}

The ordinary interval entropies are
\begin{equation}
S(A)
=
\min\left\{
\frac{c}{3}\log\frac{L_A}{\varepsilon},
\frac{c}{3}\log\frac{z_0}{\varepsilon}
\right\},
\qquad
S(B)
=
\min\left\{
\frac{c}{3}\log\frac{L_B}{\varepsilon},
\frac{c}{3}\log\frac{z_0}{\varepsilon}
\right\},
\end{equation}
and
\begin{equation}
S(C)=S(AB)
=
\min\left\{
\frac{c}{3}\log\frac{L_A+L_B}{\varepsilon},
\frac{c}{3}\log\frac{z_0}{\varepsilon}
\right\}.
\label{eq:app_asymmetric_EEs}
\end{equation}

\subsection{Genuine multi-entropy}

The genuine multi-entropy is then
\begin{equation}
\GM^{(3)}(A:B:C)
=
S^{(3)}(A:B:C)-\frac12\left(S(A)+S(B)+S(C)\right).
\label{eq:app_GM_def}
\end{equation}
Its phase diagram is naturally classified into three regimes according to the ratio $L_{\max}/L_{\min}$.

\paragraph{Case I:}
\begin{equation}
1\le \frac{L_{\max}}{L_{\min}}<\frac{\sqrt{849}-9}{18}.
\end{equation}
Then
\begin{equation}
\GM^{(3)}(A:B:C)=
\begin{dcases}
\frac{c}{6}\log\left(\frac{8}{3\sqrt3}\right),
& z_0>L_{\min}+L_{\max},
\\[6pt]
\frac{c}{6}\log\left(\frac{8}{3\sqrt3}\frac{L_{\min}+L_{\max}}{z_0}\right),
& \left(\frac{8}{3\sqrt3}L_{\min}L_{\max}(L_{\min}+L_{\max})\right)^{1/3}<z_0\le L_{\min}+L_{\max},
\\[6pt]
\frac{c}{6}\log\left(\frac{z_0^2}{L_{\min}L_{\max}}\right),
& L_{\max}<z_0\le \left(\frac{8}{3\sqrt3}L_{\min}L_{\max}(L_{\min}+L_{\max})\right)^{1/3},
\\[6pt]
\frac{c}{6}\log\left(\frac{z_0}{L_{\min}}\right),
& L_{\min}<z_0\le L_{\max},
\\[6pt]
0,
& z_0\le L_{\min}.
\end{dcases}
\label{eq:app_GM_case1}
\end{equation}

\paragraph{Case II:}
\begin{equation}
\frac{\sqrt{849}-9}{18}\le \frac{L_{\max}}{L_{\min}}<\frac{8}{3\sqrt{3}}.
\end{equation}
Then
\begin{equation}
\GM^{(3)}(A:B:C)=
\begin{dcases}
\frac{c}{6}\log\left(\frac{8}{3\sqrt3}\right),
& z_0>L_{\min}+L_{\max},
\\[6pt]
\frac{c}{6}\log\left(\frac{8}{3\sqrt3}\frac{L_{\min}+L_{\max}}{z_0}\right),
& \sqrt{L_{\max}(L_{\min}+L_{\max})}<z_0\le L_{\min}+L_{\max},
\\[6pt]
\frac{c}{6}\log\left(\frac{8}{3\sqrt3}\frac{z_0}{L_{\max}}\right),
& \frac{8}{3\sqrt3}L_{\min}<z_0\le \sqrt{L_{\max}(L_{\min}+L_{\max})},
\\[6pt]
\frac{c}{6}\log\left(\frac{z_0^2}{L_{\min}L_{\max}}\right),
& L_{\max}<z_0\le \frac{8}{3\sqrt3}L_{\min},
\\[6pt]
\frac{c}{6}\log\left(\frac{z_0}{L_{\min}}\right),
& L_{\min}<z_0\le L_{\max},
\\[6pt]
0,
& z_0\le L_{\min}.
\end{dcases}
\label{eq:app_GM_case2}
\end{equation}

\paragraph{Case III:}
\begin{equation}
\frac{L_{\max}}{L_{\min}}\ge \frac{8}{3\sqrt3}.
\end{equation}
Then
\begin{equation}
\GM^{(3)}(A:B:C)=
\begin{dcases}
\frac{c}{6}\log\left(\frac{8}{3\sqrt3}\right),
& z_0>L_{\min}+L_{\max},
\\[6pt]
\frac{c}{6}\log\left(\frac{8}{3\sqrt3}\frac{L_{\min}+L_{\max}}{z_0}\right),
& \sqrt{L_{\max}(L_{\min}+L_{\max})}<z_0\le L_{\min}+L_{\max},
\\[6pt]
\frac{c}{6}\log\left(\frac{8}{3\sqrt3}\frac{z_0}{L_{\max}}\right),
& L_{\max}<z_0\le \sqrt{L_{\max}(L_{\min}+L_{\max})},
\\[6pt]
\frac{c}{6}\log\left(\frac{8}{3\sqrt3}\right),
& \frac{8}{3\sqrt3}L_{\min}<z_0\le L_{\max},
\\[6pt]
\frac{c}{6}\log\left(\frac{z_0}{L_{\min}}\right),
& L_{\min}<z_0\le \frac{8}{3\sqrt3}L_{\min},
\\[6pt]
0,
& z_0\le L_{\min}.
\end{dcases}
\label{eq:app_GM_case3}
\end{equation}
The fourth line is the intermediate plateau characteristic of the strongly asymmetric regime.

\subsection{Large-asymmetry limit}

In the limit $L_{\max}\to\infty$ with $L_{\min}$ fixed, the plateau extends all the way to infinity, and one finds
\begin{equation}
\GM^{(3)}(A:B:C)=
\begin{dcases}
\frac{c}{6}\log\left(\frac{8}{3\sqrt3}\right),
& \frac{8}{3\sqrt3}L_{\min}<z_0,
\\[6pt]
\frac{c}{6}\log\left(\frac{z_0}{L_{\min}}\right),
& L_{\min}<z_0\le \frac{8}{3\sqrt3}L_{\min},
\\[6pt]
0,
& z_0\le L_{\min}.
\end{dcases}
\label{eq:app_GM_large_asymmetry}
\end{equation}
This reproduces the $BC$-symmetric benchmark discussed in the main text.

\subsection{Figures}

\begin{figure}[htbp]
	\centering
	\includegraphics[width=0.5\textwidth]{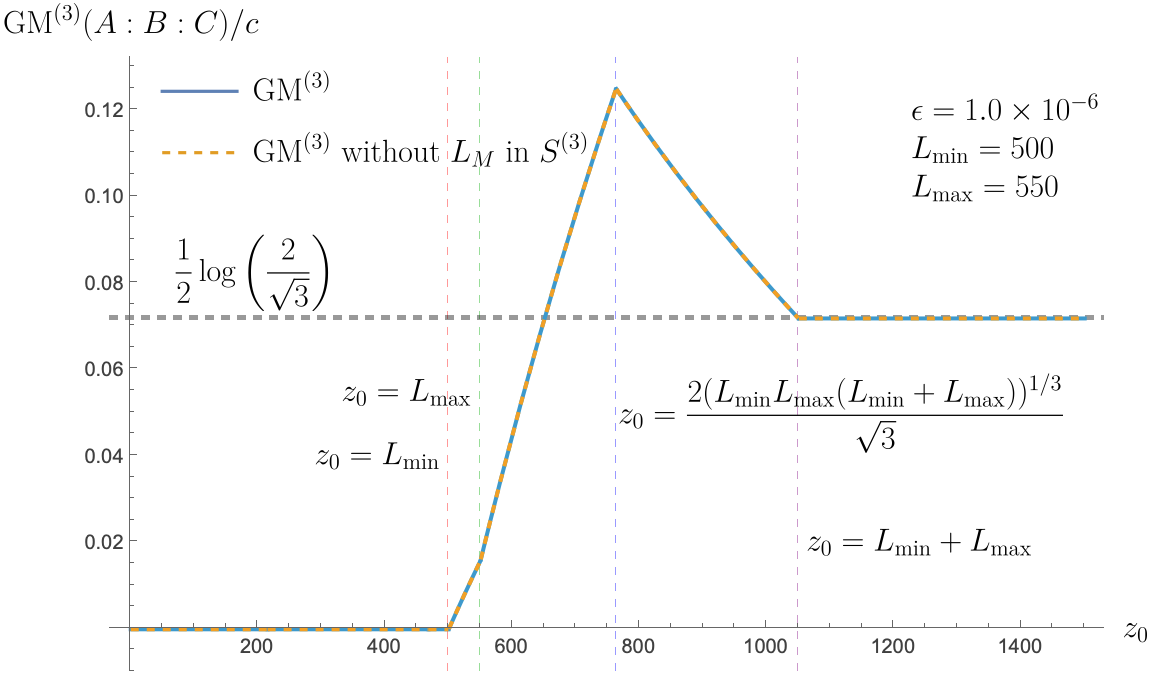}\\[0.5em]
	\includegraphics[width=0.5\textwidth]{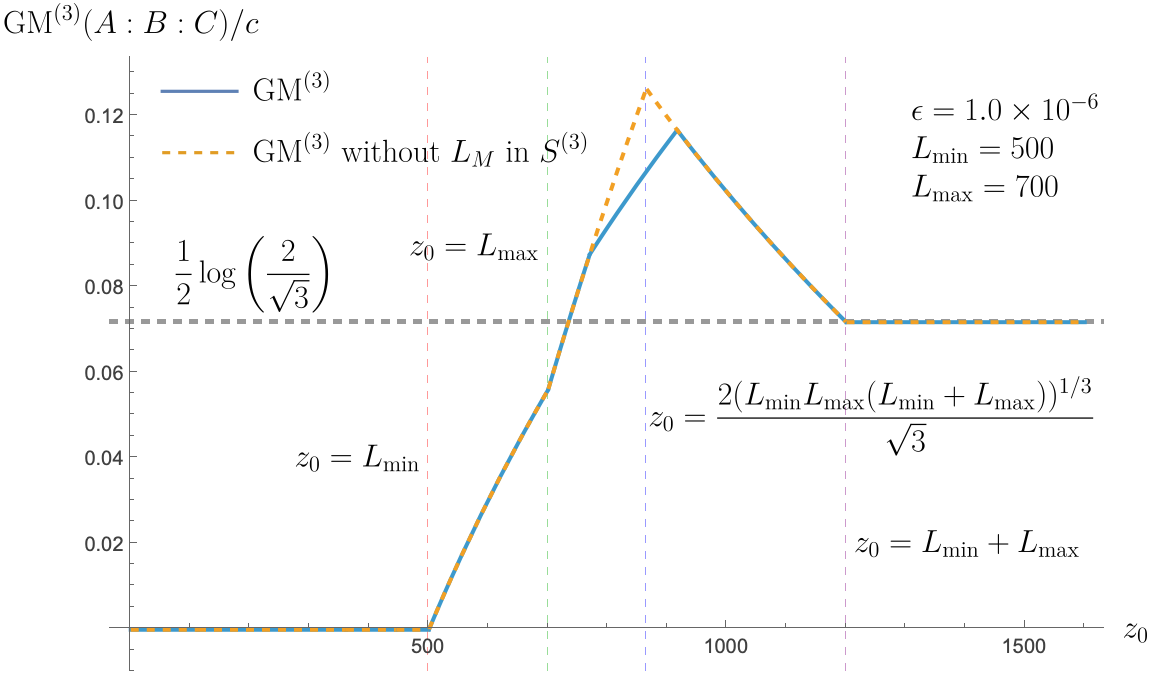}\\[0.5em]
	\includegraphics[width=0.5\textwidth]{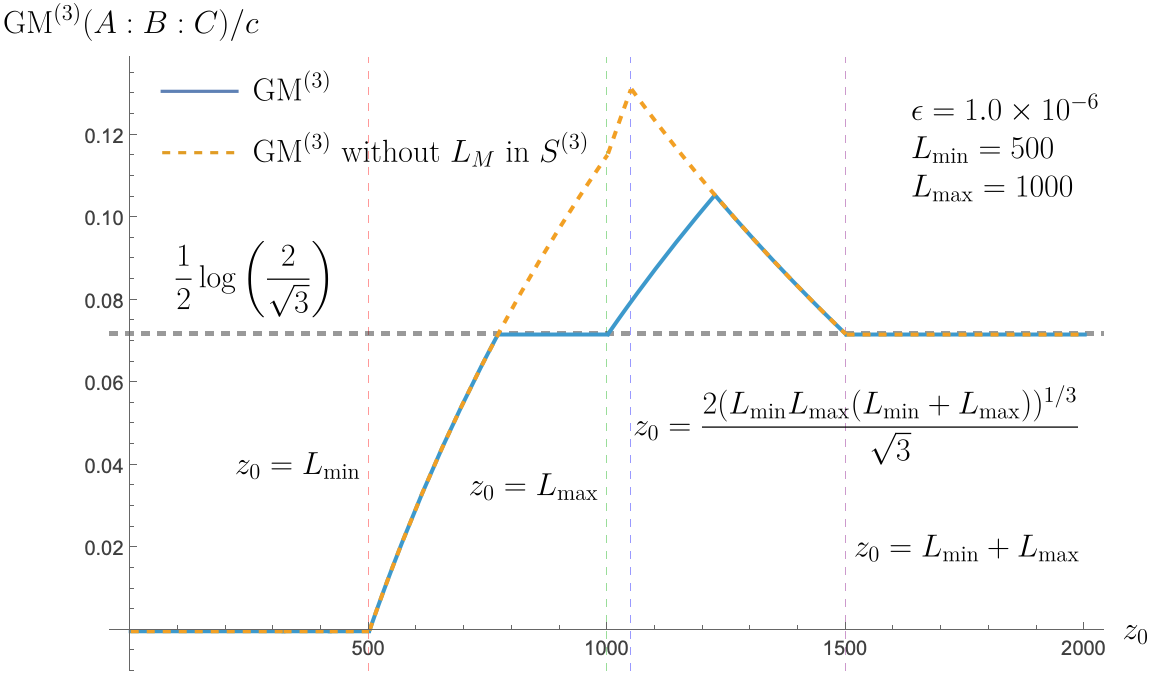}\\[0.5em]
	\includegraphics[width=0.5\textwidth]{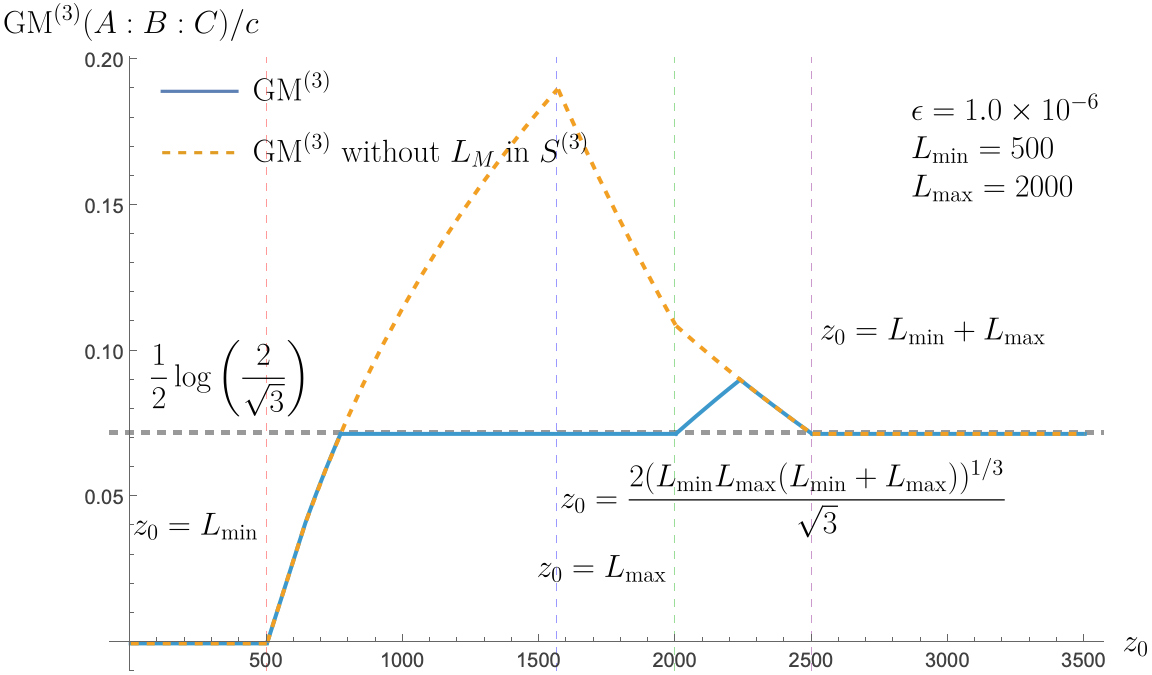}
	\caption{Plots of the rescaled genuine multi-entropy $\GM^{(3)}(A:B:C)/c$ as a function of the hard-wall depth $z_0$ for asymmetric configurations. The solid blue curve denotes the true minimum including the mixed saddle $M$. The dashed orange curve denotes the result obtained if one artificially restricts the candidates to the fully connected $Y$ saddle and the fully disconnected $W$ saddle. The inclusion of the mixed saddle smooths the spike and, in sufficiently asymmetric cases, produces an intermediate plateau.}
	\label{fig:GM_plots-Asymmetric-CABC}
\end{figure}

We illustrate these behaviors in Fig.~\ref{fig:GM_plots-Asymmetric-CABC}, which shows the phase structure of the genuine multi-entropy as a function of the hard-wall depth $z_0$ for several asymmetric configurations.

A universal conclusion is again
\begin{equation}
\GM^{(3)}(A:B:C)=0
\qquad
\text{for } z_0<L_{\min}.
\end{equation}
This is the asymmetric hard-wall realization of the effective junction law.

\FloatBarrier

%%%%%%%%%%%%%%%%%%%%%%%%%%%%

\bibliographystyle{JHEP}
\bibliography{reference}

\end{document}